\documentclass[acmsmall, nonacm]{acmart}
\usepackage{graphicx}
\usepackage{latexsym}
\usepackage{listings}
\usepackage{xcolor}
\usepackage{amsthm}
\usepackage{todonotes}
\usepackage{xspace}
\usepackage{changepage} 
\usepackage{colortbl}
\usepackage{multirow}
\usepackage{enumerate}
\usepackage{textcomp}
\usepackage{graphicx}
\usepackage{latexsym}
\usepackage{listings}
\usepackage{xcolor}
\usepackage{amsthm}
\usepackage{todonotes}
\usepackage{lineno,hyperref}
\usepackage{acro}
\usepackage[capitalise]{cleveref}
\usepackage{caption}
\usepackage{subcaption}

\AtBeginDocument{%
  \providecommand\BibTeX{{%
    \normalfont B\kern-0.5em{\scshape i\kern-0.25em b}\kern-0.8em\TeX}}}

\setcopyright{none}
\copyrightyear{2017}
\acmYear{2017}
\acmDOI{}

\DeclareAcronym{joana}{
	short= JOANA,
	alt = JOANA's,
	long = Java Object-sensitive Analysis
}
\DeclareAcronym{ifc}{
	short= IFC,
	long = Information Flow Control
}
\DeclareAcronym{jvm}{
	short= JVM,
	long = Java Virtual Machine
}
\DeclareAcronym{jre}{
	short= JRE,
	long = Java Runtime Environment
}
\DeclareAcronym{csv}{
	short= CSV,
	long = Comma-Separated Values
}
\DeclareAcronym{wala}{
	short= WALA,
	alt = WALA's,
	long = T.J. Watson Libraries for Analysis
}
\DeclareAcronym{sdg}{
	short= SDG,
	alt = SDG's,
	long = System Dependence Graph
}
\DeclareAcronym{api} {
	short = API,
	long = Application Programming Interface
}
\DeclareAcronym{ssa} {
	short = SSA,
	long = Static Single Assignment
}
\DeclareAcronym{pdg} {
	short = PDG,
	long = Procedure Dependence Graph
}
\DeclareAcronym{vcs} {
	short = VCS,
	long = Version Control System
}
\DeclareAcronym{sca} {
	short = SCA,
	long = Semantic Conflicts Analyzer
}
\DeclareAcronym{cfg} {
	short = CFG,
	long = Control Flow Graph
}
\DeclareAcronym{csg} {
	short = CSG,
	long = Change Sequence Graph
}
\DeclareAcronym{scm}{
	short = SCM, 
	long = Software Configuration Management
}
\begin{document}

\title{USING INFORMATION FLOW TO ESTIMATE INTERFERENCE BETWEEN DEVELOPERS SAME METHOD CONTRIBUTIONS}

\author{Roberto Souto Maior de Barros Filho}
\email{rsmbf@cin.ufpe.br}
\orcid{}
\affiliation{%
  \institution{Centro de Informática, Universidade Federal de Pernambuco}
  \streetaddress{Av. Jorn. Aníbal Fernandes}
  \city{Recife}
  \state{Pernambuco}
  \country{Brazil}
  \postcode{50740-560}
}

\author{Paulo Borba}
\email{phmb@cin.ufpe.br}
\orcid{0000-0002-0381-2843}
\affiliation{%
  \institution{Centro de Informática, Universidade Federal de Pernambuco}
  \streetaddress{Av. Jorn. Aníbal Fernandes}
  \city{Recife}
  \state{Pernambuco}
  \country{Brazil}
  \postcode{50740-560}
}

% \renewcommand{\shortauthors}{Galileu Santos, et al.}
\renewcommand{\shortauthors}{} 

\begin{abstract}
  In a collaborative software development environment, developers often implement their contributions (or tasks) independently using local versions of the files of a system. However, contributions from different developers need to be integrated (merged) to a central version of the system, which may lead to different types of conflicts such as syntactic, static semantic or even dynamic semantic conflicts. The first two types are more easily identifiable as they lead to syntactically incorrect programs and to programs with compilations problems, respectively. On the other hand, dynamic semantic conflicts, which may be caused by subtle dependencies between contributions, may not be noticed during the integration process. This type of conflict alters the expected behaviour of a system, leading to bugs. Thus, failing to detect dynamic semantic conflicts may affect a system's quality. 
%Previous works use tests to detect this conflict category, an approach which is limited by the quality of the project's test suite.
Hence, this work's main goal is to understand if \ac{ifc}, a security technique used for discovering leaks in software, could be used to indicate the presence of dynamic semantic conflicts between developers contributions in merge scenarios. However, as defining if a dynamic semantic conflict exists involves understanding the expected behaviour of a system, and as such behavioural specifications are often hard to capture, formalize and reason about, we instead try to detect 
%a necessary but not sufficient condition for dynamic semantic conflict: 
a code level adaptation of the notion of interference from Goguen and Meseguer. We limit our scope to interference caused by developers contributions on the same method. More specifically, we want to answer if the existence of information flow between developers same-method contributions of a merge scenario can be used to estimate the existence of interference. 
Therefore, we conduct an evaluation to understand if information flow may be used to estimate interference. In particular, we use \ac{joana} to do the \ac{ifc} for Java programs. \ac{joana} does the \ac{ifc} of Java programs by using a \ac{sdg}, a directed graph representing the information flow through a program. As \ac{joana} accepts different options of \ac{sdg}, we first establish which of these \ac{sdg} options (instance based without exceptions) is the most appropriate to our context%by running \ac{joana} and comparing the results for different options
. Additionally, we bring evidence that information flow between developers same-method contributions occurred for around 64\% of the scenarios we evaluated. Finally, we conducted a manual analysis, on 35 scenarios with information flow between developers same-method contributions, to understand the limitations of using information flow to estimate interference between same-method contributions. From the 35 analysed scenarios, for only 15 we considered that an interference in fact existed. We found three different major reasons for detecting information flow and no interference: cases related to the nature of changes, to excessive annotation from our strategy and to the conservativeness of the flows identified by \ac{joana}. We conclude that information flow may be used to estimate interference, but, ideally, the number of false positives should be reduced. In particular, we envisage room for solving around three quarters of the obtained false positives.
\end{abstract}

	

\maketitle

\section{Introduction}
\label{chap:intro}
In a collaborative development environment, developers often perform their contributions (or tasks) independently, using individual copies of project files. Nevertheless, generally, at some point, those contributions need to be integrated to a central repository of a system. Furthermore, when two different developers modify system files in similar periods of time (in parallel), it is necessary to merge the different contributions to generate an integrated version of the system. This process of merging contributions may lead to different types of conflicts. For instance, the merge may lead to a syntactically incorrect program, to a program with compilation failures, or even cause changes on the expected behaviour of a system. From these, we call the first conflict category as syntactic conflicts, the second as static semantic conflicts, and, the third as dynamic semantic conflicts.

To learn about the occurrences of these conflicts created by merging different parallel contributions, previous studies investigated how often each of these conflicts occur. More precisely, although they use a different terminology, Kasi and Sarma \cite{kasi2013cassandra} found that, depending on the project, conflicts similar to syntactic conflicts (merge conflicts identified by a merge tool that treats software artefacts as text) occurred for between 7.6\% and 19.3\% of the merge scenarios (a set consisting of a common ancestor and its derived revisions) evaluated by them. Furthermore, this study also found that conflicts similar to static semantic conflicts (called by them build conflicts), in their sample, ranged from 2.1\% to 14.7\% of the evaluated scenarios. Finally, they found indication that conflicts similar to dynamic semantic conflicts (called by them test conflicts) ranged from 5.6\% to 35\%. Similarly, Brun et al. \cite{brun2013early} found evidence of conflicts similar to syntactic conflicts, static semantic conflicts and dynamic semantic conflicts ranging from 7\% to 42\%, from 0.1\% to 10\%, and from 3\% to 28\% of the merge scenarios, respectively. 

Although these studies have some imprecisions and considered only a small number of projects (Brun et. al considers nine projects for syntactic conflicts and three for semantic conflicts categories, while Kasi and Sarma considers four projects), they indicate that the three types of conflicts may occur in practice. Nevertheless, we argue that dynamic semantic conflicts are harder to detect than the others. More specifically, while syntactic conflicts lead to syntactically incorrect programs and static semantic conflicts to compilation issues, dynamic semantic conflicts cause only variations on the runtime behaviour of a system. As a result, they may pass unnoticed during the integration process, possibly leading to the insertion of bugs on a system, affecting its quality. Furthermore, we argue that the existing strategies have limitations to detect this type of conflict.

Thus, we investigate a new strategy for detecting dynamic semantic conflicts between developers contributions from merge scenarios. More specifically, our idea is checking if \ac{ifc}, a security technique used for discovering leaks in software, could be used to indicate the presence of dynamic semantic conflicts. To be more precise, given a merge scenario with contributions from two different developers, we check if there is information flow between these contributions as an approximation to the existence of interaction between these contributions. Our intuition is that if there is no information flow, then there is no interaction, and, consequently, no dynamic semantic conflict. In contrast, the existence of information flow indicates interactions between the contributions, and, consequently, possible dynamic semantic conflicts. In particular, we use \ac{joana}\footnote{JOANA website - \url{http://pp.ipd.kit.edu/projects/joana/}} \cite{graf2013using,joana14it,graf2015checking}  to do the \ac{ifc} for Java programs. \ac{joana} does the \ac{ifc} of Java programs by using a structure called \ac{sdg}. Basically, a \ac{sdg} is a directed graph representing information flow through a program.

However, defining if a dynamic semantic conflict actually exists involves knowledge of a system's expected behaviour, before and after integrating two developers contributions from a merge scenario. Since such behavioural specifications are often hard to capture, formalize and reason about, we instead try to detect a necessary but not sufficient condition for dynamic semantic conflict: a code level adaptation of the simpler notion of interference, first defined by Goguen and Meseguer \cite{goguen1982security}. In particular, %we argue that interference may be divided in two categories: \emph{desired} and \emph{undesired}. Furthermore, we argue that dynamic semantic conflicts correspond to \emph{undesired} interference, while \emph{desired} interference correspond to situations of desired (and planned) situations of behaviour change. Hence, using the notion of interference we may also detect situations that are not actual dynamic semantic conflicts, they are desired interference. Additionally, 
to reduce the scope of our problem, we focus in interference caused by developers same-method contributions. More specifically, we focus in cases where both merge scenario contributions edit the same (Java) methods. 

Hence, we check if the existence of information flow between developers same-method contributions (in merge scenarios) may be used to estimate interference. More precisely, we want to check if when there is information flow there is also interference. Nonetheless, as \ac{joana} has different options to create \acp{sdg}, we also need to choose one. In summary, we investigate the following research questions:
\begin{itemize}  
	\item{\textbf{Research Question 1 (RQ1)} - \emph{Configuration}: Which \ac{sdg} option is the most appropriate to identify information flow between merge scenario same-method contributions?}
	\item{\textbf{Research Question 2 (RQ2)} - \emph{Severity}: Is there direct information flow between merge scenario same-method contributions? How often?}
	\item{\textbf{Research Question 3 (RQ3)} - \emph{Limitations}: In which situations is there information flow and no interference?}
\end{itemize}

The first question (RQ1) establishes a \ac{sdg} option to execute \ac{joana}. The second question (RQ2) helps us understand if, in practice, information flow occurs between same-method contributions. Finally, with the third question (RQ3) we aim to understand the limitations of using information flow to estimate interference. 

To answer those questions we conducted two different types of analysis: automatic and manual. The first aims to answer which \ac{sdg} configuration (option) is the most appropriate to our context (RQ1) and to find the frequency of information flow between developers same-method contributions (RQ2). Furthermore, we select, from this step, cases containing information flow, to manually analyse in order to answer if there is also interference, and, if not, for what reasons (RQ3). More precisely, our manual analysis focuses in information flow true/false positives of interference. In contrast, we do not analyse true/false negatives. More specifically, we are aware that our analysis may miss valid cases of interference, however, evaluating true/false negatives is out of our scope and is left as future work. 

The remainder of this work is organised as follows. 
 \cref{chap:chap3} presents our motivation, discusses theoretical foundation and details our strategy. \cref{chap:eval} details the evaluation of our strategy. Finally, \cref{chap:conclusion} draws our conclusions, summarises the contributions of this work, and discusses related and future work.

\section{Running Information Flow Control to estimate same-method interference}
\label{chap:chap3}
When two different developers (we call them here \emph{left} and \emph{right}) modify files in common in similar periods of time, it is necessary to merge the different contributions to generate an integrated revision. Once these contributions from different developers are merged into a single version, the integrated version may contain conflicts. More specifically, previous studies bring evidence about the occurrence of dynamic semantic conflicts, and from other types of conflicts such as syntactic and static semantic \cite{kasi2013cassandra,brun2013early}. We focus here on the first type. In particular, these studies use tests to detect this type of conflict, finding occurrences ranging from 3\% to 35\% of the evaluated merge scenarios depending on the project. Nevertheless, we argue, in \cref{chap3:sec1}, that the currently available merge tools and strategies contain limitations to detect this type of conflict. Hence, we investigate a new strategy. As discussed in \cref{chap3:sec2}, deciding if a dynamic semantic conflict exists involves understanding a system's expected behaviour before and after integration of merge scenario contributions from different developers. So, we actually use the concept of interference, first defined by Goguen and Meseguer \cite{goguen1982security}, to predict dynamic semantic conflicts. Additionally, to reduce the scope of our problem, we focus on interference caused by developers same-method contributions. More specifically, we focus on cases where both merge scenario contributions edit the same (Java) methods. Finally, as interference itself is in general not computable, in \cref{chap3:strategy}, we detail our strategy of using \ac{joana} to check for information flow between developers same-method contributions of merge scenarios to estimate interference.

\subsection{The problem of detecting dynamic semantic conflicts}
\label{chap3:sec1}
As pointed out by previous studies \cite{kasi2013cassandra,brun2013early} dynamic semantic conflicts may occur when different contributions from merge scenarios are integrated. Furthermore, failing to detect a dynamic semantic conflict may affect a system's quality, as this type of conflict inserts unexpected behaviour on a system, possibly causing the insertion of bugs. Moreover, we argue that the current existing strategies have limitations to detect this type of conflict. 

\subsubsection*{Dynamic Semantic Conflicts}
As discussed, there are situations where an integrated revision is both syntactically correct and successfully compiling, but due to subtle dependencies between the contributions, there is unexpected behaviour. To illustrate this situation, consider \cref{fig:dynamSemanConf2}. In particular, initially, there is a method \emph{generateBill} that sums the prices of a list of items and places the result in a variable \emph{total} (\cref{base:dynamSemanConf2}). Left changes the initial calculation to add a discount of 10\% of the price of items costing more than 100 (lines 5 and 6 from  \cref{integ:dynamSemanConf2}), while right adds code to calculate the mean of the prices of the items (line 9 from \cref{integ:dynamSemanConf2}).

In the example in question, there is a difference in behaviour after integrating left and right. Specifically, left changed the calculation of the sum of the prices to include the discounts, while right uses the result of this calculation to calculate the mean of the prices. In particular, left changes \emph{total}, and right uses \emph{total}. For instance, consider that the list of items has 3 items, with the following prices: 10, 50 and 300. In right's revision, the calculated \emph{pricesMean} for this case would be 120 ($(10 + 50 + 300) / 3$). However, in the integrated revision, the \emph{pricesMean} would be 110, due to the discount inserted by left. In particular, in the integrated revision, \emph{total} would contain the value 330 ($10 + 50 + 300 - 300 \times 0.1$) after the \emph{for} execution (lines 3-7 from \cref{integ:dynamSemanConf2}), while, in right's revision, \emph{total} would contain 360 ($10 + 50 + 300$) in the same point. As \emph{pricesMean} is calculated based on the value of \emph{total}, the value of \emph{pricesMean} is different in right and integrated.

\subsubsection{Limitations of existing strategies}
\begin{figure*}[t!]
	\begin{subfigure}{\linewidth}
		\centering
\begin{lstlisting}[language=Java,numbers=left,xleftmargin=20pt]
generateBill(List<Item> items) {
    double total = 0;
    for(Item item : items) {
    	total += item.getPrice();
    }  
    
    //...
}
\end{lstlisting}
		\vspace{-1em}
		\caption{Base}
		\label{base:dynamSemanConf2}
	\end{subfigure}
		
	\begin{subfigure}{\linewidth}
		\centering
\begin{lstlisting}[language=Java, numbers=left, xleftmargin=20pt]
generateBill(List<Item> items) {
    double total = 0;
    for(Item item : items) {
        total += item.getPrice();}
        if(item.getPrice() > 100) //left
            total -= item.getPrice() * 0.1; //left
    }  

    pricesMean = items.size() > 0 ? total / items.size() : 0; //right
    //...
}
\end{lstlisting}

		\vspace{-1em}
		\caption{Integrated}
		\label{integ:dynamSemanConf2}
	\end{subfigure}
	\caption{Merge scenario containing dynamic semantic conflict}
	\label{fig:dynamSemanConf2}
\end{figure*}
We divide the existing strategies in two groups with respect to how they deal with dynamic semantic conflicts: \emph{unaware} and \emph{aware}. In the first group, we put strategies such as textual and structured (semi-structured) merge tools. We consider such merge strategies to be syntactic, as they have no knowledge of the behaviour of a program. Hence, they do not detect dynamic semantic conflicts, which may lead this type of conflict to pass unnoticed during the integration process, possibly causing the insertion of bugs in a system.
%Both of them, may identify some dynamic semantic conflicts by chance, but are not focused on this type of conflict. More specifically, as a textual merging tool treats the software source code as text (and has no awareness of the program semantics, or even syntax), it is only able to identify dynamic semantic conflicts in very specific cases where the editions that lead to the conflict occurred in the same areas of the text. For example, if left and right edit the same line and those editions lead to a dynamic semantic conflict, then a textual merging tool would be able to identify this specific conflict. Similarly, although structured tools have knowledge about a program's syntax, they have little knowledge about its dynamic semantics. Specifically, because of their knowledge about a program's syntax, they are able to identify some static semantic conflicts, such as when two methods with the same signature are added to a Java program. Nevertheless, with respect to identifying dynamic semantic conflicts they have little use. In particular, in the best case, as the textual tools, they may only identify some specific cases by chance. 
For instance, consider again the example of \cref{fig:dynamSemanConf2}. None of these tools would identify the dynamic semantic conflict discussed in \emph{pricesMean}. For this case, a textual tool would not be able to identify the conflict, because left and right changed different areas of the text (left changed lines 5 and 6 and right line 9 from \cref{integ:dynamSemanConf2}). Similarly, a structured tool would not identify the conflict because different elements were edited. To be more specific, left included an \emph{if} statement (in lines 5 and 6 from \cref{integ:dynamSemanConf2}) while right included an assignment to \emph{pricesMean} (line 9 from \cref{integ:dynamSemanConf2}), which is a different statement from the \emph{if} added by left.

In the second group, we place strategies such as code review, tests and semantic merging, discussed below. 

\paragraph{Code review}
The first of these, code review, although useful for detecting some dynamic semantic conflicts, is time-consuming and error-prone, specially for large software. Particularly, it is a good strategy to be used as a complement to other strategies, but may miss a considerable number of conflicts if used as the only strategy for detecting dynamic semantic conflicts. 

\paragraph{Tests}
Similarly, tests are useful for detecting dynamic semantic conflicts, but the amount of conflicts detected depends on the test suite quality and coverage. More precisely, projects with good test suites will tend to detect more conflicts than projects using tests with bad coverage and/or quality. For example, consider again \cref{fig:dynamSemanConf2}. For this case, the detection of the dynamic semantic conflict in question depends on the existence of a test which checks not only the \emph{pricesMean}, but which checks it with at least one item from the list with the price above 100. To be more specific, the modified behaviour caused by the conflict manifests only when there are items with prices higher than 100. Therefore, a test which checks the \emph{pricesMean} only for items with prices equal or less than 100, will pass and, as a consequence, will not be able to detect the conflict. So, only a test which includes items above 100 will be able to detect the conflict from this example. 

\paragraph{Semantic merging}
Finally, semantic merging tools exist specifically to detect this kind of conflict. Nonetheless, these tools are still theoretical, and to the best of our knowledge there is no existing merging tool of such kind available (although some syntactic ones claim to be semantic). Specifically, most tools of such kind have scalability issues \cite{mens2002state}. For instance, one of the problems from strategies such as the one first proposed by Horwitz, Prins and Reps \cite{horwitz1989integrating} and later extended by Binkley, Horwitz and Reps \cite{binkley1995program} is to scale the structure proposed by them (\ac{pdg} and \ac{sdg}, respectively), specially because in their case they propose to create a full program \ac{sdg} for each revision of the merge scenario (base, left and right) and then merge the \acp{sdg} to generate a fourth one (the integrated one). 

%Similar to semantic merging, our strategy also uses \acp{sdg} to represent the information flow through a program. Nonetheless, we try to mitigate part of the performance issues by using a single \ac{sdg} (instead of four),

\subsection{Using interference to predict dynamic semantic conflicts}
\label{chap3:sec2}
%%voltar aqui
In the previous section we argued that dynamic semantic conflicts occur and that the existing strategies contain limitations to detect them. More precisely, a dynamic semantic conflict occurs when part of a program's behaviour was correct before integrating a merge scenario and it is not correct any more in the integrated revision. Therefore, it is necessary to understand the expected behaviour of a system and the code resulting from a merge scenario contributions in order to establish if a dynamic semantic conflict exists. We instead try to detect a necessary but not sufficient condition for dynamic semantic conflicts: a code level adaptation of the simpler notion of interference, first defined by Goguen and Meseguer \cite{goguen1982security}.% where
%\begin{itemize}
%	\item[]{``one group of users, using a certain set of commands, is \underline{noninterfering} with another group of users if what the first group does with those commands has no effect on what the second group of users can see.'' \cite{goguen1982security}}
%\end{itemize}

We chose that definition because dynamic semantic conflicts are precisely caused by unplanned interference between contributions from different developers. In fact, interference among developers tasks might be desired and planned, but, when it is not, we have dynamic semantic conflicts. So, detecting interference might be a good predictor for detecting dynamic semantic conflicts. We adapt this notion to our context by considering that a merge scenario has always two ``groups of users'': \emph{left} and \emph{right}. Additionally, we consider that the ``set of commands'' used by each group corresponds to the contributions from each of them on the merge scenario. In summary, we want to check if there is interference between the two sets of merge scenario contributions and use this information as a predictor of dynamic semantic conflicts. In particular, we consider that \emph{left} interferes with \emph{right} if \emph{left's} contribution affects or changes the behaviour of \emph{right's} contribution. Similarly, \emph{right} interferes with \emph{left} if \emph{right's} contribution affects the behaviour of \emph{left's} contribution. In a more general way, the notion of interference identifies situations of side effects on the behaviour of the integrated revision of a merge scenario caused by the interaction between the contributions.

Since interference is itself non-computable, the general idea of this work is using program analysis \cite{nielson2005principles} techniques to estimate interference and therefore predict dynamic semantic conflicts. With the goal of using interference to predict dynamic semantic conflicts in mind, it is necessary to better understand the relation between interference and dynamic semantic conflicts, and then establish the types of interference we focus and their characteristics. 

\subsubsection{The connection between interference and dynamic semantic conflicts}
\label{interference}
\begin{figure*}[t!]
%	\begin{subfigure}{\linewidth}
%		\centering
%		\lstinputlisting[language=Java,numbers=left,xleftmargin=20pt]{snippets/baseDominates.java}
%		\vspace{-1em}
%		\caption{Base}
%		\label{base:interfNoconf}
%	\end{subfigure}
	
%	\begin{subfigure}{\linewidth}
		\centering
\begin{lstlisting}[language=Java,numbers=left,xleftmargin=20pt]
static boolean dominates(State thisState, State other) {
    % Multi-state - no domination for different states
    if(thisState.isBikeRenting() != other.isBikeRenting())
        return false;
    if(thisState.isCarParked() != other.isCarParked())
        return false;
    if(thisState.isBikeParked() != other.isBikeParked()) //left
		return false; //left
    // ...
        
    return updatedCalculation(thisState, other); //right 
}
\end{lstlisting}
		\vspace{-1em}
%		\caption{Integrated}
%		\label{merged:interfNoconf}
%	\end{subfigure}
	\caption{Merge scenario containing interference, but not necessarily dynamic semantic conflicts}
	\label{fig:interfNoconf}
\end{figure*}

The notion of interference represents side effects on the behaviour of the integrated revision of a merge scenario caused by the interaction between the contributions, while dynamic semantic conflicts represent situations of side effects on the behaviour of the integrated revision that causes it to diverge from the expected behaviour.
In summary, we argue that every dynamic semantic conflict involves an interference, but not every interference necessarily leads to a dynamic semantic conflict.

We name these situations of interference with no dynamic semantic conflict as \emph{desired interference} and the situations where interference and dynamic semantic conflicts coincide as \emph{undesired interference}. More specifically, if an interference occurs on a merge scenario, but the integrated revision behaves as expected, then there is no dynamic semantic conflict associated and the interference is \emph{desired}. On the other hand, if an interference occurs and the integrated revision behaviour diverges from the expected, then a dynamic semantic conflict also occurred and the interference is \emph{undesired}. 

For instance, \cref{fig:interfNoconf} illustrates a situation, inspired on a merge commit from OpenTripPlanner,\footnote{OpenTripPlanner GitHub page - \url{http://github.com/opentripplanner/opentripplanner}} where there is interference, but not necessarily a dynamic semantic conflict. In this example, there is a method called $dominates$, which compares two states ($thisState$ and $other$) and returns a $boolean$ informing if one of them dominates the other. In the illustrated scenario, left added a conditional with $return$  $false$ (lines 7 and 8), while right updated the final domination calculation by calling a new method (line 11). 

%For instance, \cref{fig:interfNoconf} illustrates a situation, inspired on a merge commit from OpenTripPlanner,\footnote{OpenTripPlanner GitHub page - \url{http://github.com/opentripplanner/opentripplanner}} where there is interference, but not necessarily a dynamic semantic conflict. In this example, there is a method called $dominates$, which compares two states ($thisState$ and $other$) and returns a $boolean$ informing if one of them dominates the other. In the illustrated scenario, left added a conditional with $return$  $false$ (lines 7 and 8 from \cref{merged:interfNoconf}), while right updated the final domination calculation by calling a new method (line 9 from \cref{base:interfNoconf} and 11 from \cref{merged:interfNoconf}). 

In this example, there is interference from left on right, because left's control flow interferes with right's. In particular, when the $return$ $false$ added by left (line 8) is executed, the method returns and right's contribution (line 11) is not executed. In other words, there are situations where the method call modified by right ($updatedCalculation$) was executed before the integration with left, and is not executed after the integration with left's contribution. 

Nonetheless, if the overall method requirement (expected behaviour) is that there is no domination calculation for different states, then, even though there is a side effect on the method's behaviour (there is interference), there is no dynamic semantic conflict (the interference is \emph{desired}), as left's contribution conforms to this requirement. More precisely, there were two tests of different states (lines 3-4 and 5-6 for fields $isBikeRenting$ and $isCarParked$ respectively), and left only added a third test for a new field ($isBikeParked$). 

Alternatively, the method requirement could be that for the added field ($isBikeParked$), domination calculation should be done even for different states. In that case, left would be inappropriately affecting right's contribution, as the method call changed by right to do the updated calculation (\emph{updatedCalculation} in line 11) is not executed for different states due to the conditional added by left (lines 7 and 8).

To conclude, given an interference exists, defining if there is also a dynamic semantic conflict (if it is a \emph{desired} or \emph{undesired} interference) requires knowledge of the expected behaviour of the system being evaluated, its requirements, and the specification of the developers tasks that will create contributions to be later integrated. Since such behavioural specification are often hard to capture, formalize and reason about, we only focus on detecting interference to predict the existence of dynamic semantic conflicts, but we do not evaluate if an interference is \emph{desired} or \emph{undesired}. In other words, we detect interference, but we do not check if it actually leads to a dynamic semantic conflict, as this information requires knowledge of the requirements for each system being evaluated. However, we argue that existing automatic strategies of dynamic semantic conflicts detection, work similarly. In particular, tests and semantic merge detect interference, not dynamic semantic conflicts directly. More precisely, both of these strategies may also detect \emph{desired} interference. %In summary, given an interference, to conclude if a dynamic semantic conflict in fact exists, code review from system's experts is necessary.

\subsubsection{Selecting same-method interference from different interference patterns}
\label{sameMethod}
As previously discussed, we focus on detecting interference instead of dynamic semantic conflicts. Nevertheless, we expect the existence of different patterns of interference for different program languages. For example, for Java, we expect at least two patterns: 
\begin{itemize}
	\item{Same-method - Both contributions edit the same method.}
	\item{Dependency of modified method  - One contribution, directly or indirectly, depends on a method edited by the other. For instance, left adds a call to a method modified by right.}
\end{itemize}
\cref{fig:dynamSemanConf2,fig:interfNoconf} illustrate situations where both contributions edited the same method (\emph{generateBill} and \emph{dominates}, respectively). An example of the second pattern may be illustrated in a situation where \emph{left} adds a call to a method \emph{foo} inside method \emph{bar}, while \emph{right} modifies the calculation of the value returned by \emph{foo}. More specifically, imagine a situation where the value returned by \emph{foo} used to be 1 and is changed to 2 due to right's modification. In such a situation there is an interference from right on left, as the value returned by \emph{foo} used to be 1, for this situation, when left added the call to \emph{foo} and is changed to 2 after the integration.

It is important to notice that the described patterns occurrences do not represent the actual cause of an interference. More specifically, the described patterns only represent general situations which may lead to an interference. In fact, both patterns may include more specific situations as the actual cause that leads to interference. For instance, both patterns may actually happen because left's contribution reads a value from a variable (or field) modified by right or because right overwrites a value also modified by left. More precisely, the first situation may occur when left changes the value of a variable, $x$ for instance, while right uses the modified value after the integration. There is interference in such a situation because the value expected by right was changed by left. The second may occur when left changes the value of a variable on a point 1 of the integrated revision used by a statement on point 3, while right modifies the value of the same variable on a point 2 between left modification from point 1 and the statement from point 3. There is interference in such a situation because right overwrites the value also modified by left affecting the value of the variable on point 3. Similarly, interference may also happen due to different situations such as editions to \emph{annotations} or \emph{modifiers}.

To prioritize and reduce the scope of our problem of detecting interference, we decided to individually search for occurrences of one of these patterns instead of trying to deal with the concept as a whole. In particular, we here focus on the \emph{same-method} pattern to try to detect interference arising from developers contributions to Java programs. 

We decided to start with this pattern for a number of reasons. Firstly, it was the simpler one to automatically identify. It requires only identifying the methods edited by each contribution and then checking for methods edited by both contributions. In contrast, the second pattern would require not only identifying methods edited by each contribution, but also checking the call graph \cite{ryder1979constructing,grove2001framework} of these methods. That is, the second pattern involves checking if a method edited by one contribution is in the call graph from the other. Secondly, each \emph{same-method} pattern occurrence involves only a single method, which may be easily isolated from the rest of a system. More precisely, it only requires a single method as the starting point for performing interference detection. This aspect is important to reduce the scope of our analysis. Lastly, Accioly \cite{paolaPhd} provided evidence that this pattern (same-method contributions) occurs in practice. That is, using a sample of 128 Java projects, this study found considerable occurrences of same-method contributions.

%As the \emph{same-method} pattern may be isolated from the rest of a system (it requires only a single method as the starting point for performing interference detection), we here focus on this pattern to try to detect interference arising from developers contributions to Java programs. 
%Marcelo - Melhorei explicação de merge conflicts que não tinha ficado clara
Nonetheless, Accioly also shows, using a semi-structured tool, that same-method contributions occur on around 80\% of merge conflicts in Java files \cite{paolaPhd}. Because this is a high frequency and also because we are interested in detecting dynamic semantic conflicts, which are not included in those merge conflicts detected by a typical merge tool (such as a textual or structured/semi-structured tool), we decided to check if there was also a considerable number of this pattern occurrences without merge conflicts detected by a typical merge tool. 

Thus, with the purpose of better understanding the selected pattern, we estimated the percentage of scenarios with merge conflicts, identified by a semi-structured tool, when there are \emph{same-method} contributions. Using a sample of 64128 scenarios from 119 GitHub Java projects (used sample is detailed in \cref{subsect:sample}) we found the pattern (both contributions edited one method in common) in 4239 scenarios. From these scenarios with occurrence of the pattern, 3028 contained merge conflicts due to editions on the same areas of the same method (syntactical conflicts), which represents around 71\% of the scenarios found with the pattern (3028 from 4239). Therefore, although editions to the same method with syntactical conflicts are frequent, a considerable number of scenarios with the pattern occurrence (1211 from 4239 or 29\%) have no syntactical conflicts but could have dynamic semantic conflicts and interference. More precisely, those scenarios with no merge conflicts identified by a typical tool are potential candidates for our analysis of interference between same-method contributions. It is important to mention that we did not count non-conflicting occurrences of the pattern in conflicting scenarios with other patterns of merge conflicts. More specifically, there are situations where left and right edit the same method \emph{foo} and there is no merge conflict identified by a typical tool on this method, but the tool may identify other merge conflicts outside the method for the scenario in question. We did not count these situations as occurrences of the pattern and, as a result, the obtained percentage of 71\% may be viewed as a ceiling of the percentage of merge conflicts detected by typical tools on scenarios with the same-method pattern.

\subsubsection{Perspective of interference: open-world vs closed-world}
\label{perspective}
%\begin{figure*}[t!]
%	\begin{subfigure}{\linewidth}
%		\centering
%		\lstinputlisting[language=Java,numbers=left,xleftmargin=20pt]{snippets/baseDecoder.java}
%		\vspace{-1em}
%		\caption{Base}
%		\label{base:openNotclosed}
%	\end{subfigure}
%	
%	\begin{subfigure}{\linewidth}
%		\centering
%		\lstinputlisting[language=Java,numbers=left,xleftmargin=20pt]{snippets/finalDecoder.java}
%		\vspace{-1em}
%		\caption{Integrated}
%		\label{integ:openNotclosed}
%	\end{subfigure}
%	\caption{Interference from open-world perspective, but not from closed-world}
%	\label{fig:openNotclosed}
%\end{figure*}

As just explained, with the intention of reducing the scope of our problem of detecting interference, we focus on same-method interference. Nevertheless, while Goguen and Meseguer's notion of interference is focused on side effects on the global behaviour of a system, there are also situations of local side effects that do not affect the global behaviour of a system. %For instance, consider the example from \cref{fig:openNotclosed} inspired on a merge scenario from Netty.\footnote{Netty GitHub page - \url{http://github.com/netty/netty}} For this example, there is no interference according to Goguen and Meseguer's notion, but there is a possible local side effect on the behaviour of method \emph{decode} (lines 7-13 from \cref{base:openNotclosed} and lines 12-21 from \cref{integ:openNotclosed}). 

Hence, with the view of also covering these local situations, we associate the notion of interference with the concept of perspective of the analysis. More precisely, we present two different perspectives of interference: \emph{open-world} and \emph{closed-world}. The first perspective assumes the changed method might be called in arbitrary contexts, not restricted to the calling context of an specific system. It is viewed as part of an \ac{api}, an open-world, not of a single system and the closed-world it imposes. On the other hand, the second perspective is restricted to the calling context of the system being evaluated. In summary, an open-world perspective considers every side effect on the behaviour of a method as interference, while a closed-world, as Goguen and Meseguer's notion, considers only side effects that affect behaviour on the system level. %Therefore, for our example, from an open-world perspective there is interference, while from a closed-world there is no interference. In particular, while an open-world perspective would consider interference because of potential side effects on the behaviour of method \emph{decode} when \emph{lengthAdjustment} is \lstinline[basicstyle=\listingsfont\itshape]|true|, a closed-world perspective would not consider interference as \emph{lengthAjustment} is always \lstinline[basicstyle=\listingsfont\itshape]|false| in the calling context of this program and consequently there will never be any actual side effect on global behaviour. 

To conclude, a closed-world perspective requires a deeper analysis and establishing the expected behaviour of a system, while an open-world perspective classifies every side effect on a method's behaviour as interference. Given the fact that a closed-world perspective is more expensive (computationally speaking), as it considers the global behaviour of a system, we detect interference using an open-world perspective even though it has the potential of also identifying local side effects on behaviour that might be not actually relevant.

\subsection{Executing Information Flow Control to estimate same-method interference}
\label{chap3:strategy}
As discussed in the previous section, we want to check for same-method interference using an open-world perspective. However, as interference itself is in general not computable, we need a way to estimate interference. Our general strategy is to use \ac{ifc} to check for information flow between merge scenario same-method contributions, and use this information to estimate interference between the contributions. Our intuition is that the existence of information flow indicates interactions between the contributions, and, consequently, possible interference. 

To illustrate our strategy, in \cref{fig:interfNoconf}, as method \emph{dominates} was edited by left and right, we want to check if there is information flow between these contributions (between lines 7 and 8, and line 11). More specifically, we check if left (lines 7 and 8) has information flow to right (line 11) and also if right (line 11) has information flow to left (lines 7 and 8). In this example, we detect flow between left and right and as a result we establish that this method contains information flow between its contributions and a possible interference.

%Marcelo - Porque não intraprodecural?
%Uirá - Porque o JOANA? Porque não data-flow?
%Marcelo - terminologia... IFC versus static dependency tracking (tainting)
We use \ac{joana} to do the \ac{ifc} of Java programs. As a tool from a security context, \ac{joana} is conservative and tries to avoid false negatives as much as possible by presenting some false positives. \ac{joana} is \textbf{interprocedural}, which means that methods calls are taken in account when analysing a program. That type of analysis is more advanced than an \textbf{intraprocedural} analysis, which considers only a single method/procedure and does not take in account methods/procedures called from it. Additionally, %as explained in \cref{chap:background}, 
\ac{joana} considers data and control flow in its \acp{sdg}. We could use more basic approaches, such as doing only intraprocedural analysis or considering only data flow (instead of also considering control flow). However, we would tend to present more false negatives using such simplified approaches. That is, we would miss more valid cases of information flow. Furthermore, \ac{joana} is able to handle full Java programs of up to 100KLOC with arbitrary threads (without reflection). In addition, it is important to mention that a taint analysis is probably more appropriate to describe the type of analysis that we do here as we only deal with one security level (high/low, or, more precisely, left/right). Nevertheless, as we use \ac{joana} and as \ac{joana} does the \ac{ifc} of Java programs, we reefer in this work to \ac{ifc}.

\subsubsection{Strategy}
\label{chap3:strat}
%\begin{figure}
%	\centering
%	\includegraphics[scale=0.5]{imgs/strategy_flow.png}
%	\caption{Strategy order}
%	\label{fig:strategyFlow}
%\end{figure}
Given a merge scenario with no merge conflicts identified by a syntactical merge tool and a group of methods edited by both merge scenario contributions, we run \ac{joana} to check if there is information flow between contributions and use the result to estimate the existence of interference. In summary, if \ac{joana} detects information flow between contributions of a merge scenario we consider that there is interference between them, as information flow between the contributions indicates that they interact. In other words, the existence of information flow between contributions indicates that one contribution may affect the other. In contrast, if \ac{joana} does not detect any information flow then we conclude that there is no interference between the contributions. 

More precisely, \ac{joana} is executed as an approximation of the actual set of information flow existence. In particular, \ac{joana} may present false positives (and negatives) with respect to information flow existence. As a matter of fact, as a conservative tool, \ac{joana} tends to present more false positives than false negatives. That is, in general, if there is information flow, \ac{joana} is usually able to detect it. Nevertheless, to do so, \ac{joana} also finds a number of spurious cases that are not information flow. Furthermore, as previously discussed, our intention is using the existence of information flow to estimate the existence of interference, which may or may not represent a dynamic semantic conflict. 

%As can be noticed in \cref{fig:strategyFlow}, our 
Our strategy is divided in three parts: a \emph{merge and contribution identification step}, a \emph{build step} and an \emph{IFC step}. The first is responsible for merging the scenario and generating an integrated revision, which may contain merge conflicts. %For instance, in the example from \cref{fig:openNotclosed}, this step obtains \cref{integ:openNotclosed} after merging the scenario (after integrating base, left and right). 
The second step compiles the integrated revision because it is necessary to build (compile) the system since \ac{joana} needs the resultant \emph{class} files to run. As a result, when the compilation fails, the final step is not executed. Lastly, the third step executes \ac{joana} (\ac{ifc}) on the integrated revision with the objective of estimating the occurrence of interference.

%Marcelo - qual seria o workflow para usar a ferramenta? 
%motivar no final… rodar em backgorund
%Uirá - discutir possível aplicação na prática
In practice, we envisage our strategy as a complementary verification that would be activated in the merge process. That is, for each merge scenario integrated with no conflicts, our strategy could be automatically executed in a server to check for information flow. After the verification was completed, an email would be sent detailing if any information flow was identified between the contributions of the scenario and if so, detailing the obtained flows. Given information flow is reported, a developer could review the code looking specifically the points where information flow was identified to reason if the identified flows lead or not to dynamic semantic conflicts (unexpected variations on the system behaviour). 

\subsubsection*{Merge and contribution identification step}
\label{subsection:mergeStep}
The merge and contribution identification step merges contributions using a non-semantic strategy, for example textual or \textit{semi-structured/\newline structured}. We choose a non-semantical approach at this stage because the semantic part is checked at the \emph{IFC step}. In particular, we use FSTMerge \cite{apel2011semistructured} as our merge tool. As any merge tool, this tool is responsible for integrating contributions from a merge scenario and checking if they conflict. Additionally, previous work \cite{cavalcanticomparing} has found advantages on using semi-structured merge tools, such as FSTMerge, when compared to textual merging. For instance, this work shows that semi-structured merge tends to detect less false positives and negatives of merge conflicts. Additionally, due to our focus on methods edited by both contributions, we also need to know the methods in which this situation occurs. FSTMerge is able to find these methods, because it is a semi-structured tool and thus understands the program structure and represents each method as a node in the parse tree. 

Nevertheless, in order to execute our \ac{ifc} analysis we also need to know which lines, from methods modified by both contributions, were edited by each of the merge scenario contributions. To obtain these lines, for each method edited by left and right, we modified FSTMerge to call a routine which executes Git Blame\footnote{Git Blame documentation - \url{http://git-scm.com/docs/git-blame}} on the integrated revision and checks for each method line if the last commit that edited it was left or right. More precisely, for each method line our invocation of Git Blame returns the last commit that edited it. For example, for the method \emph{generateBill} from \cref{integ:dynamSemanConf2}, lines 5 and 6 were last edited by left, line 9 by right and the rest by base or previous commits.  %\emph{decode} from \cref{integ:openNotclosed}, lines 14 and 16 were last edited by left, line 19 by right and the rest by base or previous commits. 
Next, if the last commit associated to a line is left (or right) our routine will associate this line to a list of lines edited by left (right). Consequently,  after the execution of this routine, we obtain the list of lines edited by left and the list of lines modified by right for a specific method.
%For instance, continuing our example from \cref{integ:openNotclosed}, the left list of lines would contain lines 14 and 16, while the right list would contain line 19. 

It is important to notice that this approach of using Git Blame on the integrated revision to identify the edited lines has limitations. The main one is that, because we check the integrated revision, removed lines are not present any more. They were already removed and as a result we do not identify them. Additionally, there are situations of identical lines added by both left and right in the exact same place. These situations of identical lines are a corner case of this approach, as Git Blame typically \emph{blames} one of them, either left or right. However, as both left and right added the line, \emph{blaming} only one of them is incorrect and may affect the analysis. Hence, we also specifically check for situations of identical lines and when they occur we do not include those lines in the lists of edited lines. We are aware that the more intuitive possibility would be considering lines on this situation on both lists of edited lines. Nevertheless, as we check for information flow between source code lines, this would not make sense. More specifically, information flow checks for flows between high and low elements, and, marking a line as edited by left and right would lead to marking it as both high and low, which is not possible. In particular, this would correspond to checking for information flow from the line in question to itself. In summary, a line can only be on the list of edited lines of left or right but not both. Consequently, we decided to not include lines on this situation on the lists. For instance, consider a hypothetical example where left added a new line on line 2 of the integrated revision, right changed line 6, and both added line 4. In this situation, we do not consider line 4, as both contributions added an identical line on the same position. We assume both developers are aware of flows between 4 and the other lines, not causing interference. As a result, we consider that left edited only line 2 and right only line 6. 

At the end of this phase, the merge scenario is integrated, existing merge conflicts are reported, and methods edited by both left and right, as well as the respective edited lines, listed. %For instance, considering \cref{fig:interfNoconf}, \cref{integ:interfNoconf} is obtained, no merge conflicts exist, and method \emph{dominates} was edited by left and right (left edited lines 7 and 8, and right line 11). 

\subsubsection*{Build step}
As already explained, \ac{joana} needs a system's class files to execute. Thus, at this step, we build the scenario integrated revision, to compile and generate the class files from the source files. 

Nevertheless, this step is only executed when the merge and contribution identification step identifies that the scenario being analysed has no merge conflicts but has methods edited by left and right. We only execute this step for scenarios without merge conflicts, because we are interested in verifying situations which are not identified by traditional merging strategies. That is, as previously explained, our final goal is detecting dynamic semantic conflicts. In other words, we focus on analysing if cases which are integrated with no problems may contain interference. Furthermore, since we focus on detecting same-method interference, we only try to build (and later run the \ac{ifc} step) the integrated revision when the merge scenario contains methods edited by both contributions.

\subsubsection*{IFC step}
\label{ifcStep}
The \ac{ifc} step is only executed when the build step is able to successfully compile the integrated revision of the scenario. More precisely, the integrated revision class files are necessary because \ac{joana} needs them for its analysis and, as a result, when the integrated revision compilation fails, it is not possible to execute this step.

%\begin{figure}
%	\centering
%	\includegraphics[scale=0.5]{imgs/ifcStep.png}
%	\caption{IFC step order}
%	\label{fig:ifcStep}
%\end{figure}

%As illustrated in \cref{fig:ifcStep}, 
This step can be divided in three parts. First, when the integrated revision is successfully compiled, its \ac{sdg} representation is created or loaded. Since the \ac{sdg} is the representation of the program information flow, when it is not successfully created/loaded our process is interrupted. Then, when the \ac{sdg} is successfully obtained, for each method edited by left and right, we execute the \ac{ifc} analyses. Finally, when information flow exists it is necessary to translate the obtained flows between elements of \aca{joana} internal representation to flows between source code lines.  %%One to check direct information flow between contributions (from left to right and from right to left) and the other to verify if there is indirect information flow from the contributions to common points. %%Finally, we map existing flows between \ac{sdg} nodes to flows between source code lines. 

\paragraph{Create/load \ac{sdg}}
\label{createSdg}
Before executing any \ac{ifc} analysis with \ac{joana} it is necessary to create a \ac{sdg} representing the program information flow and as a matter of fact this phase is the performance bottleneck of our strategy. Therefore, for each integrated revision of a merge scenario, we work with a single \ac{sdg} and run different \ac{ifc} analyses on it. Moreover, as we previously mentioned, there are cases where the \ac{sdg} representation of a program may fail to be created. This may happen for a number of reasons, but the most common of them is when the \ac{sdg} being created is too heavy, with respect to memory or time, to be computed. These situations are a limitation of our approach, since, when they occur, the corresponding scenario can not be analysed as there is no \ac{sdg} representing its information flow.

Semantic merging using \acp{sdg} tend to deal with four \acp{sdg} (for base, left, right and integrated) \cite{horwitz1989integrating,binkley1995program}. However, as this approach is too heavy, we chose to only create a single \ac{sdg}: for the integrated revision. For this reason, our analysis may miss cases potentially identified by a semantic merging. For example, removed behaviour, that is, situations where instructions and their associated behaviour are removed, may also cause interference because it may cause difference on the information flow. For instance, removing a variable increment affects future uses of the variable. Nevertheless, since we only create a \ac{sdg} for the integrated revision and since every removed instruction is not present any more at that point, we do not identify this type of situation.

%As mentioned in \cref{chap:background}, 
To create a \ac{sdg} using \ac{joana}, a group of options needs to be defined such as \emph{use of concurrency}, \emph{use of exceptions}, \emph{pointer analysis}, \emph{dependencies} and \emph{entry-points}.
%The process of \ac{sdg} creation in \ac{joana} includes a set of options. One of these is the \emph{entry-point}, which represents the starting point of the analysis. Additionally, there are other aspects to be considered such as \emph{inclusion of exceptions (and concurrency)}, \emph{pointer analysis} and information \emph{regarding the program's dependencies}. 

%\paragraph{Entry-point}
%To create the call graph (and then the \ac{sdg}) of a program, \ac{joana} (and \ac{wala}) requires the definition of an entry-point for that program. An entry-point is basically a method representing the starting point for the analysis. For Java programs this entry-point will typically be the \emph{main} method. %For instance, consider \cref{fig:cgExamp}. This example, shows a small program in \cref{fig:cgProg} and its corresponding call graph in \cref{fig:cg}. The creation of a call graph starts by the entry-point. In particular, the \emph{main} method in our example. This method, calls 3 methods (\emph{foo} and \emph{bar} in line 3 and \emph{print} in line 4) and as a result there are edges from \emph{main} to each of them in \cref{fig:cg}. Furthermore, as there is a call in \emph{bar} to \emph{bar2}, there is also an edge between these methods. 

For performance reasons we do not use concurrency, as using it is computationally expensive \cite{hammer2009flow}. The positive side of this decision is that we avoid excessively increasing the size of the \ac{sdg}. The negative one is that we may miss some cases which would be identified using this option. 

As discussed by Hammer and Snelting \cite{hammer2009flow}, dynamic runtime exceptions can alter the control flow of a program and thus lead to implicit flow, in case the exception is caught by some handler on the call-stack. Alternatively, exceptions may create a covert channel in case the exception is propagated to the top of the stack, yielding a program termination with stack trace. Hence, \ac{joana} has options to integrate exceptions. However, we decided to use \acp{sdg} without expections. The reason for this decision is later detailed in \cref{chap:eval}.

In case a Java program creates objects, any precise program analysis (such as \ac{joana}) must run a pointer analysis (or points-to-analysis) \cite{smaragdakis2015pointer} first. The goal of pointer analysis is to compute an approximation of the set of program objects a pointer variable or expression can refer to during runtime. \ac{joana} has eight different pointer analyses: \emph{type based}, \emph{instance based}, \emph{n1 object sensitive}, \emph{object sensitive}, \emph{unlimited object sensisitive}, \emph{n1 call stack}, \emph{n2 call stack} and \emph{n3 call stack}. From this, we selected \emph{instance based} as our pointer analysis and we also detail this decision later in \cref{chap:eval}.  

%Hence, \ac{joana} has options to integrate exceptions. To do that, basically, \ac{joana} adds control flow edges from instructions that may throw exceptions to an appropriate exception handler.

%Additionally, we decided to use \acp{sdg} without exceptions and with an \emph{instance based} pointer analysis. The reasons for these decisions are later detailed in \cref{chap:eval}. 

Another important aspect to take in consideration when creating a \ac{sdg} is the dependencies. \aca{joana} process of \ac{sdg} computation involves creating a class hierarchy. This hierarchy contains system compiled classes, but classes from dependencies are also added to the class hierarchy created by \ac{joana}. Basically, there are two types of dependencies to take in consideration: external and native dependencies. 

By external dependencies, we refer to external systems/libraries which a system relies on. More precisely, these external dependencies are not part of the source code of a system, but without them the system does not work. To be able to perform a complete analysis, \ac{joana} accepts \emph{jar} files with dependencies code. Nonetheless, it is important to mention that, generally, \ac{joana} is able to perform the analysis without external dependencies. However, in that case, the analysis is incomplete, as it does not consider the dependencies code.

In addition, \ac{joana} accepts \emph{stubs} for different \acp{jre}. Basically, these stubs include predefined models of native methods deep down in the Java standard library. There are stubs for \ac{jre} 1.4 and 1.5. Additionally, it is also possible to opt for a configuration without stubs. As in the case of external dependencies, in general, \ac{joana} is able to create the \ac{sdg} without stubs, but the analysis in question would be incomplete.

Finally, the entry-point is basically a method representing the starting point of the analysis. In particular, the natural entry-point of a Java program is its \emph{main} method.  Nevertheless, for some programs a \emph{main} does not exist or multiple \emph{mains} exist. For these cases, a different approach is necessary. As discussed in \cref{sameMethod}, we are initially focused on same-method interference, and therefore, we select all methods edited by both contributions as entry-points, even when a single main exists. This option solves the problem of the \emph{main} existence and also helps to limit the scope of the analysis, more specifically the \ac{sdg} size. %% However, this decision has a trade-off associated between improving performance (by limiting the \ac{sdg} size) and loosing some cases of information flow (and indirectly of interference) as we do not analyse the whole program. Actually, the effect of this decision depends on the methods position in the call hierarchy of the program. For instance, if a method edited by both contributions is on top of the hierarchy (the \emph{main} of the program for example) the \ac{sdg} will be more detailed and miss fewer cases of information flow. On the other hand, lower methods in the hierarchy will tend to have smaller \ac{sdg} sizes and consequently miss more cases of information flow (and of interference).
More precisely, we focus on same-method information flow to detect same-method interference, which means that we may miss cases of information flow, and consequently of interference, which are not caused by editions to the same method. In particular, if on a merge scenario a method \emph{foo} is edited by both left and right, while left also edits method \emph{bar} and right edits method \emph{bar2}, we only consider the editions to method \emph{foo} as it was the only method edited by both. Consequently, we do not identify information flow involving left's edition on \emph{bar} or right's edition on \emph{bar2}. Hence, our strategy tends to be incomplete, but more scalable compared to analysing the whole program.

Furthermore, most of the scenarios with occurrences of same-method editions contain a single method on this situation. Nevertheless, some of the scenarios contain more than one occurrence of this pattern. Given a scenario with \emph{n} methods with occurrences of this pattern, there are two possibilities: passing all the \emph{n} occurrences of the pattern on a scenario to create a single \ac{sdg} with \emph{n} entry-points (one for each method) or creating \emph{n} \acp{sdg} with a single entry-point each (each \ac{sdg} with a different method as entry-point). 

The advantage of the first approach is that it is simpler and more intuitive, since every scenario is always associated to a single, richer, \ac{sdg} representing its information flow. On the other hand, the advantage of the second approach is that, because a different \ac{sdg} is created for each method with the pattern, if one method's \ac{sdg} is too heavy to be computed, other \acp{sdg} may still be computed for other methods and interference may be detected on them. In contrast, with the first approach the single \ac{sdg} would not be computed because of the method in question. Nonetheless, cases where all \emph{n} methods have \acp{sdg} too heavy to be computed will tend to take more time to fail to create than failing to create a single \ac{sdg} as in the first approach. For instance, if we consider that the \aca{sdg} creation fails after reaching a timeout the first approach would fail after the timeout, while the second would fail after $n \times timeout$ as it tries to create \emph{n} \acp{sdg}. Additionally, the first approach is more easily extensible for analysing other patterns as it works with a single richer \ac{sdg}. To keep the correspondence between scenario and \acp{sdg} created and facilitate future extensions to analyse different patterns, we decided to adopt the first approach.

%%To conclude, as creating the \ac{sdg} is computationally expensive, we save a recently created \ac{sdg}. That way, if the same \ac{sdg} is necessary in the future for further checks, it can be loaded instead of being created again from scratch.
\paragraph{IFC analyses}
%\begin{figure}
%	\centering
%	\includegraphics[scale=0.25]{imgs/ifc_analyses.png}
%	\caption{IFC analyses order}
%	\label{img:ifcAnalyses}
%\end{figure}
After the \ac{sdg} is created, or loaded from a previously saved one, we start running the actual \ac{ifc} analyses. In particular, different \ac{ifc} analyses are individually executed for each method edited by both contributions of a merge scenario. %\cref{img:ifcAnalyses} depicts the order of the \ac{ifc} analyses. 
For each method edited by both contributions, first we execute \ac{ifc} analyses to check for direct information flow  between contributions, from left to right and from right to left. Then, when there is no direct information flow, we also execute \ac{ifc} analyses to check for indirect information flow from the contributions to common target points. In other words, we want to check if there are points (instructions) with information flow from (affected by) both left and right. In particular, the indirect part first executes an analysis from left to base, then from right to base, and finally checks for common target points between the flows of information identified by those analyses.

\paragraph{Direct IFC analyses between contributions}
The intention of the analyses at this stage is to check if there is information flow from left to right and/or from right to left. To obtain these answers, we execute two \emph{ifc} analyses: one checks information flow from left to right, and the other from right to left. 

%As mentioned in \cref{subsect:joana}, after
After a \ac{sdg} is created, an \ac{ifc} analysis is formed by two stages: \emph{annotation} and \emph{execution}. The first corresponds to the part of defining \emph{high} (sources) and \emph{low} (sinks) nodes on the \ac{sdg}, while the second looks for paths between \emph{high} and \emph{low} nodes.

%was decode 
Starting by the left to right analysis, the annotation part takes as input the created \ac{sdg} and a method edited by left and right (with the line numbers edited by each). For instance, for \cref{fig:interfNoconf}, a \ac{sdg} representation of \cref{fig:interfNoconf} and in method \emph{dominates} left edited lines 7 and 8, and right line 11. We use the fact that each \ac{sdg} node is associated to a source code line to annotate the \ac{sdg} nodes. For each \ac{sdg} node from the method being analysed, we check if the line associated to the current node was edited by left, right or none. Nodes from lines edited by left are marked as \emph{high}, while nodes from lines edited by right are marked as \emph{low} and nodes from the rest of the lines are not marked. %For the example from \cref{integ:openNotclosed}, nodes associated to lines 14 and 16 are marked as \emph{high} and nodes associated to line 19 as \emph{low}. 
Finally, after the annotation part is concluded, the execution part looks for information flow from left to right.% In the example from \cref{fig:interfNoconf}, flows is detected between a node associated to line 14 and a node associated to line 19.

Once the left to right analysis execution finishes, the annotation part from the right to left analysis just needs to invert the nodes classification from the previous one. In particular, \emph{high} nodes of the first analysis are marked as \emph{low} on the second, and similarly \emph{low} nodes of the first become \emph{high} on the second. %For the example from \cref{integ:openNotclosed}, nodes associated to lines 14 and 16 are now marked as \emph{low}, while nodes from line 19 are marked as \emph{high}. 
After the inversion on the annotation part, the execution part looks for information flow from right to left. %Considering the example from \cref{integ:openNotclosed}, there is no information flow in this way (from right to left).

\paragraph{Indirect IFC analyses from contributions to common target points}
\begin{figure*}[t!]
%	\begin{subfigure}{\linewidth}
%		\centering
%		\lstinputlisting[language=Java,numbers=left,xleftmargin=60pt]{snippets/outerHtmlBase.java}
%		\vspace{-1em}
%		\caption{Base}
%	\end{subfigure}
	
%	\begin{subfigure}{\linewidth}
		\centering
\begin{lstlisting}[language=Java,numbers=left,xleftmargin=60pt]
public void outherHtmlHead(StringBuilder accum) {
	String html = getWholeText();
	if(a && a2)//left
		html = normaliseWhitespace(html);
	if(b || b2) //right
		accum.append(x);
	accum.append(html); //Affected by both
}
\end{lstlisting}
		\vspace{-1em}
%		\caption{Integrated}
%		\label{merged:both-flow}
%	\end{subfigure}
	\caption{Merge scenario with indirect information flow}
	\label{both-flow}
\end{figure*}

When there is no direct information flow between the contributions, we also check if the contributions flow to common target points. To illustrate this situation, consider the example from \cref{both-flow}, which was inspired on a merge scenario from Jsoup.\footnote{Jsoup GitHub page - \url{http://github.com/jhy/jsoup}} In this example, left added a second condition to the \emph{if} from line 3 ($\&\&$ $a2$), while right added a second condition to the \emph{if} from line 5 ($||$ $b2$). There is no direct information flow between those contributions. However, both of them affect line 7. More specifically, left directly affects variable \emph{html} on line 4, right affects the method parameter \emph{accum} on line 6 and both \emph{html} and \emph{accum} are used on line 7. In particular, there is a common target point (\emph{accum}) which is affected by both left and right. In other words, there is no direct side effect on behaviour from left on right, or from right on left, but the integration of left and right may cause a side effect on the behaviour of a common target point (\emph{accum}). That is, left and right may indirectly interfere with each other. We call situations of flows to common target points as potential indirect interference, because the common points have potentially different behaviour on left, on right and after the integration. 

To be able to check this situation we need to execute two extra \ac{ifc} analyses and then check for common target points in the flows identified by each. Firstly, we check for information flow between instructions from \emph{left} and \emph{base} instructions. \emph{Base} corresponds to instructions from lines not edited by left or right in the analysed scenario. %For instance, in \cref{both-flow}, we classify instructions from line 3 as from left, from line 5 as from right and from the other lines as from base. 
For our example, the \ac{ifc} analysis from left to base identifies flows from instructions on line 3 to instructions on lines 4 and 7 respectively. Secondly, we check for information flow between instructions from right and base. Considering our example, this second \ac{ifc} analysis would identify flows from instructions on line 5 to instructions on lines 6 and 7. Finally, we check if there are instructions affected by both \ac{ifc} analyses. In our example, the instruction from line 7 has flows on both analyses. Hence, line 7 is a common target point.

\paragraph{Translating identified flows}
Given the \ac{ifc} analyses were executed, when there is information flow, those flows are between \ac{sdg} nodes. %For instance, in \cref{integ:openNotclosed} something like ``Illicit flow from SDGNode 15951 ($v35 = lengthIncluded$) At SDGNode 16121 ($this.failIfNecessary()$)''. 
The final step consists of translating the flows between \ac{sdg} nodes to flows between source code lines in order to give the exact location at the source code that caused an information flow. %For the example from \cref{integ:openNotclosed} the previous flow between \ac{sdg} nodes is translated to something like ``Illegal flow from Decoder.decode() (line 14) to Decoder.decode() (line 19)''. 

\paragraph{Defining if a scenario contains information flow between contributions}
After we execute our strategy for a specific merge scenario, \ac{joana} returns all flows identified between source code lines of methods with same-method contributions. In particular, a scenario contains information flow between the contributions if \ac{joana} identifies at least one flow between source code lines. On the other hand, if no flow is identified, then there is no information flow between the contributions of this scenario.

\section{Evaluation}
\label{chap:eval}
Given the strategy proposed in \cref{chap:chap3}, it is necessary to evaluate it to understand if information flow may be used to estimate interference. This chapter details such evaluation. First, we reproduce merge scenarios from the development history of different projects hosted on GitHub and execute \ac{joana}, with different options, on their integrated revisions, to measure performance and check for information flow between the contributions. Then, we manually analyse part of the scenarios with information flow to establish if there is also interference.

\subsection{Research Questions}
\label{sec:rq}
Considering the motivation and strategy described in \cref{chap:chap3}, we need to understand the frequency of information flow between contributions from merge scenarios, and when the existence of information flow is not related to interference. However, as \ac{joana} has different options to create a \ac{sdg}, we also need to choose one. In summary, the following research questions are investigated:
\begin{itemize}  
	\item{\textbf{Research Question 1 (RQ1)} - \emph{Configuration}: Which \ac{sdg} option is the most appropriate to identify information flow between merge scenario same-method contributions?}
	\begin{itemize}
		\item{\textbf{Research Question 1A (RQ1A)} - \emph{Exceptions}: Should the \ac{sdg} include exceptions analysis?}
		\item{\textbf{Research Question 1B (RQ1B)} - \emph{Pointer analysis}: Which pointer analysis should be used?}
	\end{itemize}
	\item{\textbf{Research Question 2 (RQ2)} - \emph{Severity}: Is there direct information flow between merge scenario same-method contributions? How often?}
	\item{\textbf{Research Question 3 (RQ3)} - \emph{Limitations}: In which situations is there information flow and no interference?}
\end{itemize}

The answer to the first question (RQ1) will give us a \ac{sdg} configuration to execute \ac{joana}. Since a \ac{sdg} configuration involves different aspects, we focus on two of them, exceptions and pointer analysis, and consequently split the original question in two parts (RQ1A and RQ1B). The first part (RQ1A) aims to answer if \acp{sdg} will include exceptions. The second (RQ1B) establishes which pointer analysis should be used. Furthermore, as previously mentioned in \cref{ifcStep}, concurrency is not included for performance reasons.

The second question (RQ2) aims to check if direct information flow between same-method contributions actually occurs. For this question, high frequencies of information flow between contributions indicate that it is common to detect direct information flow between merge scenario contributions. In contrast, low frequencies indicate that this situation does not generally occur and, as a result, our strategy would identify fewer cases and be less relevant. It is important to mention that we focus in cases of direct information flow only, because the detection of indirect flows to common target points was only implemented at a later stage of the work. Hence, the cases of flows to common target points (explained in \cref{chap3:strat}) are classified as without information flow here.

Finally, the third question (RQ3) aims to understand the relation between the existence of information flow and interference. We want to check if when there is information flow there is also interference, and, in particular, in which situations there is information flow and no interference. More precisely, we focus on information flow true/false positives of interference. In contrast, we do not cover true/false negatives. More specifically, we are aware that our analysis may miss valid cases of interference, however, we are interested in first understanding if information flow existence indicates interference existence. Additionally, we are interested in the reasons for information flow existence without interference (false positives).

\subsection{Design}
Our study involves two different types of analysis: \emph{automatic} and \emph{manual}. The first aims to answer which \ac{sdg} configuration should be used (RQ1) and find out the frequency of information flow between contributions (RQ2). Furthermore, the first generates cases to the second which analyses a subset of the cases with information flow from the first to check if there is interference when there is information flow. Additionally, the manual analysis also checks in which cases there is information flow and no interference (RQ3).

\subsubsection{Sample}
\label{subsect:sample}
In order to select subjects to our sample, we initially selected the 128 Java projects from a previous work \cite{paolaPhd, Accioly2017}. This work's selection criterion involved only selecting Java projects with more than 500 stars on GitHub, which helps to avoid the selection of toy projects as each star is given by a different GitHub user. In particular, a project with 500 stars is ``stared'' (or considered interesting) by 500 different GitHub users. Additionally, this sample also contains projects from different domains such as databases, search engines and frameworks. Furthermore, the selected projects also have varying sizes and number of collaborators. More precisely, there are projects with sizes varying from 4KLOC (SimianArmy\footnote{SimianArmy GitHub page - \url{http://github.com/Netflix/SimianArmy}}) to 1653KLOC (Elasticsearch\footnote{Elasticsearch GitHub page - \url{http://github.com/elastic/elasticsearch}}) and with the number of collaborators varying from 10 (droidparts)\footnote{droiodparts GitHub page - \url{http://github.com/droidparts/droidparts}} to 403 (tachyon).\footnote{tachyon GitHub page - \url{http://github.com/amplab/tachyon}} 

Nevertheless, as we need to compile the integrated revisions in order to run \ac{joana}, we filtered the initial 128 projects sample to contain only projects with a build script of specific types, with the goal of making the process of compilation automatic. More specifically, we filtered this initial list to contain only projects with Ant,\footnote{Ant website - \url{http://ant.apache.org}} Gradle\footnote{Gradle website - \url{http://gradle.org}} or Maven\footnote{Maven website - \url{http://maven.apache.org}} build scripts, as these three build systems were the more common in the sample and the ones that were familiar to us. This step reduced our projects list from 128 to 119 (OG-Platform\footnote{OG-Platform GitHub page - \url{http://github.com/OpenGamma/OG-Platform}} was actually discontinued). 

%Marcelo - didn't understand
Furthermore, we looked for merge scenarios in those 119 projects with same-method contributions and no merge conflicts identified by FSTMerge (as explained in \cref{subsection:mergeStep} we use this tool as our merge tool). As previously mentioned, we do not consider scenarios with merge conflicts identified by FSTMerge, because we are interested in verifying situations which are not identified by typical merging strategies. %Specifically, we only check scenarios that are free of conflicts according to FSTMerge. 
Hence, we look for scenarios without merge conflicts according to FSTMerge from 2006 to 2016 (a period of almost eleven years, as the selection was realised on 2016). All of the 119 projects had merge scenarios in the selected period (total of 64128 scenarios), but only 97 had scenarios with occurrences of the same-method contributions pattern and no merge conflicts (other 14 projects only had scenarios with occurrences of the pattern with merge conflicts).

Finally, from 97 projects containing scenarios with occurrences of the pattern and no merge conflicts identified by FSTMerge, for 52 projects at least one scenario was successfully compiled. As it can be noticed, it was not possible to successfully compile any scenario for a significant number of projects (45 of 97). This situation may occur for a number of reasons. The more direct one is when the integrated code is in fact wrong and can not be compiled. This may occur because some of the contributions introduced code that caused the compilation problem, but also because of a static semantic conflict. Furthermore, we also noticed some cases where the compilation problems were actually caused by the used merge tool (FSTMerge). More precisely, FSTMerge only fully works for Java 5 resources and, as a result, files with newer elements (such as the diamond operator) fail during the parse and are not included in the integrated revision, which leads to compilation problems. However, there are other reasons for a compilation failure different than the integrated code being incorrect, such as when it is necessary to manually download additional tools to execute a build. For example, multiple projects required specific versions of the Android SDK to run a build. Another common reason was failing to download the dependencies specified in the build script, which could happen due to outdated links to the dependencies for instance. These cases require manual intervention in the compilation process, for instance by manually downloading the required tools (dependencies) and, as a result, we decided to not consider them because it would be excessively time-consuming. %A summary of the applied reductions on the project list is illustrated in \cref{fig:projSelection}, and the 
Our final sample can be found from \cref{links}.

So far, we focused on discussing how our project list was selected. Nonetheless, one project may have multiple scenarios satisfying our requirements, scenarios which were successfully compiled and contain same-method editions without merge conflicts identified by FSTMerge. In fact, we analyse more than one scenario from the same project, but we avoid analysing too many scenarios from the same project as this could bias our results. For example, if we had 100 scenarios from 52 projects, but 40 were from a single project, this could bias our data towards this project's characteristics. More specifically, we avoided running \ac{joana} for more than 7 scenarios from the same project. The only exception to this rule, was the project orientdb,\footnote{orientdb GitHub page - \url{http://github.com/orientechnologies/orientdb}} which we ran 11 scenarios because for 3 of them no \ac{sdg} was successfully created, and for other 5 the \ac{sdg} was successfully created for a single configuration. So, if a project contains too many scenarios satisfying our criteria, we prioritise the scenarios by date (the more recent ones) and by the methods matching the pattern (was the method with the pattern already analysed in a different scenario?). Therefore, if a project has more scenarios than we intend to analyse, we select the more recent scenarios and also avoid selecting scenarios with the same methods with the pattern as analysing the same method multiple times in different scenarios may also bias our data with the method characteristics. On the other hand, if a project has only a few scenarios satisfying the compilation and pattern existence with no merge conflicts criteria, we do not look for date or repetition. %For each project in \cref{table:sample} we detail the total number of scenarios executed and the number of scenarios where it was possible to create a \ac{sdg} with at least one configuration. In particular, it was not possible to create \acp{sdg} for four projects (astyanax, go-lang-idea-plugin, groovy-core and hive), independently of the configuration used.

\subsubsection{Automatic analysis}
\label{sec:desAutom}
This part of the analysis involved executing \ac{joana} with different configurations for 157 merge scenarios integrated revisions from 52 projects. This is done reproducing the merge scenarios from the development history of those projects, and then executing \ac{joana} with different configurations on the integrated revisions to check for information flow between same-method contributions of those scenarios. 

As just mentioned, we compare different configurations of \ac{joana}. More specifically, these configurations vary with respect to two aspects: use of exceptions (RQ1A) and pointer analysis (RQ1B). Regarding exceptions, from four available options, we consider \aca{joana} most precise option of exceptions (integrate exceptions with interprocedural optimizations), and, the option that ignores exceptions. Basically, here, we consider them as include or not exceptions. We use the most precise exceptions option as our option of including exceptions, because, this option, when possible, does not consider impossible exceptions. Regarding pointer analysis, we consider the eight pointer analyses available in \ac{joana}. Thus, considering the eight pointer analysis, and the use or not of exceptions (2 options), there is a total of 16 ($8 \times 2 = 16$) possible configurations to be analysed.

%Marcelo - Do ponto de vista tecnico, o enfoque pareceu bem maior no exercicio de descobrir parametros do WALA/JOANA...
We could select \aca{joana} default configuration and avoid the effort involved in selecting a proper configuration to execute \ac{joana}. Nevertheless, as we considered that the configuration is important to check the applicability of our strategy and as we considered that this decision may affect our conclusions, we decided to first select the most appropriate configuration to our context.

To establish which \ac{sdg} configuration should be used (RQ1) we executed \ac{joana} with different configurations and compared the results with respect to quantity of direct information flow, size of the created \ac{sdg}, and also with respect to the number of successful \ac{sdg} creations. In summary, we want a configuration that is precise (has few false positives of information flow), but with an acceptable size and, more importantly, with a reasonable number of successful creations. 

%Marcelo - eh estranho reportar, em alguns experimentos, a metrica "numero de fluxos" isoladamente (sem saber se isto eh bom ou ruim).  No limite, uma analise que reporta todos os fluxos (extremamente imprecisa) ira ser vencedora sempre.
First, we consider the number of direct flows identified between the contributions source code lines and, more importantly, the existence of direct information flow between the scenarios contributions. In other words, for each configuration, we consider the quantity of direct flows identified and if at least one flow was identified, since at least one flow needs to be detected to classify a scenario as with direct information flow between the contributions. Additionally, for the exceptions part (RQ1A), as \acp{sdg} without exceptions may miss information flow \cite{hammer2009flow}, we further investigate scenarios that contain information flow with exceptions and do not contain without. This is important to understand if cases of information flow due to exceptions are relevant to detect interference. We do not do the same for the pointer analysis part (RQ1B), as a more precise pointer analysis is designed to avoid spurious cases of information flow detected by a less precise one, not to detect extra cases missed by the other, as with exceptions. In other words, the inclusion of exceptions has the goal of reducing information flow false negatives, while more precise pointer analyses have the goal of reducing false positives. Thus, due to the fact that more precise pointer analyses focus on minimizing false positives, when for a given scenario one pointer analysis finds information flow and other does not, we consider that the one which did not find information flow is the correct one. 

%\begin{figure*}[t!]
%	\centering
%	\includegraphics[scale=0.45]{imgs/iterations.png}
%	\caption{Study iterations}
%	\label{fig:iterations}
%\end{figure*}

Then, we consider that the \ac{sdg} size is defined by the number of nodes, and by the number of edges. Lastly, we consider the number of successful creations of each configuration. This last aspect is important, because there are scenarios that lead to successful \ac{sdg} creation with one configuration but not with another. More precisely, \ac{joana} may take too long, or even break for some configurations. In particular, we consider a \ac{sdg} is too heavy if it takes more than one day to create (timeout), if it needs more than 120 GB of heap to compute (out of memory), or if it exceeds the stack natural limit of calls (8192 KB) during its construction (stack overflow). All the configurations were executed on a machine with 40 Intel(R) Xeon(R) CPU E5-2660 v2 @ 2.20GHz processors, 251 GB of RAM memory, running Ubuntu 14.04.4 LTS (GNU/Linux 3.13.0-79-generic x86\_64) as operating system.

Because all combinations of pointer analysis (8 possibilities) and use of exceptions (2 possibilities) involve 16 ($8 \times 2 = 16$) different configurations, and because creating a \ac{sdg} is expensive, we divided our automatic analysis in iterations. The basic idea is to progressively increase the sample and remove some configurations at each iteration until only one (the selected configuration) is left. An image with this idea is accessible from \cref{links}%.\cref{fig:iterations}. 
. In particular, we answer which configuration should be used (RQ1) after three iterations. First, after iteration 1 we answer if exceptions should be included (RQ1A) and remove one pointer analysis, leaving 7 possible configurations. Then, after iteration 2 we remove other 4 configurations, leaving 3 possibilities. Finally, after iteration 3 we answer which pointer analysis should be used (RQ1B), finishing answering RQ1. 

Furthermore, as iteration 1 involves 16 configurations and including external dependencies tend to increase the \ac{sdg} size, \acp{sdg} are created without the dependencies (external jars) at this iteration, while iterations 2 and 3 also include the dependencies (as they involve fewer configurations). Additionally, iteration 1, and 20 scenarios from iteration 2, were executed using stubs for the \ac{jre} 1.4 (the default), while the rest of the scenarios were executed using stubs for the \ac{jre} 1.5 as we realised at this point this option is more appropriate because it is more updated. Thus, once we select a configuration to be used (RQ1), we rerun, in iteration 4, the scenarios from iteration 1 and those from iteration 2 which used stubs for \ac{jre} 1.4, passing their correspondent dependencies and using stubs for \ac{jre} 1.5 to obtain the appropriate frequency of direct information flow between merge scenario contributions (RQ2). Additionally, it is important to mention that although we varied passing or not the dependencies and the stubs version (for \ac{jre} 1.4 or 1.5) between the scenarios, we never varied these between different configurations from the same scenario as this would harm our comparison. In summary, for each scenario the configurations only vary with respect to use of exceptions (RQ1A) and pointer analysis (RQ1B) and aspects such as use of dependencies and stubs version are the same, although different scenarios varied with respect to use of dependencies and stubs version. 

In addition, it is important to notice that the data used at each iteration is incremental. For example, iteration 1 includes 83 scenarios from 34 projects, iteration 2 uses those scenarios and adds 36 scenarios to the initial data (from projects with scenarios in iteration 1 and also from 16 new projects). Similarly, iteration 3 increments the data from iteration 2 by adding 38 scenarios. A part of these scenarios is from two new projects, but another one is from projects with scenarios in iteration 2. 

Finally, all the iterations from the automatic analysis generate possible cases to be analysed by the manual analysis, which aims to answer in which situations there is information flow and no interference (RQ3).

\subsubsection*{Infrastructure}
\label{infra}

The built infrastructure (see \cref{links}) can be divided in four parts: \emph{mining}, \emph{build}, \emph{IFC} and \emph{data analysis}. The first is responsible for finding merge scenarios with the occurrence of same-method contributions, the second for finding which of these scenarios may be compiled, the third for detecting information flow between the scenarios contributions, and finally the last for handling the data and generating plots and reports. 

\paragraph{Mining}
Our mining step is similar to the ones used in previous work \cite{paolaPhd,kasi2013cassandra,cavalcanticomparing,Accioly2017}. First, we use GitMiner,\footnote{GitMiner GitHub page - \url{http://github.com/pridkett/gitminer}} which converts the history from a GitHub project into a Neo4j\footnote{Neo4j website - \url{http://www.neo4j.org/}} graph database. GitMiner receives a GitHub project (or a particular user), connects to GitHub via GitHub’s \ac{api} and loads all the metadata available about the project (or user) and stores it in a Neo4j database. 

Subsequently, we implemented scripts that query the Neo4j database to retrieve a list of merge commits ids, and their parent ids, ordered by the date of the merge. Our scripts uses Gremlin,\footnote{Gremlin GitHub page - \url{http://github.com/tinkerpop/gremlin/}} a graph traversal language to perform the queries. Basically, the graph database represents commits as nodes with a \emph{isMerge} field to indicate if a commit is a merge commit. To identify merge scenarios we traverse the commits and query which commits have the \emph{isMerge} field with a \textit{true} value. Then, we use JGit,\footnote{JGit page - \url{http://www.eclipse.org/jgit/}} a Java library implementing the Git version control system, to clone the project locally and, for each merge commit, we checkout and copy the revisions involved in the merge scenario (left, base and right). 

Next, after obtaining the merge scenarios, we merge the revisions using FSTMerge, check which of the scenarios contain occurrences of the same-method contributions pattern, and save these in a \ac{csv} file. The saved file contains information related to the scenario, the method signature in the format expected by \ac{joana}, and the list of lines edited by left and right. 
\paragraph{Build}
Subsequently, before running \ac{joana} on a scenario, it is necessary to compile this scenario. Therefore, we check which of the merge scenarios with the pattern occurrence are successfully compiled on the build step. With the goal of doing this verification, we implemented a python\footnote{python website - \url{http://www.python.org}} script which runs the build process (Ant, Maven or Gradle) for every scenario with the pattern of a specific project. It writes a \ac{csv} file specifying for each scenario if the build was successful or not. If the build was successful, it also writes which build system was used (Ant, Maven or Gradle). For projects with Maven as the build system, running the build process for compilation is more straightforward than the others (Ant and Gradle) as there is the command \textit{mvn compile} to perform such task. However, for projects with Ant and Gradle build systems, there is no equivalent default command and the commands (targets in Ant and tasks in Gradle) are different for each project. Hence, each project requires a manual analysis on its build script in order to define which ``compilation command'' should be used by our python script.

\paragraph{IFC}
Then, we run \ac{joana} to check for information flow between the contributions. As we run \ac{joana} with different configurations, we implemented a python script to iterate through the desired configurations and invoke \ac{joana} with the appropriate arguments. We decided to invoke each different configuration from a python script, instead of also doing this logic in Java, to avoid problems of one configuration using elements from another one due to shared resources in the same \ac{jvm} execution. Hence, for each desired configuration from a scenario, our python script invokes \ac{joana} (forcing a different \ac{jvm} execution) and waits for \aca{joana} execution to finish to call the next configuration. We established a timeout of 24 hours, in our python script, as the limit for waiting for \aca{joana} execution, as \aca{joana} computation may become too expensive computationally. If this timeout is reached and \aca{joana} computation with the specific configuration is not finished, this configuration is considered too expensive for the particular case and the script tries the next configuration. For each scenario, this step produces a \ac{csv} file containing data informing for which configurations the \ac{sdg} was successfully created and which of them contained information flow between the contributions. 

It is important to mention that cases of indirect flows to common target points (see \cref{ifcStep}) were only identified and implemented at a later stage of the work and, therefore, they are not included in the generated \ac{csv} files. Hence, although our \ac{joana} implementation also checks for indirect flows, those cases are not saved in the generated \ac{csv} files and, as a result, they are not considered by the automatic analysis. More precisely, the automatic analysis only considers that there is information flow if there is a direct information flow between the contributions. In contrast, in the manual analysis we also know if an indirect flow to a common target point is identified as, every time we execute \ac{joana}, we save \emph{txt} files with the detailed execution.

\paragraph{Data Analysis}
Finally, we implemented a R\footnote{R project website - \url{http://www.r-project.org/}} script to analyse our results. With this script, we read the \ac{csv} files generated after \aca{joana} execution of each scenario and generate the necessary plots and reports. In summary, we are able to manipulate (and summarise) the data generated after the different executions of \ac{joana} to obtain information such as information flow frequency and number of successful \ac{sdg} creations for different \ac{sdg} configurations. 

\subsubsection{Manual analysis}
A manual analysis was also conducted to understand the limitations of using information flow to estimate interference between contributions in order to answer our third question (RQ3). 

Since we are interested in understanding if there is interference when there is information flow, our analysis focus on methods, from the automatic analysis, that have direct information flow between the contributions. Thus, we do not cover interference true/false negatives. More specifically, we are aware that our strategy may miss valid cases of interference, however, we are interested in first understanding if information flow existence indicates interference existence. That is, we first want to check if our strategy may help to detect some cases of interference.

Furthermore, we prioritised methods with less than 100 lines, in our manual analysis, as bigger methods have more information and as a result tend to take more time to manually analyse. 

Additionally, we do not analyse test code in our manual analysis. More precisely, we consider main source code more important than test code. Thus, we focus in main source code and do not manually analyse test code.

Lastly, we only manually analyse one method per scenario to avoid biasing our data with information from a single scenario. Similarly, we avoid selecting too many scenarios from a single project. In particular, we only manually analysed at most three scenarios from the same project.

In summary, given a method selected to be manually analysed we know it contains information flow between its contributions and we manually analyse it using an open-world perspective to establish if there is also interference. If there is also interference, then information flow and interference coincide for the given method. On the other hand, if there is no interference, we need to understand why there is information flow but no interference. 

%Marcelo - como vc. estabeleceu ground truth para identificar interferencia (a partir de flows)?
Basically, we use the location of the identified information flow in the source code to guide our manual analysis. That is, the identified flows show possible points of interference. We manually analyse those points in order to establish if there is interference involving them. 

Using these criteria, from 57 methods (from 48 scenarios and 28 projects) with information flow, we manually analysed 35 methods (from 35 scenarios and 24 projects).

\subsection{Results and discussion}
\label{results}
As previously mentioned in \cref{sec:rq} we aim to understand which \ac{sdg} configuration is the most appropriate for identifying information flow between merge scenario contributions (RQ1), what is the frequency of information flow between same-method contributions on merge scenarios (RQ2) and in which situations there is information flow between the contributions and no interference (RQ3). We detail here the obtained results from both the automatic and manual analysis and discuss their implications.
\subsubsection{Automatic Analysis}
As previously mentioned in \cref{sec:desAutom} we divided our automatic analysis in iterations, we show here the results for each iteration. Full results are available on \cref{links}.
\subsubsection*{Iteration 1}
\label{it1}
%As already illustrated in \cref{fig:iterations}, i
Iteration 1 data corresponds to \aca{joana} execution of 83 scenarios from 34 projects. For all 83 scenarios, the \acp{sdg} were created without passing dependencies and with stubs for \ac{jre} 1.4. From the 83 executed scenarios, the \ac{sdg} was successfully created for at least one configuration for 71 scenarios from 34 projects. Additionally, the \ac{sdg} was created for all configurations for 62 scenarios from 33 projects. Furthermore, from these 62 scenarios, it was possible to identify instructions for lines edited by left and right on 40 scenarios from 26 projects. Because these reductions are significant, we further investigated the reasons for their occurrence and discuss it later in this section. 

It is important to mention that, depending on the aspect being analysed, we might use a subsample of the 83 scenarios. For example, to check for the number of times the \ac{sdg} was successfully created for each configuration we consider the complete sample from iteration 1 (83 scenarios). However, to compare \ac{sdg} nodes and edges, we consider only the 62 scenarios where a \ac{sdg} was created for all configurations. More specifically, including cases where a \ac{sdg} was created only for some cases could bias our data. Similarly, to compare number of flows and information flow frequency, we consider only the 40 scenarios where a \ac{sdg} was created for all configurations and there are instructions corresponding to left and right contributions. Thus, iteration 1 data involves 83, 62 or 40 scenarios depending on the aspect being analysed: number of \ac{sdg} creations, number of nodes/edges and information flow, respectively.

This iteration considered all the possible configurations derived from combining exceptions use (2 possibilities) and pointer analysis (8 possibilities), leading to 16 possible configurations. At the end of this iteration, we answered if exceptions should be included, discarding 8 possible configurations, and also discarded one of the 8 existing pointer analyses.

\subsubsection*{RQ1A - Should the \ac{sdg} include exceptions analysis?}
\label{rq1a}
To answer this question (RQ1A), first we noticed, as expected, that \acp{sdg} with exceptions tend to have more information flow between the contributions than \acp{sdg} without exceptions. Then, we noticed that although \acp{sdg} with exceptions had more information flow their size was only slightly bigger. Finally, we investigated scenarios that contain information flow between the contributions only when considering \acp{sdg} with exceptions, to understand if those ``extra'' cases are relevant to estimate interference.

\paragraph{\acp{sdg} with exceptions contain more information flow}
\begin{figure*}[t!]
	\centering
	\begin{subfigure}[t]{0.5\textwidth}
		\centering
		\includegraphics[scale=0.3]{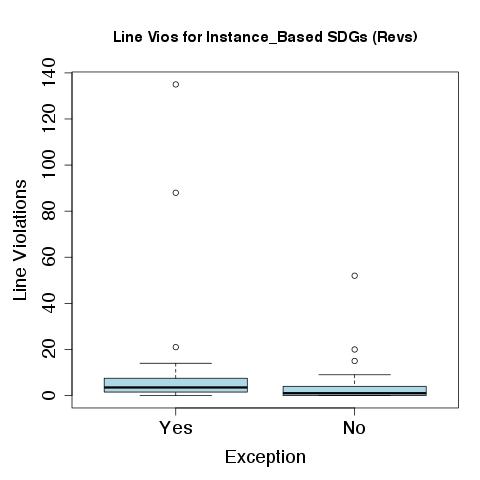}
		\caption{Original}
		\label{fig:lineViosIbOrig}
	\end{subfigure}%
	~ 
	\begin{subfigure}[t]{0.5\textwidth}
		\centering
		\includegraphics[scale=0.3]{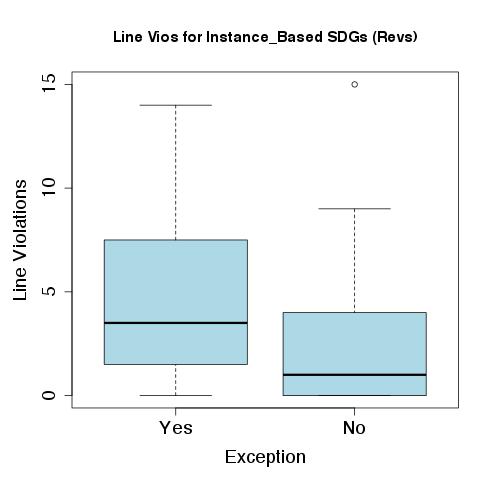}
		\caption{Amplified}
		\label{fig:lineViosIbAmplif}
	\end{subfigure}
	\caption{Original (a) and amplified (b) boxplots of number of flows (violations) identified between source code lines for Instance Based \acp{sdg} with (left boxes labeled as ``Yes'') and without (right boxes labeled with ``No'') exceptions}
	\label{fig:lineViosIb}
\end{figure*}
First, we compared the quantity of flows identified by \acp{sdg} with and without exceptions as illustrated in \cref{fig:lineViosIb}. \cref{fig:lineViosIbOrig} shows the original boxplots, but as it can be noticed it is difficult to visualize the tendencies of the boxplots in this figure due to extreme outliers. Therefore, we also generated an amplified version in \cref{fig:lineViosIbAmplif}, which uses the same data as \cref{fig:lineViosIbOrig}, but only shows values up to 15 on the y-axis. More precisely, \cref{fig:lineViosIbAmplif} may be seen as the result of cropping and zooming in \cref{fig:lineViosIbOrig} from 0 to 15 on the y-axis. As it may be noticed in \cref{fig:lineViosIbAmplif}, \acp{sdg} with exceptions (labeled with ``Yes'') contain more flows (violations) than \acp{sdg} without exceptions (labeled with ``No''). In fact, the first quartile from the former is higher than the median from the latter, and the median from the former is close to the third quartile from the latter.

\begin{figure*}[t!]
	\centering
	\begin{subfigure}[t]{0.23\textwidth}
		\centering
		\includegraphics[scale=0.2]{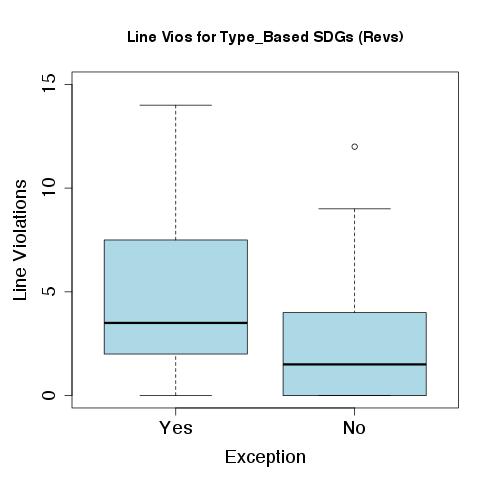}
		\caption{Type Based}
		\label{fig:lineViosTb}
	\end{subfigure}%
	~ 
	\begin{subfigure}[t]{0.23\textwidth}
		\centering
		\includegraphics[scale=0.2]{imgs/RevsExceptionsPlotLineViosINSTANCEBASEDyLim15.jpg}
		\caption{Instance Based}
		\label{fig:lineViosIbMult}
	\end{subfigure}
	~
	\begin{subfigure}[t]{0.23\textwidth}
		\centering
		\includegraphics[scale=0.2]{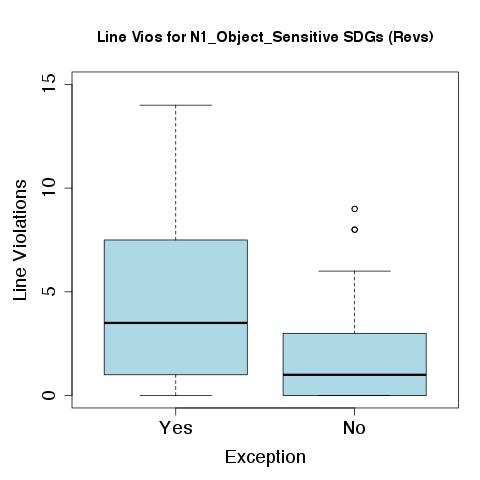}
		\caption{N1 Object Sensitive}
	\end{subfigure}
	~
	\begin{subfigure}[t]{0.23\textwidth}
		\centering
		\includegraphics[scale=0.2]{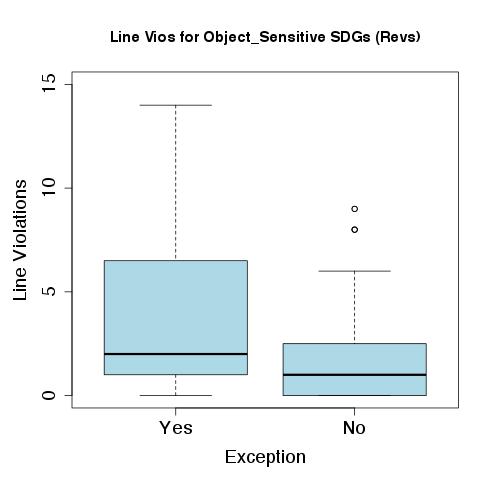}
		\caption{Object Sensitive}
	\end{subfigure}
	\par
	~ 
	\begin{subfigure}[t]{0.23\textwidth}
		\centering
		\includegraphics[scale=0.2]{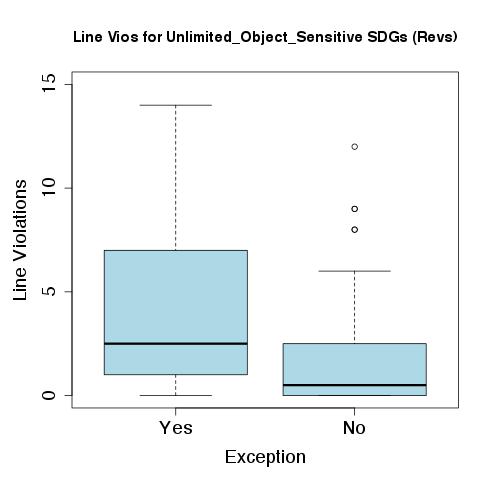}
		\caption{Unlimited Obj. Sensit.}
		\label{fig:lineViosUnlimit}
	\end{subfigure} 
	~
	\begin{subfigure}[t]{0.23\textwidth}
		\centering
		\includegraphics[scale=0.2]{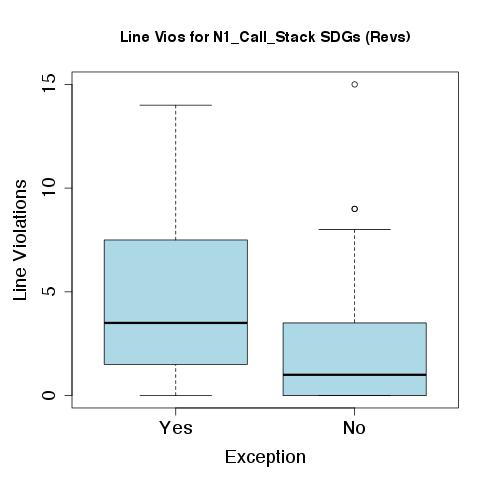}
		\caption{N1 Call Stack}
	\end{subfigure}
	~
	\begin{subfigure}[t]{0.23\textwidth}
		\centering
		\includegraphics[scale=0.2]{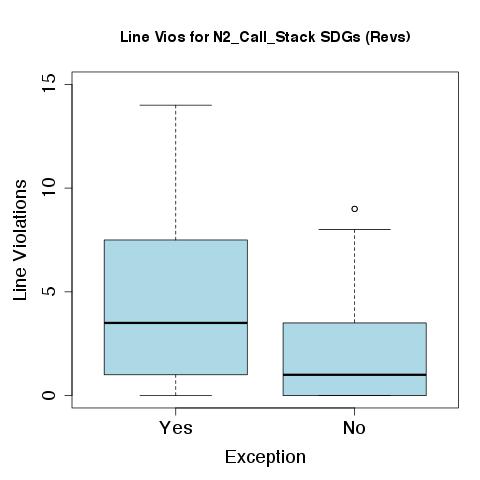}
		\caption{N2 Call Stack}
	\end{subfigure}
	~
	\begin{subfigure}[t]{0.23\textwidth}
		\centering
		\includegraphics[scale=0.2]{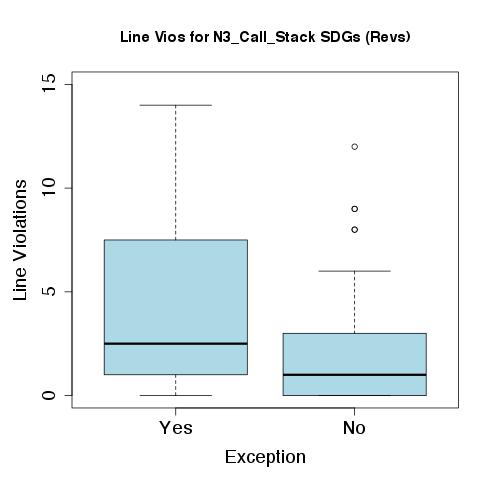}
		\caption{N3 Call Stack}
	\end{subfigure}
	\caption{Amplified boxplots of number of flows identified between source code lines for \acp{sdg} with (left boxes) and without (right boxes) exceptions for all precisions}
	\label{fig:lineViosAllPrecs}
\end{figure*}

\cref{fig:lineViosIb} shows that for \emph{instance based} \acp{sdg}, \acp{sdg} with exceptions contained more flows (violations) than \ac{sdg} without exceptions. Indeed, as it may be noticed in \cref{fig:lineViosAllPrecs}, this tendency is maintained for all pointer analyses. Although for some pointer analyses the difference is bigger, all the pointer analyses contain more flows for \acp{sdg} with exceptions. More specifically, for every pointer analysis in \cref{fig:lineViosAllPrecs} the left box (\acp{sdg} with exceptions) has a median close to the third quartile from the right one (\acp{sdg} without exceptions) and a first quartile greater than or equal to the median from the right one. 

To confirm that this tendency is in fact statistically relevant, we performed a hypothesis test. First, we tested our data for normality using Shapiro-Wilk Normality Test \cite{shapiro1965analysis}.\footnote{Shapiro-Wilk test on R - \url{http://stat.ethz.ch/R-manual/R-devel/library/stats/html/shapiro.test.html}} This test has the null hypothesis that the data is normal. Since the obtained p-value was of the order of $10^{-30}$, considering a significance level of 0.05, we can not accept the null hypothesis that the data is normal and consequently we need to consider that our data is not normal. Then, we performed a one-sided paired Wilcoxon Signed-Rank Test \cite{wilcoxon1945individual}\footnote{Wilcoxon Signed-Rank test on R - \url{http://stat.ethz.ch/R-manual/R-devel/library/stats/html/wilcox.test.html}} with the null hypothesis that the two groups (with and without exceptions) have the same quantity of flows and the alternative hypothesis that \acp{sdg} with exceptions have more flows. The obtained p-value was lower than $2.2 \times 10^{-16}$. Thus, at a significance level of 0.05, we can not accept the null hypothesis that the groups have the same amount of flows. As a result, the alternative hypothesis that \acp{sdg} with exceptions have more flows is accepted.

\begin{table}[ht]
	\scriptsize
	\centering
	\begin{tabular}{c|c|*{8}{c|}}
		\multicolumn{2}{c}{} & \multicolumn{2}{c}{Exceptions} \tabularnewline
		\cline{2-4}
		 & \bfseries Pointer analysis
		&    \bfseries Yes & \bfseries No  \tabularnewline[1 ex] 
		\cline{2-4}
		&    \bfseries Type Based & 87.5 &  62.5   \tabularnewline [1ex] 
		\cline{2-4}
		&    \bfseries Instance Based & 85 & 60  \tabularnewline [1ex] 
		\cline{2-4}
		&    \bfseries N1 Object Sensitive & 85 & 55 \tabularnewline [1ex] 
		\cline{2-4}
		&    \bfseries Object Sensitive & 85 &  52.5  \tabularnewline [1ex] 
		\cline{2-4}
		&    \bfseries Unlimited Object Sensitive & 85 &  50  \tabularnewline [1 ex]
		\cline{2-4}
		&    \bfseries N1 Call Stack & 85 &  60  \tabularnewline [1ex] 
		\cline{2-4}
		&    \bfseries N2 Call Stack & 85 &  57.5  \tabularnewline [1ex] 
		\cline{2-4}
		&    \bfseries N3 Call Stack & 85 &  52.5  \tabularnewline [1 ex]
		\cline{2-4}
	\end{tabular}
	\caption{Percentages (\%) of scenarios with direct information flow occurrence in iteration 1}
	\label{table:freqsIt1}
\end{table} 

Once we established that \acp{sdg} with exceptions contain more flows between source code lines than \acp{sdg} without exceptions, we also compared the frequencies of scenarios with information flow between contributions (considering only the existence of flow instead of the quantity) to check if \acp{sdg} with exceptions contain information flow in more scenarios than \acp{sdg} without exceptions. More specifically, if for a given scenario a \ac{sdg} with exceptions of a specific pointer analysis has eight flows and another one, of the same pointer analysis, without exceptions, has one flow, then both \acp{sdg} contain information flow for the given scenario. On the other hand, if a \ac{sdg} with exceptions has one flow, while a \ac{sdg} without exceptions does not have any flow, then the \ac{sdg} with exceptions contains information flow on the scenario, while the one without exceptions does not. The obtained frequencies of information flow of the scenarios from iteration 1 are detailed in \cref{table:freqsIt1} for all possible combinations of configurations. As it may be observed, \acp{sdg} with exceptions tend to contain information flow in more scenarios. In particular, for every pointer analysis, the value on the ``Yes'' column is higher than the value on the ``No'' column. More precisely, the differences between the columns vary from 25 (Type Based, Instance Based and N1 Call Stack) to 35 (Unlimited Object Sensitive) percent. In summary, \textbf{\acp{sdg} with exceptions contain more flows and also more scenarios with information flow than \acp{sdg} without exceptions}.

\paragraph{Exceptions use did not significantly change \ac{sdg} size}
\acp{sdg} with and without exceptions were also compared with respect to their size. More specifically, we compare the \acp{sdg} number of nodes and edges. With respect to the number of nodes, \acp{sdg} with and without exceptions are almost identical. On the other hand, with respect to the number of edges, there is a slight difference. Both tendencies were kept for all pointer analyses: the number of nodes for \acp{sdg} with and without exceptions were very similar, while \acp{sdg} with exceptions had a number of edges slightly higher (see \cref{links}). In particular, for the number of nodes the median was always the same and the mean varied from 0\% (\emph{type based}) to 0.0019\% (\emph{n2 call stack}). In contrast, for the number of edges the median was from 1.08\% (\emph{n3 call stack}) to 6.13\% (\emph{type based}) higher for \acp{sdg} with exceptions and the mean from 2.20\% (\emph{unlimited object sensitive}) to 6.14\% (\emph{n1 call stack}).

\paragraph{Exceptions tend to be irrelevant for estimating interference}

Once we established that \acp{sdg} with exceptions contained more information flow, we decided to manually investigate a few cases where only \acp{sdg} with exceptions contained information flow, to understand if those cases are relevant for estimating interference. One of the reasons we observed that led \acp{sdg} with exceptions to have more flows than \acp{sdg} without exceptions is that a line containing a statement that may raise an exception leads to flows to all lines that could be executed after this line. More specifically, if for a method of 100 lines, there is a statement that may raise an exception in line 2, a \ac{sdg} with exceptions leads to flows from line 2 to all lines from 3 to 100 with instructions that may be executed if the exception is not raised. Moreover, we observed two types of flows caused by exceptions: \emph{accidental} and \emph{intentional}. 

The first represents situations where there is the possibility of an exception being implicitly raised due to an unexpected situation. For instance, a possible \emph{NullPointerException}. It is important to mention that a \ac{sdg} with exceptions may include exceptions that never actually occur in practice. More precisely, if a specific field is never instantiated with \emph{null} as parameter, then a \emph{NullPointerException} will never actually occur. 

In contrast, the second represents situations where an exception may be explicitly raised because of a predicted situation. In other words, while in the first case an exception may be raised because of an unexpected situation, here an intentional exception is explicitly raised by introducing a \textit{throw new}. 
In the original security context from \ac{joana}, it is important to consider exceptions, as the implicit flows caused by them may lead to information leak. Nevertheless, we use \ac{joana} in a different context. More specifically, our intention is estimating interference between the contributions of a merge scenario.
In this context, we claim that information flow caused by accidental exceptions is less relevant for estimating interference between contributions of a merge scenario than information flow caused by intentional exceptions (and than information flow not involving exceptions). More specifically, accidental exceptions identify flows due to possibly incorrect code of one of the contributions, not due to actual intentional behaviour changes of the contribution itself. To be more precise, while intentional exceptions capture intentional modification of the system behaviour, accidental exceptions capture possible flaws of one of the contributions that, in practice, are identified too often by \ac{joana} but may not even occur. 

In summary, we argue that including accidental exceptions significantly increases the number of interference false positives. Thus, we consider this type of exception less relevant for estimating interference. In contrast, intentional exceptions are relevant, as they capture intentional behaviour change. 

In summary, as we consider accidental exceptions less relevant to our context, we would like to detect intentional exceptions, but not the accidental ones. Nonetheless, \ac{joana} does not provide an option to include intentional exceptions and ignore accidental, it does only provide options including both or ignoring both. Although is possible to modify \ac{joana} to include only the intentional exceptions, we decided to use one of the available options. Hence, if we include exceptions in our \acp{sdg} we identify both, while if we do not, we miss both. Therefore, we need to decide if it is better to avoid accidental exceptions and miss the intentional ones, or if it is better to consider intentional exceptions and as a result also include accidental ones. 

We expect accidental exceptions to happen considerably more than intentional exceptions, as the latter requires explicit code to raise an exception in the application, while the former may be potentially identified in any statement of a program. To confirm this, we analysed 10 scenarios (from 9 different projects) which have information flow only with exceptions and looked for the types of exceptions contained in each. Only 2 of the 10 scenarios contained intentional exceptions, while at least 9 of the 10 scenarios contained accidental. Therefore, as expected, these numbers suggest that accidental exceptions tend to happen significantly more than intentional exceptions. As a result, \textbf{we conclude that extra flows contained only in \acp{sdg} with exceptions tend to be less relevant for estimating interference, as they are generally due to accidental exceptions, and we consider flows due to accidental exceptions less relevant}.

\paragraph{\acp{sdg} should not include exceptions}
To conclude, we observed that, depending of the pointer analysis, for 25 to 35 percent of the analysed scenarios, \acp{sdg} with exceptions had information flow while \acp{sdg} without exceptions did not. Then, we further investigated some cases that contain information flow between contributions only when considering exceptions, to understand if exceptions are relevant to estimate interference. The conclusion was that, in general, they tend to be irrelevant (as they tend to be accidental exceptions). Hence, \textbf{we decided to not include exceptions analysis in our \acp{sdg}}. That way, we focus on reducing the number of irrelevant cases (due to accidental exceptions) with the drawback of missing some actually relevant cases (such as cases of intentional exceptions).

\subsubsection*{RQ1B - Which pointer analysis should be used?}
Regarding the pointer analysis, our first finding at iteration 1 was that the pointer analysis has more effect on information flow occurrence for \acp{sdg} without exceptions. Then, we discarded one of the pointer analysis (unlimited object sensitive) as we considered it too heavy, leaving 7 remaining pointer analyses for iteration 2.

\paragraph{Pointer analysis had more effect on information flow occurrence for \acp{sdg} without exceptions}
As detailed in \cref{table:freqsIt1}, in our sample, the frequencies of information flow occurrence varied more in the pointer analyses for \acp{sdg} without exceptions. More precisely, for \acp{sdg} with exceptions (``Yes'' column) only the frequency of Type Based \acp{sdg} was slightly different (87.5\%), while the rest of the pointer analyses converged to the same frequency (85\%), with a maximum difference of 2.5\% (87.5\% - 85\%) and standard deviation of 0.88. On the other hand, for \acp{sdg} without exceptions (``No'' column) the results varied more, from 62.5\% (on \emph{type based}) to 50\%(on Unlimited Object Sensitive), with a maximum difference of 12.5\% (62.5\% - 50\%) and standard deviation of 4.43. The intuition is that the results of the pointer analyses for \acp{sdg} with exceptions vary less because \acp{sdg} with exceptions, independently of the pointer analysis, already have a high frequency of information flow due to possible exceptions. In contrast, \acp{sdg} without exceptions do not identify those extra flows and thus the differences between the pointer analyses are augmented.  

These numbers indicate that, if we had decided to use \acp{sdg} with exceptions, a basic pointer analysis, such as \emph{instance based}, would be a good candidate as it had no extra scenarios with false positives of information flow compared to more precise analysis such as \emph{object sensitive} and \emph{n1-3 call stack}. In fact, the least precise pointer analysis (\emph{type based}) identified only 2.5\% more scenarios. Nonetheless, as we use \acp{sdg} without exceptions, we need more data before deciding which pointer analysis to use.

\paragraph{Unlimited Object Sensitive was too heavy and added little extra precision}
\begin{figure*}[t!]
	\centering
	\begin{subfigure}[t]{0.5\textwidth}
		\centering
		\includegraphics[scale=0.3]{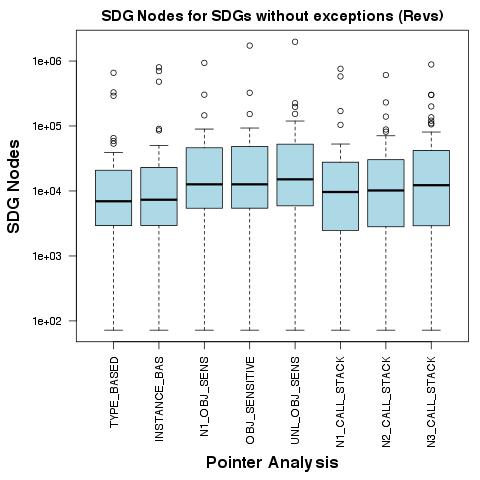}
		\caption{\ac{sdg} Nodes}
		\label{fig:sdgNodesPAIt1}
	\end{subfigure}%
	~ 
	\begin{subfigure}[t]{0.5\textwidth}
		\centering
		\includegraphics[scale=0.3]{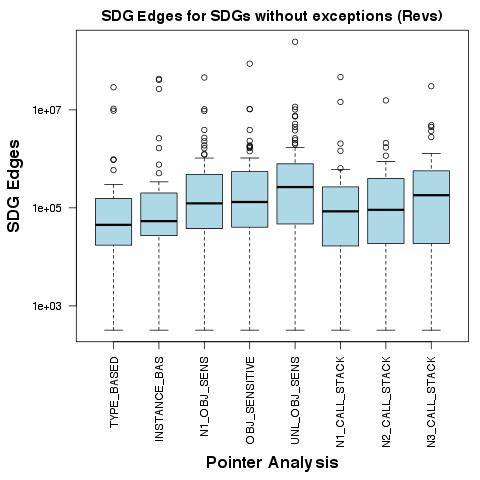}
		\caption{\ac{sdg} Edges}
		\label{fig:sdgEdgesPAIt1}
	\end{subfigure}
	\caption{Boxplots of \ac{sdg} Nodes (a) and Edges (b) for \acp{sdg} without exceptions of all pointer analyses}
	\label{fig:sdgSizePAIt1}
\end{figure*}
As previously discussed, the pointer analysis affected the frequency of information flow occurrence, using \acp{sdg} without exceptions, for 12.5\% of the scenarios from iteration 1. In particular, \emph{unlimited object sensitive} \acp{sdg} contained information flow in 12.5\% less scenarios than \emph{type based} \acp{sdg} (50\% vs 62.5\%). In other words, the increased precision of \emph{unlimited object sensitive} \acp{sdg} avoided false positives in 12.5\% of the scenarios. Nonetheless, we also need to understand how this gain in precision affects the \ac{sdg} size. With that goal, \cref{fig:sdgSizePAIt1} compares the \ac{sdg} sizes of all pointer analyses for \acp{sdg} without exceptions. More specifically, \cref{fig:sdgNodesPAIt1} compares the number of nodes and \cref{fig:sdgEdgesPAIt1} the number of edges.

As it may be noticed in \cref{fig:sdgSizePAIt1}, \acp{sdg} with \emph{unlimited object sensitive} pointer analysis (unl\_obj\_sens) are considerably heavier than the other pointer analyses. In particular, their median numbers of nodes and edges are 15093 and 265458 respectively, which is 22.96\% and 46.33\% bigger than the number of nodes and edges of the second heavier pointer analysis (\emph{n3 call stack}), and 117\% and 490\% bigger than the lightest pointer analysis (\emph{type based}), respectively. More precisely, \emph{unlimited object sensitive} \acp{sdg} are 12.5\% more precise than \emph{type based} \acp{sdg}, but at the cost of a \ac{sdg} with 117\% more nodes and 490\% more edges. 

\begin{figure*}[t!]
	\centering
	\begin{subfigure}[t]{0.5\textwidth}
		\centering
		\includegraphics[scale=0.3]{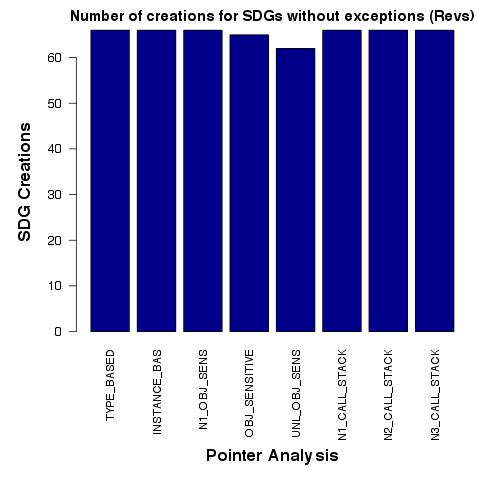}
		\caption{Number of \acp{sdg} Created}
		\label{fig:sdgCreationsBars}
	\end{subfigure}%
	~ 
	\begin{subfigure}[t]{0.5\textwidth}
		\centering
		\includegraphics[scale=0.3]{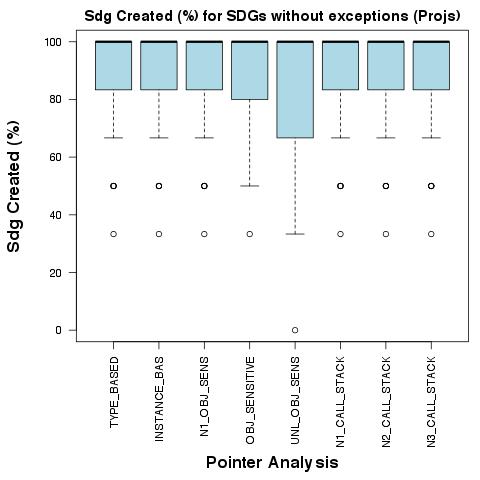}
		\caption{Boxplots with  \% of \acp{sdg} created per project}
		\label{fig:sdgCreationProjs}
	\end{subfigure}
	\caption{Different views of \acp{sdg} created by each pointer analysis}
	\label{fig:sdgCreations}
\end{figure*}

Additionally, we considered not only the size of successfully created \acp{sdg}, but also for the number of times \ac{joana} failed to create the \ac{sdg} for each pointer analysis, as illustrated in \cref{fig:sdgCreations}. \cref{fig:sdgCreationsBars} shows the total number of successful creations for each pointer analysis, and \cref{fig:sdgCreationProjs} shows boxplots, for each pointer analysis, with the percentages of successful creations per project. As it can be viewed in \cref{fig:sdgCreationsBars}, \emph{unlimited object sensitive} was the pointer analysis with the smallest number of successful \ac{sdg} creations. Moreover, this smaller number of successful creations is true for multiple projects from iteration 1 sample, as the first quartile of the \emph{unlimited object sensitive} boxplot from \cref{fig:sdgCreationProjs} is noticeably lower than the first quartile from the other boxplots, indicating a lower percentage of successful creations for 25\% of the projects. 

\textbf{In summary, \emph{unlimited object sensitive} \acp{sdg} are considerably heavier (22.96\% to 117\% on the number of nodes and 46.33\% to 490\% on the number of edges) than the other pointer analyses and in some cases this leads to \acp{sdg} too heavy to be computed. Given the precision gains of this pointer analysis only varied from 2.5\% (compared to \emph{object sensitive} and \emph{n3 call stack}) to 12.5\% (compared to \emph{type based}), we decided to discard it as it is too heavy}. In particular, compared to other pointer analyses, the precision gains of this pointer analysis affected the result for between 1 ($1 / 40 = 2.5\%$) and 5 ($5 / 40 = 12.5\%$) scenarios (from 5 different projects).

\paragraph{Failures to create \ac{sdg} and their reasons}
\begin{figure}[t]
	\centering
	\includegraphics[scale=0.5]{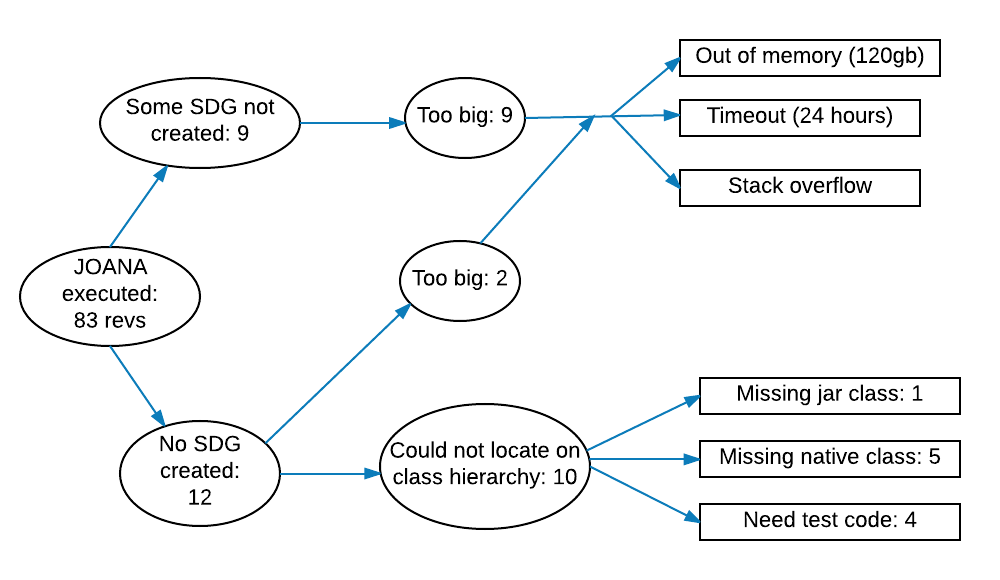}
	\caption{Summary of \ac{sdg} failures}
	\label{fig:sdgFail}
\end{figure}
As previously discussed, there are situations where the \ac{sdg} is not successfully created for some configurations, or even for all. As illustrated in \cref{fig:sdgFail}, there are different reasons for that. The scenarios were divided according to if a \ac{sdg} was created for some configuration, but not for others (9 scenarios) or if the \ac{sdg} failed to be created for all configurations (12). As it can be noticed, when the \ac{sdg} was not created only for some configurations, the reason was always because the \ac{sdg} was too heavy (or big). This may be identified by an out of memory or stack overflow errors or by a timeout of 24 hours.

On the other hand, only 2 of the 12 scenarios which the \ac{sdg} was not created for all configurations failed because the \ac{sdg} was too heavy. As previously mentioned in \cref{chap3:strategy}, \aca{joana} process of \ac{sdg} computation involves creating a class hierarchy. In particular, most of these scenarios (10 out of 12) failed because one class necessary to create the \ac{sdg} was not found in the class hierarchy created by \ac{joana}. Ideally, this class hierarchy contains the compiled classes from a system and also the classes from external dependencies and native classes. As detailed in \cref{fig:sdgFail}, the first reason for failing to locate a class in this class hierarchy is when a necessary external dependency, represented by a class from a jar file, is missing. Similarly, the second reason is when a necessary native class is missing. More precisely, \ac{joana} only analyses up to Java 1.5 and its default is version 1.4. As a result, classes from more recent Java versions raise this problem. 

Finally, the last reason is when the method passed as entry point is actually from a test and as a result it would be also necessary to compile test code, which is not always the case in our infrastructure. Regarding these failures due to lack of test compilation, as previously discussed (in \cref{infra}), the build commands used to compile projects vary depending on the project used and its build system. In particular, projects compiled with \emph{Maven} are the exception. In the case of these projects, the compilation command used by us (\textit{mvn compile}) does not compile tests and a different command is necessary to also compile them (\textit{mvn test-compile}). In contrast, for projects compiled with \emph{Ant} or \emph{Gradle}, the used commands were project specific, and, as a result, they may or may not include test compilation, depending on the command itself. 

\paragraph{Reasons for not identifying instructions}
\begin{figure}
	\centering
	\includegraphics[scale=0.25]{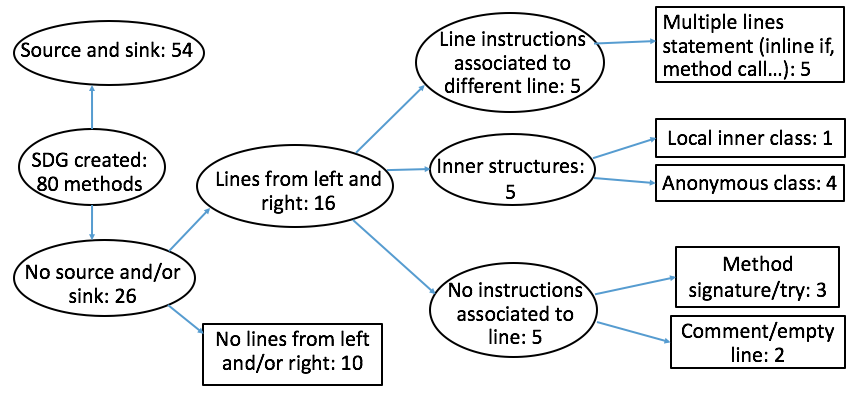}
	\caption{Summary of reasons for not identifying instructions (no source and/or sink)}
	\label{fig:noSrcAndSink}
\end{figure}

Similarly, as mentioned in the beginning of this iteration, we also found situations, where, although left and right edited a method, we were not able to identify instructions corresponding to those editions for at least one of these contributions (left or right). In particular, for 26 methods from 22 scenarios that had changes from both left and right, we were not able to identify instructions corresponding to those editions for at least one of the contributions (left or right). 

As detailed in \cref{fig:noSrcAndSink}, this may occur for a number of reasons. For example, in 10 of the 26 methods there was no actual source code line edited by left (or right). As discussed in \cref{subsection:mergeStep}, one explanation for these cases is: left (or right) only removes lines, since the removed lines (and equivalent instructions) will not exist any more in the integrated revision. Another one is when all the editions done by left (or right) were also identically edited by right (or left), as we do not consider identical lines in the list of edited lines (as discussed in \cref{subsection:mergeStep}). Nevertheless, for 16 of the 26 methods there were source code lines edited by left and right, but no corresponding instructions were found for at least one of the contributions (left or right). 

For 5 methods, there was no instructions associated to the edited lines. One of the explanations for this is: the edited line corresponds to a comment (comments have no instructions associated to them). Similarly, this might also happen to lines corresponding to a method signature or annotation.

Additionally, for other 5 methods, the edited lines were inside inner structures such as local inner classes and anonymous classes. The problem with these inner structures is that, generally, only the inner element instantiation is directly included in the list of instructions from the method, while its internal code is not included.
Furthermore, for other 5 methods the reason is that, sometimes, all the instructions from statements broken in multiple lines are associated to the first line; as a result other lines from the statement have no instruction directly associated. More precisely, one of the occurrences of this situation was an inline if statement broken in three lines: one for the condition, one to be executed if the condition is true and other with the else part. For this case, although the statement was broken in three lines, all the instructions corresponding to this statement were associated to the first line (line 1), as if the statement was written in a single line. Hence, if lines 2 and 3 are edited but line 1 is not changed, no instructions are identified as all the instructions were associated to line 1. As in the case of the inner structures, we could specifically look for these situations (we could mark line 1 in our example). However, for simplicity, we also do not do that.

Finally, for one of the methods, the problem was actually that the signature passed to \ac{joana} was different from expected. More specifically, \ac{joana} expects inner classes in the format $OutterClass\$InnerClass$ and we passed $OutterClass.InnerClass$. This problem was actually fixed for later iterations.

\subsubsection*{Iteration 2}
This iteration involved the addition of 36 new scenarios to what was analysed in iteration 1, resulting in a total of 119 scenarios where we executed \ac{joana}. Compared to iteration 1 data, 16 new projects were also included, giving a total of 50 analysed projects. For all the 36 added scenarios, the \acp{sdg} were created considering external dependencies. Furthermore, for the first 20 of the 36 added scenarios, we passed stubs for the \ac{jre} 1.4. However, after executing the scenarios from iteration 1 and those 20 scenarios we realised the more appropriate option was using stubs for the \ac{jre} 1.5, as it considers a more recent version. Thus, the last 16 scenarios were executed with this option. Additionally, of the 119 scenarios (from 50 projects), the \ac{sdg} was created for at least one configuration in 98 scenarios (from 47 projects) and for all configurations, from iteration 2, in 84 scenarios (from 47 projects). Furthermore, from the 84 scenarios, 54 scenarios (from 36 projects) had instructions to annotate from left and right. Thus, iteration 2 data involves 119, 84 or 54 scenarios depending on the aspect being analysed: number of \ac{sdg} creations, number of nodes/edges and information flow, respectively.
\subsubsection*{RQ1B - Which pointer analysis should be used?}
Regarding the pointer analysis, iteration 1 discarded one possibility leaving 7 possible pointer analyses to be considered by iteration 2. This iteration discards 4 other pointer analyses, leaving 3 possibilities to iteration 3.

\paragraph{Grouping pointer analyses by similarity and selecting one pointer analysis from each group}
\begin{table}
    \scriptsize
    \centering
    \begin{tabular}{|c|c|c|}
        \hline 
        \textbf{Group} & \textbf{Pointer Analysis} & \textbf{\%} \\
        \hline 
        \multirow{2}{*}{G1} & Type Based & 62.96 \\
        & Instance Based & 61.11 \\
        \hline 
        \multirow{2}{*}{G2} & N1 Object Sensitive & 55.56 \\
        & Object Sensitive & 53.70 \\
        \hline 
        \multirow{3}{*}{G3} & N1 Call Stack & 61.11 \\
        & N2 Call Stack & 59.26 \\
        & N3 Call Stack & 53.70 \\
        \hline
    \end{tabular}
    \caption{Percentages (\%) of scenarios with direct information flow occurrence in iteration 2}
    \label{table:freqsIt2}
\end{table}
\begin{figure*}[t!]
	\centering
	\begin{subfigure}[t]{0.5\textwidth}
		\centering
		\includegraphics[scale=0.3]{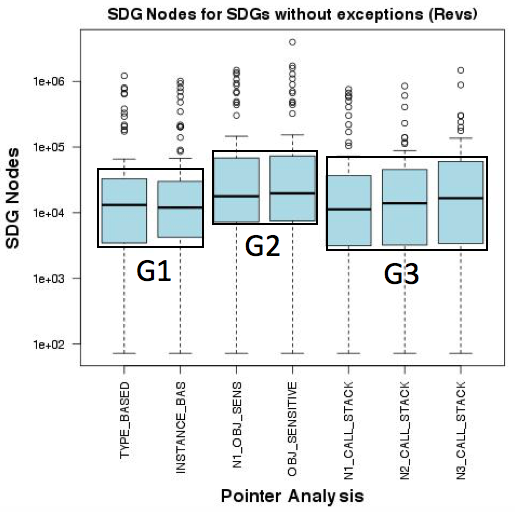}
		\caption{\ac{sdg} Nodes}
		\label{fig:sdgNodesIbIt2}
	\end{subfigure}%
	~ 
	\begin{subfigure}[t]{0.5\textwidth}
		\centering
		\includegraphics[scale=0.3]{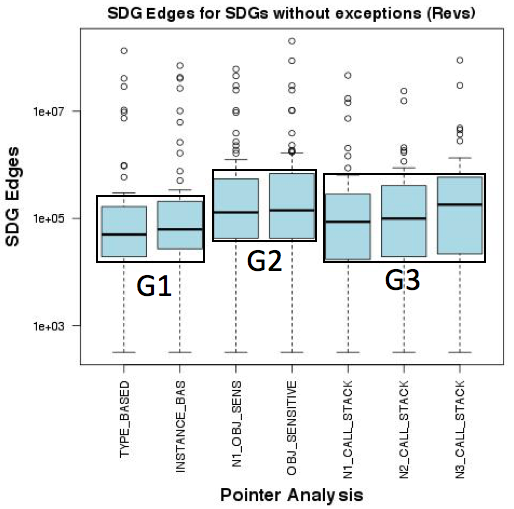}
		\caption{\ac{sdg} Edges}
		\label{fig:sdgEdgesIbIt2}
	\end{subfigure}
	\caption{Grouped boxplots of \ac{sdg} Nodes (a) and Edges (b) for Instance Based \acp{sdg} without exceptions of iteration 2 pointer analyses}
	\label{fig:sdgSizeIbIt2}
\end{figure*}
After analysing the data from iteration 2, we decided to divide the pointer analyses in three groups: Basic (G1), Object Sensitivity (G2) and Nx Call Stack (G3). This division is explicit in \cref{table:freqsIt2} and \cref{fig:sdgSizeIbIt2}. As it may be noticed in \cref{fig:sdgSizeIbIt2}, pointer analyses in the same group have similar numbers of nodes and edges, with more precise pointer analyses tending to have more nodes and edges. For example in the Nx Call Stack group (G3), N1 Call Stack tends to contain fewer nodes and edges than N2 which tends to contain fewer than N3. In contrast, the numbers of nodes and edges tend to vary more for pointer analyses in different groups. Similarly, as it can be viewed in \cref{table:freqsIt2}, in our sample, the frequencies of scenarios with information flow tended to be close for pointer analyses in the same group. For instance, the difference between \emph{instance based} and \emph{type based} in group 1 (G1), and \emph{object sensitive} and \emph{n1 object sensitive} in group 2 (G2), is less than 2\% (62.96 - 61.11 and 55.56 - 53.70).

\textbf{After dividing the pointer analyses in three groups, we select one pointer analysis to represent each group}. As a result, after iteration 2, three pointer analyses are passed as input to iteration 3. Since the differences were small within each group, the selection of a pointer analysis within each group did not indicate significant changes on the results compared to other pointer analyses from the same group. However, to reduce false positives, we decided to select the more precise pointer analysis from each group. So, \textbf{we select \emph{instance based} from group 1, \emph{object sensitive} from group 2 and \emph{n3 call stack} from group 3 as these are the ones that contained information flow in less scenarios in their respective groups.}
\subsubsection*{Iteration 3}
Finally, iteration 3 contains the complete sample. Compared to iteration 2, 38 new scenarios are executed, and there are two new projects, giving a total of 157 executed scenarios from 52 projects. From the 157 executed scenarios, the \ac{sdg} is created for at least one configuration in 123 scenarios (from 48 projects) and for all the configurations of iteration 3, in 91 scenarios (from 48 projects). Additionally, only 58 scenarios (from 36 projects) had instructions to annotate. Thus, iteration 3 data involves 157, 91 or 58 scenarios depending on the aspect being analysed: number of \ac{sdg} creations, number of nodes/edges and information flow, respectively.
\subsubsection*{RQ1B - Which pointer analysis should be used?}
After discarding one pointer analysis from the initial eight in iteration 1 and discarding other four in iteration 2, we select the pointer analysis to be used to detect information flow between contributions from merge scenarios in iteration 3.
\paragraph{Instance Based \acp{sdg} were only slightly less precise}
\begin{table}[t!]
%	\medium
	\scriptsize
	\centering
	\begin{tabular}{|c|c|c| }
		%\tabularnewline
		\hline \bfseries \centering Pointer Analysis & \bfseries Percentages (\%) \\
		\hline Instance Based & \multicolumn{1}{c|}{58.62} \\
		\hline Object Sensitive & \multicolumn{1}{c|}{51.72} \\
		\hline N3 Call Stack & \multicolumn{1}{c|}{51.72} \\\hline
	\end{tabular}
	\caption{Percentages (\%) of scenarios with direct information flow occurrence in iteration 3}
	\label{table:freqsIt3}
\end{table} 

As it can be noticed in \cref{table:freqsIt3}, \textbf{\emph{instance based} \acp{sdg} were about 7\% (58.62 - 51.72 = 6.9) less precise than \emph{object sensitive} and \emph{n3 call stack} \acp{sdg}}. In other words, in around 7\% of the scenarios an \emph{instance based} \ac{sdg} contained information flow, while the more precise ones, \emph{object sensitive} and \emph{n3 call stack}, did not. It is important to mention here that, as expected, there were no scenarios where the \emph{instance based} pointer analysis did not contain information flow while one of the others did. This is expected, because as a less precise analysis compared to the others, an \emph{instance based} analysis is expected to contain information flow for more cases, not less.% (see Pointer analysis in \cref{chap:background}). 

In contrast, although \emph{object sensitive} and \emph{n3 call stack} present the same final percentage (51.72\%), there were scenarios where they differed with respect to the existence of information flow. More precisely, there were two scenarios where a \emph{n3 call stack} contained information flow while an \emph{object sensitive} did not, and also two scenarios where \emph{n3 call stack} did not contain information flow and \emph{object sensitive} did. In particular, the percentages matched because the occurrences of the former and the latter manifested in the exact same quantities (two of each). Therefore, although \emph{object sensitive} and \emph{n3 call stack} differed for some scenarios, their final percentage of frequency of information flow was the same. 

\paragraph{Instance Based \acp{sdg} were successfully created more often}

\textbf{Using an \emph{instance based} pointer analysis, \ac{joana}
	successfully created the \ac{sdg} for more scenarios than 
	with 
	the others (\emph{object sensitive} and \emph{n3 call stack})}. In particular, the \emph{instance based} approach created the \ac{sdg} for 118 scenarios, the \emph{object sensitive} for 105 and \emph{n3 call stack} for 93. Moreover, this tendency was kept across the project list %as demonstrated in \cref{fig:sdgCreationsBoxIt3}
	. More specifically, the first quartile, of percentage of successful creation, from \emph{instance based} is close to 80\%, while from \emph{object sensitive} and \emph{n3 call stack} are below 60\%.
\paragraph{Although Instance Based \acp{sdg} have less precision, they bring the potential of analysing more scenarios}
As just discussed \emph{instance based} \acp{sdg} are less precise, but were created more often. More precisely, we showed that \emph{instance based} \acp{sdg} had about 7\% more scenarios with information flow. In contrast, we also showed that \emph{instance based} \acp{sdg} were successfully created more times. In particular, instance based \acp{sdg} were successfully created for around 75.16\% (118/157) of the executed scenarios, \emph{object sensitive} for around 66.88\% (105 / 157) and \emph{n3 call stack} for around 59.24\% (93 / 157). Thus, \emph{instance based} \acp{sdg} are around 7\% less precise, but are successfully created for around 8\% more scenarios compared to \emph{object sensitive} (75.16 - 66.88 = 8.28) and 16\% more scenarios compared to \emph{n3 call stack} (75.16 - 59.24 = 15.92). 

Considering the percentage of creations and precision, \emph{n3 call stack} is the worst option from the remaining pointer analyses, as it showed the same precision as \emph{object sensitive} (51.72\%), but with a lower rate of successful creations (about 8\% lower than \emph{object sensitive} and 16\% than \emph{instance based}). 

\textbf{Finally, between \emph{instance based} and \emph{object sensitive}, we decided to select \emph{instance based} as the 7\% of scenarios which \ac{joana} is less precise with an \emph{instance based} analysis, in our opinion, are compensated by about 8\% of scenarios which only with this approach \ac{joana} is able to create a \ac{sdg}. Nonetheless, for those who value more precision, \emph{object sensitive} may also be selected, as with this pointer analysis, compared to \emph{instance based}, \ac{joana} was more precise for around 7\% of the scenarios (although \ac{joana} failed to create the \ac{sdg} for 8\% more scenarios with this pointer analysis)}. Alternatively, \emph{n1 call stack} or \emph{n1 object sensitive} could also be used as alternative options in the precision/creations scale between \emph{instance based} and \emph{object sensitive}. To be more precise, theoretically, with both these pointer analyses, \ac{joana} tends to be less precise than \emph{object sensitive} (and more than \emph{instance based}) and to contain more successful creations than \emph{object sensitive} (and less than \emph{instance based}).

\subsubsection*{Iteration 4}
\label{it4}
Since we created \acp{sdg} without dependencies in the first iteration and with stubs for \ac{jre} 1.4 for iteration 1 and part of iteration 2, we decided to rerun these scenarios for our final configuration (instance based without exceptions) with dependencies and stubs for \ac{jre} 1.5. Nevertheless, as both including dependencies and using stubs for \ac{jre} 1.5 tend to increase the \ac{sdg} size, as both include more classes in the class hierarchy, we do not try to rerun cases which the \ac{sdg} was too heavy to be created. In fact, we only rerun cases where the \ac{sdg} was originally created and there were instructions identified from both contributions (there was source and sink%, see \cref{subsect:joana}
), as we noticed this hardly changed between the different configurations we executed. Hence, from the 62 scenarios satisfying these conditions in iterations 1 and 2, we needed to rerun 56, as the other 6 were already executed with dependencies and stubs for \ac{jre} 1.5. Similarly, iteration 3 had 17 additional scenarios which satisfied the conditions but were already executed correctly and as a result our sample to find the frequency of information flow between contributions (RQ2) involved a total of 79 scenarios (from 38 projects).

\subsubsection*{RQ2 - What is the frequency of direct information flow between same-method contributions?}
Once we selected a configuration to be used (instance based without exceptions), we moved on to check the frequency of direct information flow between the contributions from merge scenarios. As previously explained, we wanted to understand if it was common to exist direct information flow between contributions from merge scenarios (high percentages).
\paragraph{It was common to exist direct information between the merge scenarios contributions}
Before the rerun, considering the data from iteration 3, 60.76\% of the scenarios (48 out of 79) contained direct information flow between the contributions. From the 56 re-executed scenarios, for 7 the \ac{sdg} with the final option for dependencies and stubs was too heavy. If we consider the results from the old options for these cases, there is direct information flow for 64.56\% (51 out of 79) of the scenarios. Similarly, if we do not consider them (as the final option was too heavy for them), the obtained frequency for direct information flow is 63.89\% (46 out of 72). Either way, around 64\% of the scenarios contained direct information flow between their contributions. 

As a result, \textbf{we conclude that the occurrence of direct information flow between merge scenario contributions was common, as it occurred on around 64\% of the scenarios. Since we intend to use the existence of information flow to estimate interference, this indicates a considerable number of scenarios which would be identified by our strategy}. More specifically, all the analysed scenarios have no syntactic conflicts. Nevertheless, the existence of information flow (on around 64\% of the scenarios) indicates potential existence of interference (and dynamic semantic conflicts). Furthermore, as we did not consider indirect flows to common target points, this number (64\%) could be even higher.

%Marcelo - reportar custo das análises…
\subsubsection*{Cost of \ac{sdg} creation}
To give a rough idea of the cost of the \acp{sdg} created by \ac{joana}, we measured time to create the \ac{sdg} and memory allocated in this process for scenarios where the \ac{sdg} was successfully created.
\paragraph{Comparison of costs for instance based and object sensitive}
To give an estimative of how different options may affect the cost of a \ac{sdg} we compare time and memory to create a \ac{sdg} for our last two options of configuration: instance based without exceptions and object sensitive without exceptions. We do that for 77 scenarios where the \ac{sdg} was successfully created for both options.

In this sample of 77 scenarios, \ac{joana} was able to create instance based \acp{sdg} considerably faster than object sensitive \acp{sdg}. In particular, the median (about 40 seconds) and the third quartile (around 1 minute and 38 seconds) from the first are lower than the median from the second (about 2 minutes and 26 seconds). Furthermore, the third quartile from object sensitive was more than 20 minutes, which is more than ten times the third quartile obtained using instance based (1 minute and 38 seconds). As a matter of fact, except for one outlier that needed around two hours and a half to compute, instance based \acp{sdg} successful creations took up to 20 minutes. In contrast, there are multiple outliers where using object sensitive, \ac{joana} took hours to compute the \ac{sdg}. To be more specific, 11 cases took more than 1 hour to compute, 5 took more than 4, and 1 took more than 13 hours.

%As it may be seen in \cref{fig:memory}, with respect to memory we observed a similar tendency. 
We observed a similar tendency when evaluating memory. That is, using instance based \acp{sdg}, \ac{joana} tends to consume less memory than using object sensitive \acp{sdg}. In particular, the obtained medians were 6.55 GB and 13.77 GB, respectively. As a matter of fact, the third quartile from instance based (10.7 GB) is also lower than the median from object sensitive (13.77 GB). Additionally, the third quartile for object sensitive (20.18 GB) was almost two times the third quartile for instance based (10.7 GB). Except for one outlier, instance based \ac{sdg} consumed up to 22 GB. In contrast, there are 18 occurrences where using object sensitive \acp{sdg} more than 22 GB was necessary, 13 occurrences where more than 40 GB of memory was necessary, and 4 occurrences where more than 60 GB was necessary. 

%\begin{figure}
%	\centering
%	\includegraphics[scale=0.3]{imgs/Revs_PrecisionsPlot_Memory_noExcep.jpg}
%	\caption{Boxplots of memory allocated (in GB) to successfully create the \ac{sdg} for instance based and object sensitive \acp{sdg} without exceptions}
%	\label{fig:memory}
%\end{figure}

\paragraph{Costs of instance based considering extra scenarios that \ac{sdg} creation fails with other configurations}
Additionally, we extend the initial sample of 77 scenarios, with 23 other scenarios where \ac{joana} was able to create the \ac{sdg} using our final configuration (instance based without exceptions), but it failed to create one using an object sensitive pointer analysis. Thus, this sample includes 100 scenarios where it was possible to create a \ac{sdg} using our final configuration. We hope, with this extended sample, to bring a more accurate idea of cost of \ac{sdg} creation, by also considering potentially heavier cases that \ac{joana} fails to create the \ac{sdg} using other configurations. 

%As it may be seen in \cref{fig:timeIb}, 
We observed that even when also considering cases that fail using other configurations, \ac{joana} is generally able to compute the \ac{sdg} for our final configuration in a matter of minutes. More precisely, \textbf{the time of computation of the \acp{sdg} from the extended sample was at most 1 minute and 5 seconds for 50\% of the cases, at most 8 minutes and 35 seconds for 75\% of the cases, and at most 1 hour for 90\%. To conclude, 98\% of the cases took at most 4 hours}. 

Including heavier cases had some impact in the overall amount of memory consumed. However, most of the cases may still be computed using less than 30 GB of memory. In particular, \textbf{50\% of the cases used at most 8.2 GB, 75\% of the cases used at most 16.1 GB, and 90\% of the cases used at most 29 GB of memory}.

\subsubsection{Manual Analysis}

Using the automatic analysis, we established a \ac{sdg} configuration to run \ac{joana} (RQ1) and then identified that it was common to exist direct information flow between merge scenarios contributions (RQ2). However, as we intend to use the existence of information flow to estimate interference, we still need to understand in which situations there is information flow and no interference (RQ3) to be able to determine if the occurrence of information flow may help to identify interference. In other words, to conclude our study we conducted this manual analysis to understand if a considerable number of scenarios with information flow indicates a considerable number of scenarios with interference. Furthermore, we also investigate in which situations the existence of information flow does not represent the existence of an interference. 

\subsubsection*{RQ3 - In which situations there is information flow and no interference?}
\label{rq3}
\begin{figure*}[t!]
	\centering
	\includegraphics[scale=0.42]{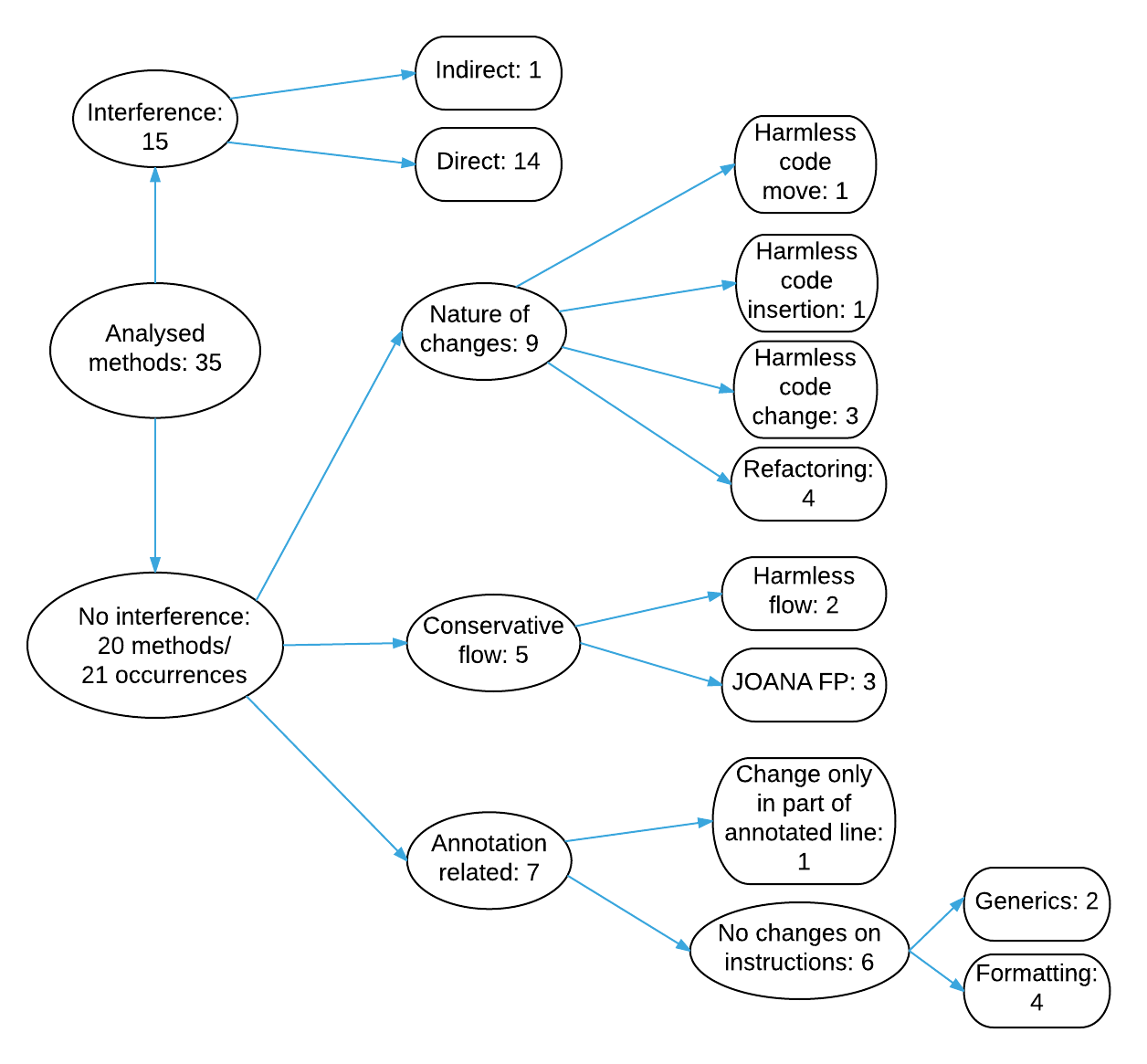}
	\caption{Manual analysis overview}
	\label{fig:manualAnalysis}
\end{figure*}

The results from the manual analysis are accessible from \cref{links} and \cref{fig:manualAnalysis} gives an overview of the obtained results
. Furthermore, it is important to mention in \cref{fig:manualAnalysis} the category with no interference has one extra occurrence compared to the number of methods (21 occurrences of no interference in 20 methods) due to a method (from revision 4a97b-ebb08 from tachyon) where left did a refactoring while right did a change only on formatting. More precisely, for this case, individually, both the refactoring and change on formatting would already be enough for defining that there is no interference. However, as both happen, for this case, we consider two occurrences for a single method, one of refactoring and one of formatting. 

As it may be noticed in \cref{fig:manualAnalysis}, from the 35 analysed methods (from 35 different scenarios) we considered that there was interference only in 15 (42.86\%). From the 15 cases with interference, 14 had direct interference and one had an indirect one. From these, some were already discussed in this work. In particular, the example used in \cref{fig:interfNoconf} was inspired by a case of direct information flow from the manual analysis (revision aa201-4585d from opentripplanner). 
Similarly, the case of indirect interference (revision e064b-38e20 from jsoup) inspired the example from \cref{both-flow} (discussed in \cref{ifcStep}).

Among the cases with no interference, but with information flow reported by \ac{joana}, we identified three major groups of causes for having information flow and no interference: those related to limitations of our annotation strategy, those related to the nature of the changes, and those related to the conservativeness of the flows.

\paragraph{Annotation limitations}
Regarding annotation limitations, we identified two different causes: no actual change on instruction level, and annotating an entire line while only part of it was changed. Both involve information flow that should not be detected because we annotated instructions that should not be annotated. 

First, consider the cases where there was no actual change on the instruction level. Identified flows involving unchanged instructions are not relevant for estimating interference because, if an instruction was not changed during a merge scenario, then there is no interference involving this instruction (as the instruction in question is the same for the revisions involved in the merge scenario - base, left and right). 

Two causes for an unchanged instruction being marked as contribution were observed: changes only on formatting and changes on generics.\footnote{Java generics - \url{https://docs.oracle.com/javase/tutorial/java/generics/why.html}} More precisely, changes only on a line's formatting (spacing for example) lead us to identify the line in question as changed and mark the instructions related to the line, but these instructions did not actually change; as a result there is no actual contribution with semantic effect that can cause interference. Similarly, in Java 5, changes only on generics (only inserting/changing/removing generics on existing code) do not alter the program at the bytecode level (instruction level), so we end up marking unchanged instructions that will not lead to interference.

\begin{figure*}[t!]
	\begin{subfigure}{\linewidth}
		\centering
\begin{lstlisting}[language=Java,numbers=left,xleftmargin=22pt]
ClientShell(ClientConfig clientConfig) {
      //...       
      adminClient = new AdminClient(url, new ClientConfig());
      
      //...    
}
\end{lstlisting}
		\vspace{-1em}
		\caption{Base}
		\label{fig:paramBase}
	\end{subfigure}

	\begin{subfigure}{\linewidth}
		\centering
\begin{lstlisting}[language=Java,numbers=left,xleftmargin=22pt]
ClientShell(ClientConfig clientConfig) {
      //...
      adminClient = new AdminClient(url, clientConfig); //left
      
      //...    
      strategy = createStrategy(adminClient.getUrl()); //right
}
\end{lstlisting}
		\vspace{-1em}
		\caption{Integrated}
		\label{fig:paramFinal}
	\end{subfigure}
	
	\caption{Example of change only on parameter of existing instantiation}
	\label{fig:changeParam}
\end{figure*}

With respect to the second cause for annotation problems, the actual problem was a change on a parameter of an existing method call. To illustrate, consider \cref{fig:changeParam}, which is inspired in the original scenario from our manual analysis (revision 24c82-64c90 from voldemort). In this example, left changed only the second parameter of the instantiation of \emph{AdminClient} (changed from \textit{new ClientConfig()} in line 3 from \cref{fig:paramBase} to \emph{clientConfig} in line 3 from \cref{fig:paramFinal}) and right inserted a line (line 6 from \cref{fig:paramFinal}). Although left modified only a parameter from the instantiation, our annotation strategy annotates the complete instantiation; as a result \ac{joana} detects information flow as the first parameter from the instantiation is used by right (in line 6 from \cref{fig:paramFinal}). In contrast, if we annotated correctly (only the second parameter from the instantiation), then no information flow would be detected, as right's change does not depend on the second parameter.

\paragraph{Conservativeness of flows}

\begin{figure*}[t!]
		\centering
\begin{lstlisting}[language=Java,numbers=left,xleftmargin=22pt]
BasicSimianContext(Properties properties) {
    for(Entry<Object, Object> prop : properties.entrySet()) { //left
        Object key = prop.getKey(); //left
        Object value = prop.getValue();  //left
        LOG.info(String.format("%s = %s", key, value)); //left
    this.config = new BasicConfiguration(properties);  //left

    this.account = config.getStr("simian.client.accountKey");
    this.user = config.getStr("simian.client.user"); //right
    this.password = config.getStr("simian.client.password"); //right
    //...
}
\end{lstlisting}
		\vspace{-1em}
	\caption{Example of harmless conservative flow for estimating interference}
	\label{fig:consFlow}
\end{figure*}

Furthermore, there were also situations where interference and information flow differed due to the conservativeness of the flows identified by \ac{joana}. More precisely, as a security tool \ac{joana} may be too conservative leading to false positives of information flow, or situations where the identified flows are useless to our context of estimating interference. These cases are classified as \emph{JOANA FP} (from \ac{joana} False Positives) and \emph{harmless}, respectively. In particular, cases classified as \emph{JOANA FP} correspond, in our opinion, to false positives of information flow. To be more specific, \ac{joana} detects information flow for these cases (without exceptions and with an \emph{instance based} pointer analysis), but we consider that there is no actual information flow. Additionally, for cases classified as \emph{harmless} flow, there is an actual flow but the flow is considered irrelevant for estimating interference. 

For example, in \cref{fig:consFlow} we show a situation (inspired on revision 24b39-b9e7b from SimianArmy) where the flows identified by \ac{joana} are \emph{harmless} flows for estimating interference. More specifically, left (lines 2 to 6) includes a for to iterate through all the properties and for each property logs the key and value from the respective property, while right (line 10 and 11) gets the value for two specific properties (\emph{user} and \emph{password}). For this example, \ac{joana} detects flow from lines 10 and 11) to lines 3 to 5) as the properties \emph{user} (line 10) and \emph{password} (line 11) are also logged in line 5. From a security context the identified flows are relevant as sensitive information (such as the password from line 11) is being logged in line 5 and hence may leak threatening confidentiality. In contrast, for our context of estimating interference the identified flows are not relevant as both left and right are reading from the same properties, but are not modifying; as a result, this way they can not interfere with each other. 

\paragraph{Nature of changes}

In contrast, there are cases with relevant information flow between the contributions, but due to the nature of changes there is no interference. More precisely, for cases in this category, after understanding the actual contributions made by left and right (and their scope) we observed that, although there is a relevant information flow between the contributions, there is no interference. Cases in this situation may occur for a number of reasons such as when one of the contributions is actually a refactoring. For those cases, as a refactoring does not change program behaviour \cite{fowler1999refactoring}, there is no interference.

Furthermore, there are other situations where due to the nature of changes there is no interference. More precisely, we identified cases of harmless code \emph{change}, \emph{insertion} and \emph{move} (representing harmless situations of modifying existing code, inserting new code and just moving the position of existing code, respectively). \cref{fig:changeCode} illustrates a situation of harmless code change (inspired on revision beaab-45462 from hector). In particular, \cref{fig:baseChangeCode,fig:finalChangeCode} show the evolution from class \emph{HSaslThriftClient} on this merge scenario, while \cref{fig:TFrame} details the code from class \emph{TFrameTransport} which is instantiated from \emph{HsalThriftClient} (line 8 from \cref{fig:baseChangeCode} and 12 from \cref{fig:finalChangeCode}). In this example, there is information flow from left (line 4 from \cref{fig:finalChangeCode}) to right (line 7 from \cref{fig:finalChangeCode}) because left affects the field \emph{transport} which is used by right. Furthermore, the identified flow is relevant because left's modification determines the value of \emph{transport}. Nevertheless, if we investigate the nature of the changes, we will notice that although left affects \emph{transport}, right modification does not involve \emph{transport} and as a result they do not interfere. 

\begin{figure*}[t!]
	'\vspace{0.4em}
	\begin{subfigure}{\linewidth}
		\centering
\begin{lstlisting}[language=Java,numbers=left,xleftmargin=22pt]
class HSaslThriftClient {
    TTransport transport;
    public HSaslThriftClient open() {
       transport = oldTranspCalc(socket,service,client);
	
       //...
       if(host.getUseTFramedTransport()) {
          transport = new TFramedTransport(transport);
       }
       return this;
    }
}
\end{lstlisting}
		\vspace{-1em}
		\caption{HSaslThriftClient base}
		\label{fig:baseChangeCode}
	\end{subfigure}
	\vspace{0.4em}
	\begin{subfigure}{\linewidth}
		\centering
\begin{lstlisting}[language=Java,numbers=left,xleftmargin=22pt]
class HSaslThriftClient {
    TTransport transport;
    public HSaslThriftClient open() {
       transport = newTransportCalc(socket, service); //left
	
       //...
       transport = wrapTFrame(transport); //right
       return this;
    }
    protected TTransport wrapTFrame(TTransport transp) { //right
       if (host.getUseTFramedTransport()) { //right
          return new TFramedTransport(transp,host.maxLen()); //right
       } else { //right
          return transp; //right
       } //right
    } //right
}
\end{lstlisting}
		\vspace{-1em}
		\caption{HSaslThriftClient integrated}
		\label{fig:finalChangeCode}
	\end{subfigure}
	\vspace{0.4em}
	\begin{subfigure}{\linewidth}
		\centering
\begin{lstlisting}[language=Java,numbers=left,xleftmargin=22pt]
class TFramedTransport extends TTransport{
    static final int DEFAULT_MAX_LENGTH = 0x7FFFFFFF;
    int maxLength; TTransport transport;
    public TFramedTransport(TTransport transp) {
       this.transport = transp;
       this.maxLength = TFramedTransport.DEFAULT_MAX_LENGTH;
    }
    public TFramedTransport(TTransport transp, int maxLen) {
       this.transport = transp;
       this.maxLength = maxLen;
    }
}
\end{lstlisting}
		\vspace{-1em}
		\caption{TFramedTransport}
		\label{fig:TFrame}
	\end{subfigure}
	\caption{Example of harmless code change}
	\label{fig:changeCode}
\end{figure*}

More precisely, right's modification was very close to a refactoring, it involved calling a generic method (\emph{wrapTFrame}) to do a very similar job compared to what was being done on base (lines 7 and 10-16 from \cref{fig:finalChangeCode} and 7-9 from \cref{fig:baseChangeCode}). To be more specific, if the condition \textit{host.getUseTFramedTransport()} used by both (line 7 from \cref{fig:baseChangeCode} and 11 from \cref{fig:finalChangeCode}) evaluates to \textit{false} then there is no difference in behaviour for this change as in both cases the field \emph{transport} will stay unchanged. In other words, in \cref{fig:baseChangeCode} line 8 will not be executed, staying with the original value, and in \cref{fig:finalChangeCode} line 14 will be returned with the original value of \emph{transport}. On the other hand, if \textit{host.getUseTFramedTransport()} evaluates to \textit{true} then there is a small difference in behaviour caused by right's change, but this difference does not involve \emph{transport}. As it may be noticed, for this case, the signatures of \emph{TFramedTransport} invoked in \cref{fig:baseChangeCode} and \cref{fig:finalChangeCode} are different (line 8 from \cref{fig:baseChangeCode} and 12 from \cref{fig:finalChangeCode}). The former has only \emph{transport} as parameter while the latter has also an extra value. If we observe \cref{fig:TFrame} and compare the constructors used in the base and integrated revisions (lines 4-7 and 8-11 from \cref{fig:TFrame}, respectively) it is possible to notice that the difference is actually for the field \emph{maxLength} (lines 6 and 10 from \cref{fig:TFrame}, respectively), while \emph{transport} is the same in both cases (lines 5 and 9 from \cref{fig:TFrame}, respectively). Hence, although left affects \emph{transport}, right's change has no difference on behaviour with respect to \emph{transport} and as a result there is no interference.

\subsubsection*{Discussion}
In summary, as we discussed, \textbf{a considerable number of methods with information flow and no interference were found: 20 from 35 or 57.14\%}. Nevertheless, from the 21 occurrences of information flow and no interference among these 20 methods, a third ($7 / 21 = 33.33\%$) were due to annotation limitations. Hence, a more advanced annotation system with knowledge of the difference in the instruction level (not only in the source code level) could solve this type of problem by identifying and discarding cases where one of the contributions contain no difference in the instruction level (6 occurrences) and also by identifying the actual changes made in more detail in order to annotate only parts of statements (1 occurrence) instead of only the full statements (including unchanged parts).

Furthermore, as \ac{joana} has a security context it is conservative and that was reflected in almost a quarter ($5 / 21 = 23.8\%$) of the cases with information flow and no interference. A different, less conservative \ac{ifc} tool, could be used to avoid those cases. Nevertheless, it is important to mention that the use of such a tool would need to be carefully evaluated, as the number of false negatives would potentially increase.

Lastly, for around 43\% ($9 / 21 = 42.86\%$) of the cases with no interference the difference of conclusion involved the actual nature of the changes. Solving these cases is harder as the problem with them is actually related to using \ac{sdg} as structure to represent the program semantics. In particular, although a \ac{sdg} represents the information flow through a program, it still largely depends on the syntactic components to represent its nodes. As a result almost any syntactical change is seen as a difference even when there is no semantic effect \cite{jackson1994semantic}. As a matter of fact, refactoring was the type of change with more occurrences in the nature of changes category (4 from 9 occurrences). In fact, we expect approaches that use \acp{sdg} to represent the program semantics to present similar issues and consequently a different approach would be necessary to solve this type of issue. In fact, Yang, Horwitz and Reps \cite{yang1992program} proposed an extension of the original work from Horwitz, Prins and Reps \cite{horwitz1989integrating} that specifically looks for semantics-preserving transformations. In particular, this work helps to avoid the erroneous identification of cases of refactoring, but not the other issues related to the nature of changes. Similarly, a tool such as SafeRefactor \cite{soares2010making} could be used to identify if a change preserves behaviour. In that case, ideally, we could eliminate the problems of erroneous identifications due to refactorings, as we could avoid marking changes that were only a refactoring (4 / 21 = 19.05\%), but the other issues related to the nature of changes would still occur. 

Finally, \textbf{we conclude that information flow between contributions may be used to estimate interference}. More specifically, we classified 42.86\% of the scenarios with information flow between contributions to be also interference. Furthermore, \textbf{although we obtained a considerable amount of false positives of information flow as an estimator of interference (57.14\%), as just discussed, improvements to our strategy would potentially solve most of the false positives}. In particular, a more advanced annotation system could solve annotation problems (33.33\% of the false positives). Additionally, using a less conservative \ac{ifc} tool could help to avoid cases of conservative flows (23.8\% of the false positives). Finally, a tool such as SafeRefactor could be used to check if changes are only refactorings (19.05\% of the false positives). Therefore, \textbf{there is room for solving around three quarters ($33.33\% + 19.05\% + 23.8\% = 76.18\%$) of the obtained false positives}. 

\subsection{Threats to validity}
Our evaluation naturally leaves open a set of potential threats to validity, which we explain in this section.

\subsubsection{Construct}
As previously discussed in \cref{ifcStep} we use \acp{sdg} without concurrency which is a threat because we may miss some cases of valid information flow due to this decision. In particular, as we do not consider concurrency, our analysis is incomplete. Nevertheless, as previously argued, we took this decision for performance reasons as including concurrency is computationally expensive and could significantly reduce the number of analysed cases.

In addition, we use the notion of interference with the goal of predicting dynamic semantic conflicts, but we do not evaluate the correspondence between interference and dynamic semantic conflicts. In particular, our evaluation is restricted to comparing information flow and interference, but our final goal is detecting dynamic semantic conflicts. As we argue in \cref{chap3:sec2}, we focus in interference because defining if a dynamic semantic conflict exists involves understanding the expected behaviour of a system before and after integration of contributions from different developers and such behavioural specifications are often hard to capture, formalize and reason about.

Additionally, one could argue that using an open-world perspective of interference may lead to the identification of some irrelevant interference in practice. However, as we explain in \cref{perspective}, a closed-world perspective requires considering the complete calling context of the system being evaluated and as a result is both computationally expensive and time-consuming to do manually. Thus, we use an open-world perspective even though it has the potential of flagging some cases that are not relevant in practice. In particular, from the 15 cases with interference from the manual analysis, for at least 4 the result could potentially diverge using a closed-world perspective.

Moreover, we focus on same-method interference by calculating same-method information flow. This is a threat to our work as this is only a subset of interference (and information flow respectively). In particular, our evidence of frequency of information flow between contributions is actually evidence of frequency of information flow between same-method contributions, which may be different from information flow in general. Similarly, all of our data and derived conclusions are restricted to same-method information flow (and same-method interference) which is different from information flow (and interference) in general. %For instance, our comparison between \ac{sdg} configurations and respective conclusions that lead to the selection of \emph{instance based \acp{sdg} without exceptions} was done for information flow between same-method contributions, not in general. 
Nevertheless, as discussed in \cref{sameMethod}, we decided to reduce the scope of interference (and consequently of information flow) because we needed a place to start and considered the problem too big to be entirely dealt with from the beginning.

\subsubsection{Internal}
Regarding the question of which configuration should be used (RQ1), a potential threat to internal validity is the fact that we changed two aspects of the created \acp{sdg} during the evaluation. More precisely, we initiated the study in iteration 1 of our automatic analysis without considering dependencies and using stubs for the \ac{jre} 1.4. In the start of iteration 2, we started to consider the dependencies and in the middle of this iteration we updated the stubs version to \ac{jre} 1.5. Although we did not vary those aspects (dependencies and stubs version) between different configurations (of pointer analysis and exceptions use) of a merge scenario, we varied them between different scenarios and as a result one could argue that this could affect our results. We argue this threat is only related to the selection of our final configuration (RQ1) because in iteration 4 we re-executed the scenarios with the final option (considering dependencies and stubs for \ac{jre} 1.5) for our final configuration (instance based without exceptions). 

Nevertheless, our decision to which configuration should be used (instance based without exceptions) was still threatened by the inclusion of dependencies and update of stubs version during the study. As a result, we decided to rerun the necessary scenarios with the final option considering dependencies and with stubs to \ac{jre} 1.5. To remove the threat related to the exceptions use (RQ1A), we rerun scenarios from iteration 1 (as this iteration was the one responsible for this decision) now passing dependencies and stubs for \ac{jre} 1.5 for \emph{instance based \acp{sdg} without exceptions} and also for \emph{instance based \acp{sdg} with exceptions}. After this re-execution, we obtained frequencies of information flow of 68.42\% for the former and 89.47\% for the latter, still maintaining a considerable difference of extra scenarios in which only the configuration with exceptions contained information flow (was 25\% and now was around 21\%).

In contrast, to completely remove the threat associated to the selection of the pointer analysis we would need to re-execute the scenarios for all the options of pointer analysis available (eight). Therefore, we decided to only mitigate this threat by only executing an extra pointer analysis that was only discarded in our final decision in iteration 3 and which we concluded that was more precise but also heavier. More precisely, we rerun scenarios for our selected pointer analysis (\emph{instance based} \acp{sdg}) and also for \emph{object sensitive}, as this was the last pointer analysis to be discarded. We re-execute the same 56 scenarios re-executed in \cref{it4}. After the re-execution, the difference which we observed between number of creations and precision was actually increased, reinforcing our conclusion to use \emph{instance based} over \emph{object sensitive}. More precisely, before the re-execution \emph{instance based} were less precise in about 7\% of the scenarios, but successfully created for around 8\% more scenarios. In contrast, after the re-execution the difference in precision was reduced to 3.7\% (62.96 - 59.26 = 3.7), while the difference of creations increased to 13.37\% (70.70 - 57.33 = 13.37). This occurred, because including dependencies and updating the \ac{jre} stubs version tend to increase the \ac{sdg} size and as a result the number of scenarios with \acp{sdg} too heavy to be created raised. In particular, from the 56 re-executed scenarios, for 7 extra scenarios both options were too heavy to be created and for other 8 only the \emph{object sensitive} analysis was too heavy.

%Marcelo - nao entendi por que vc. classificou algumas ameacas como "internal". eu esperava que fossem externas (por exemplo, as decisoes de estrategia de anotacao).
%Uirá - Algum caso de remoção nos 36%?
Furthermore, as discussed in \cref{chap3:strat} we miss contributions involving removed lines or identical lines. This is a threat to our work as we may analyse scenarios that have these types of contributions if they also have the other types of contributions. In particular, our analysis may erroneously classify some scenarios, that would be classified as with information flow if the removed lines have been taken in consideration, as without information flow. More precisely, there may be interference involved in those types of editions which we are not currently able to detect. As previously explained, with respect to removed lines the issue is directly related to our decision of using a single revision (integrated) to represent the contributions from the merge scenario and in particular of using a single \ac{sdg}. Therefore, we miss those cases with the goal of improving performance and obtaining a practical approach. Similarly, the issue with identical lines is related to the fact that as we check for information flow in the integrated revision between lines from left and lines from right, we cannot annotate a specific line in the integrated revision as arising from both left and right at the same time.

%Marcelo - nao entendi por que vc. classificou algumas ameacas como "internal". eu esperava que fossem externas (por exemplo, as decisoes de estrategia de anotacao).
Additionally, as discussed in \cref{it1} there are significant reductions associated to our sample, specially for considering frequency of information flow. More specifically, from the 157 scenarios (from 52 projects) which we executed \ac{joana}, the \ac{sdg} was created for our final configuration in 111 scenarios (from 47 projects) and only in 71 scenarios (from 36 projects) the \ac{sdg} was created and there were instructions to annotate from both left and right. Nevertheless, as previously detailed in \cref{fig:noSrcAndSink} of \cref{it1}, some of the cases without annotated instructions are actually due to limitations of our annotation strategy such as cases where the editions are inside inner structures (anonymous classes for example). So, since we only consider cases with instructions annotated from both left and right to calculate information flow frequency, we miss cases containing only this type of contribution, reducing the amount of analysed data. Furthermore, this may lead us to analyse scenarios considering only part of their actual contributions. More specifically, if left does an edition that we are not able to identify (such as an edition inside an anonymous class), but also another edition that we are able to identify, we annotate the second and miss the first. Hence, we may analyse scenarios with only part of their contributions, which may lead us to erroneously classify scenarios with valid information flow as without information flow.

Furthermore, there are threats involving both our automatic and manual analyses associated to the methods we selected to analyse. More precisely, 9.79\% (14 / 143 = 9.79) of the methods which the \ac{sdg} was successfully created for our selected configuration (in our automatic analysis), are analysed by more than one different scenario, which may slightly bias our data with the characteristics of those methods. In other words, for those 14 methods, the predictor of edition on the same-method occurred in different merge scenarios and as a result we analysed small variations of these methods. Additionally, in our manual analysis there is only one case of repeated method (from the 35 analysed methods).

In addition, 8.39\% (11 / 143) of the methods from the automatic analysis, where an \emph{instance based \ac{sdg} without exceptions} was created, were actually methods from test code. In particular, for those cases, a method from test code was edited by both contributions flagging  our predictor of editions on the same method. Since we consider test code less relevant than the main source code of a program, we did not analyse methods from test code in our manual analysis.

Lastly, we observed a considerable amount of constructors within our sample. In particular, 13.99\% (20 / 143) of the analysed methods from the automatic analysis, where an \emph{instance based \ac{sdg} without exceptions} was created, were actually constructors. In our manual analysis this frequency is even higher, 28.57\% (10 / 35) of the analysed methods were actually constructors.

Finally, one could argue that using only a manual analysis to understand the limitations of information flow to estimate interference is also a threat, specially because a manual analysis is error-prone. Nevertheless, we argue that such analysis was necessary to properly understand the limitations of using information flow to estimate interference. Furthermore, most of the cases were reviewed by a second person. In particular, only examples with clear repetitions of categories previously discussed in other examples, were not reviewed.

\subsubsection{External}
Our evaluation involved open-source Java projects hosted on GitHub with build scripts from \emph{Ant}, \emph{Gradle} or \emph{Maven}. Thus, generalization to other languages and types of projects is limited, and further studies are necessary to confirm our findings. In particular, \ac{joana} only works for Java, so a different tool would be necessary for analysing different programming languages. However, it is not hard to believe that some of our findings may be true for other types of projects. For instance, we also expect existence of information flow between contributions for other programming languages.

Furthermore, one could argue that our sample size is limited as it involved only 157 scenarios. We agree that further studies, with larger samples, are indeed necessary. However, there is a series of limiting factors for increasing sample size. For instance, we are restricted to integrated revisions, with no merge conflicts identified by FSTMerge, that are successfully compiled. Moreover, even with all decisions we made to improve performance, \ac{joana} may take hours to create a \ac{sdg} for a revision or as previously shown may not even be able to successfully create it. As a matter of fact, this factor is enlarged by the fact that to select a configuration of \ac{sdg} to be used we run multiple configurations for each scenario. For instance, for 83 scenarios we executed 16 different configurations. Additionally, although the number of scenarios may be considered small, we tried to provide a degree of diversity by selecting those scenarios from 52 different projects with different sizes and from different domains.

\section{Related Work}
\label{related}
Previous work discuss and provide evidence about collaborative development issues. For example, Mens \cite{mens2002state} provides a comprehensive overview of the field of software-merge by comparing different techniques and discussing the advantages and limitations from each. Furthermore, in previous sections we have already mentioned studies with evidence of the occurrence of conflicts \cite{kasi2013cassandra,brun2013early}. Specifically, these studies analysed merge conflicts frequency (using a textual merge tool) and also static semantic conflicts (measuring build failures) and dynamic semantic conflicts (using existing tests) frequency. In particular, with respect to dynamic semantic conflicts these studies used the existing tests and found evidence of this type of conflict ranging from 3\% to 28\% \cite{kasi2013cassandra} and from 6\% to 35\% \cite{brun2013early}. It is important to mention that both studies analysed only a small list of projects, containing four and nine projects, respectively; for semantic conflicts the numbers are actually four and three. In addition, as previously mentioned, using existing tests to detect dynamic semantic conflicts is dependent on the test suite quality and coverage. Additionally, as the program evolves over time, tests need to be updated regularly to keep a good coverage of the source code. Furthermore, similarly to our strategy, we argue that tests actually detect interference, not dynamic semantic conflicts directly. More specifically, tests may also detect desired interference (as discussed in \cref{interference}). In addition, both studies may have imprecisions due to the fact that they only consider test failures in the integrated revision. To be more precise, a test failure in the integrated revision may be due to an interference, but also due to a previous existing failure. Finally, it is important to mention that using existing tests and running our information flow verification may be seen as complementary strategies to detect interference (and dynamic semantic conflicts).

Regarding semantic merge, Horwitz, Prins and Reps \cite{horwitz1989integrating} were the first to propose an algorithm for merging program versions without semantic conflicts for a very simple assignment-based programming language. This original work was later extended to handle procedure calls \cite{binkley1995program} and to identify semantics-preserving transformations \cite{yang1992program}. As previously mentioned, compared to our work, in theory, those approaches are more complete, as they detect more interference. However, that was a deliberate decision from our part, as those works are too computationally expensive to be used in practice, even for medium sized systems. As a matter of fact, as previously mentioned, to our knowledge, there is no semantic merge tool of such kind available, although some existent syntactic tools claim to be semantic. Thus, while these tools deal with four full program \acp{sdg}, we create a single \ac{sdg} and just for a few methods of the integrated revision (the ones edited by both contributions). Thus, with the goal of improving performance and making our approach practical, we miss some valid cases of interference, compromising soundness. Furthermore, as previously mentioned in \cref{results}, we expect such approaches to present similar issues as ours with respect to interference falses positives regarding the nature of changes.

%Still regarding the performance problems, \ac{sca} \cite{shao2007evaluation} may also be considered a lightweight approach. 
Other approaches \cite{jackson1994semantic,shao2007evaluation} have taken a similar path as ours towards a more lightweight and practical approach that may miss some interference cases.

For instance, \ac{sca} \cite{shao2007evaluation} does def-use data dependency analysis intraprocedurally and uses slicing techniques to identify structures impacted by a change. This work may be seen as more lightweight than ours as we use interprocedural analysis and also take in consideration the control flow of a program. With respect to false positives, as in our analysis, the evaluation of \ac{sca} showed a considerable number of false positives due to refactorings %(although they also consider dynamic semantic conflicts false positives through program faults and we only look for interference false positives)
. Specifically, 8 of the 19 cases of false positives found by them were classified as variable or type renaming. That conforms to our findings where the nature of changes were the first reason for false positives, although their rate of false positives due to refactorings was higher than ours (8 from 19 versus 4 from 21). However, they do not mention other problems, beyond refactorings, in the nature of changes category. In addition, \ac{sca} also obtained false positives due to no actual changes (named as false identification of changes in their work). More specifically, 3 of the 19 false positives encountered by them were in this category (versus 6 from 21 in ours). Additionally, they obtain a type of false positive discussed by us, but not measured in our evaluation. More precisely, they found that 5 of their 19 false positives were due to intentional modifications to fix faults. In particular, this is closely related to our discussion (from \cref{interference}) of \emph{desired} and \emph{undesired} interference. To be more precise, these false positives correspond to false positives due to \emph{desired} interference. Lastly, still regarding false positives, although they also consider dynamic semantic conflicts false positives (and we consider only interference false positives), this study obtained similar findings to ours. Moreover, their false positive rate was similar to ours. Particularly, they obtained 19 false positives from 29 cases detected. If we discard the five false positives of intentional modifications (which we do not measure), that would result in 14 false positives from 24 cases, and, result in a false positive rate of 58.33\% ($14 / 24 = 58.33\%$). Indeed, this is very close to our own false positive rate (57.14\%). Finally, the evaluation of \ac{sca} also brings evidence related to the consequences of their decision of not considering control-flow. To be more specific, 8 of their 17 false negatives were due to control-flow dependencies, which could potentially be identified by our strategy. Therefore, compared to ours, this study uses a more lightweight approach, which tends to increase their number of false negatives compared to ours. Furthermore, despite the differences between the approaches, the false positive rate obtained by both studies was very similar. Nonetheless, as different samples were used, and as both samples are of limited size, the comparison of the false positives rate is limited.

Finally, B{\"o}hme, Oliveira and Roychoudhury \cite{bohme2013regression} argue that some errors introduced in software evolution can only be identified by stressing the interaction between the introduced changes; these errors are called change interaction errors by them. That is, they argue that some errors are not identifiable by only testing the changes in isolation. Although in a testing context, this idea has some similarities with our idea of checking for the interaction between different contributions from a merge scenario. In particular, in our case, we check for information flow between left and right to try to detect these interactions between left and right. However, it is important to mention that the work in question focuses on identifying change interaction errors in simple evolution scenarios, not in merge scenarios. Nevertheless, adapting this work to specifically check for change interaction errors between contributions from a merge scenario seems to be a promising idea. 

\section{Conclusion}
\label{chap:conclusion}
In this work, we propose a strategy for checking information flow between contributions of a merge scenario, and use this information as an estimator of interference between those contributions. Our final goal is supporting dynamic semantic conflicts detection. However, as deciding if a dynamic semantic conflict exists involves understanding of the expected behaviour of a system, and because such behavioural specifications are often hard to capture, formalize and reason about, we instead try to detect interference. In addition, we restrict our scope to interference caused by same-method contributions.

Given \ac{joana} has different options of \ac{sdg}, first, we focused in establishing the most appropriate option to our context. We focused on two aspects: use of exceptions and pointer analysis. We conducted an automatic analysis (divided in iterations) comparing the different options to establish the most appropriate option. After iteration 1, we decided to use \acp{sdg} without exceptions, as \ac{joana} tends to find too many false positives using \acp{sdg} with exceptions. Furthermore, after iteration 3, we decided to use \emph{instance based} as our pointer analysis, as this option allows us to analyse more cases with little loss in precision. However, for those who wish more precision, we recommend using an \emph{object sensitive} analysis (or \emph{n1 object sensitive}).

Nonetheless, it is important to notice that object sensitive \acp{sdg} are not only successfully created for less scenarios, but also tend to consume more memory and take more time to be computed than instance based \acp{sdg}.

Besides establishing a \ac{sdg} option to use (instance based without exceptions), our automatic analysis also helped to answer if information flow between same-method contributions actually occurs. We do that by calculating the frequency of direct information flow between same-method contributions in scenarios that were successfully merged. Since we consider that scenarios with information flow between contributions are potential cases of interference, the higher the frequency, the higher the number of scenarios that are identified by our strategy. We found, in our sample, direct information flow between the contributions in around 64\% of the analysed scenarios. Therefore, this frequency indicates a considerable number of potential candidates of interference that would be identified by our strategy.

%Uirá - discutir alta taxa de não aplicabilidade da abordagem por não criar o SGD; quantos casos seriam viáveis com uma das configurações escolhidas 
However, it is important to mention that our strategy has limited use when the \ac{sdg} fails to be created. More specifically, we are only able to look for information flow for a specific scenario if a \ac{sdg} is successfully created for this scenario. In particular, using our selected configuration of \ac{sdg}, we were able to successfully create the \ac{sdg} for around 71\% of the scenarios. As previously discussed in \cref{it1} and detailed in \cref{fig:sdgFail}, we found two main reasons for a failure on \ac{sdg} creation: failure to locate a class on the class hierarchy and the \ac{sdg} in question is too big. In our evaluation from iteration 1, from the 17 cases where our selected configuration failed to create the \ac{sdg}, 10 cases were related to the first reason and 7 to the second. 

%*- Marcelo - quais seriam os ganhos com análise intra-procedural?
%Marcelo - considere reduzir escopo (profundidade) para reduzir custo.
The failures related to the first reason, failures to locate a class in the class hierarchy, are related to technical problems and may be solved by also including the missing class. On the other hand, if a \ac{sdg} fails to be created because is too big, then, the only remaining solution is to try to reduce the size of the \ac{sdg} in question. There are different possible options to do that. One possibility, is to do only intraprocedural analysis (instead of interprocedural). %That is, instead of considering method calls within a method, only the method itself is taken in consideration. 
Another one, is to not consider external and/or native dependencies. All these solutions would result in more incomplete analysis compared to the original one, but with potential to at least analyse part of these cases where the \ac{sdg} with the original configuration fails to be created. %For instance, using \emph{instance based} \acp{sdg} without exceptions, we noticed that for 7 scenarios it was possible to create the \ac{sdg} using stubs for \ac{jre} 1.4 and no external dependencies, while when using our final option regarding dependencies (with stubs for \ac{jre} 1.5 and external dependencies) the \ac{sdg} creation failed.

Lastly, we conducted a manual analysis to understand if the existence of information flow is a good estimator for the existence of interference and to find out characteristics of situations that have information flow but no interference. We concluded that existence of information flow between contributions is related to existence of interference and we expect part of the detected interference to be actually dynamic semantic conflicts. More precisely, we classified 42.86\% of the scenarios with information flow between contributions to be also interference. However, as previously discussed, we obtained a considerable number of false positives of interference (57.14\%). We found three major reasons for these false positives: cases related to the nature of changes (42.86\% of the false positives), cases related to limitations of our annotation strategy (33.33\% of the false positives) and cases related to conservativeness of the flows identified by \ac{joana} (23.8\% of the false positives).

Nevertheless, although we obtained a considerable amount of false positives of information flow as an estimator of interference, improvements to our strategy would potentially solve most of the false positives. In particular, a more advanced annotation system could solve annotation problems (33.33\% of the false positives). Additionally, using a less conservative \ac{ifc} tool could help to avoid (at least part of) cases of conservative flows (23.8\% of the false positives), but with the risk of increasing false negatives. Finally, a tool such as SafeRefactor could be used to check if changes are simply refactorings (19.05\% of the false positives), avoiding part of the problems related to the nature of changes. Therefore, there is room for solving, potentially with drawbacks in 23.8\%, around three quarters ($33.33\% + 19.05\% + 23.8\% = 76.18\%$) of the obtained false positives. 

Furthermore, as previously discussed in \cref{chap:eval}, we do not cover interference true/false negatives in our manual analysis. More specifically, we are aware that our strategy may miss valid cases of interference, however, we are interested in first understanding if information flow existence indicates interference existence.

%Marcelo - qual seria o workflow para usar a ferramenta? 
%motivar no final… rodar em backgorund
%Uirá - discutir possível aplicação na prática
%, questão do tempo e tal, alertas no dia seguinte
In practice, we envisage our strategy as a complementary verification in the merge process. That is, for each merge scenario integrated with no conflicts, our strategy could be automatically executed in a server to check for information flow. After the verification was completed, an email would be sent detailing if any information flow was identified between the contributions of the scenario and if so, detailing the obtained flows. Such strategy involves the use of a \emph{timeout} establishing the limit of time to try to create the \ac{sdg} (and consequently to receive the email). We used a timeout of 24 hours in our study, but this information could be configurable and changed depending of the context. In fact, we observed, in a sample of 100 scenarios, that only in 2 scenarios the \ac{sdg} was successfully created after 4 hours using our final configuration. Hence, one could decide to use this timeout instead of 24 hours. As a matter of fact, using our final configuration, for most of the scenarios the \ac{sdg} was created in a matter of minutes. %(the median was 1 minute and 5 seconds and the third quartile 8 minutes and 35 seconds).
Similarly, the limit of memory used could also be configurable. In our study, we used 120 GB as the limit. Nevertheless, that could be tailored to fit the amount of available memory. In particular, for our final configuration, we observed that around 20 GB would already cover most of the cases of successful creations from the 100 analysed scenarios. %(8.2 GB covered 50\%, 16.1 GB covered 75\% and 29 GB covered 90\%).

To conclude, we see our strategy as complementary to the use of system tests and code reviews to detect dynamic semantic conflicts. And, as an alternative to semantic merging. Specifically, we expect to detect part of the interference missed by a test-based strategy, but also miss some cases that could be identified using tests. In particular, compared to using tests, our strategy is generic in the sense that does not depend on the existence of tests or on their quality and coverage to check for interference. In contrast, a test suite needs constant maintenance to keep good coverage through software evolution. Nevertheless, we do not expect a strategy using tests to present as many issues with interference false positives as the proposed approach. More precisely, if a specific test presents a different result before and after an integration, in general, we expect it to be an interference. In contrast, as already discussed, our strategy has a number of issues with interference false positives. Additionally, our strategy has the limitation that it is not able to analyse a scenario if the associated \ac{sdg} is too heavy.

Similarly, we see our strategy as complementary to code review. In particular, our strategy may help to guide code review by showing possible points of interference when information flow is found. In summary, a flow between contributions is a possible point of interference and thus it should be closely investigated. Nonetheless, it is important to emphasize that, when our strategy does not detect information flow between contributions, it does not imply that there is no interference or dynamic semantic conflict. For instance, as previously discussed, interference involving removed lines is not identified by our strategy. Similarly, there is no information flow for cases of overwriting (mentioned in \cref{sameMethod}), but they may generate an interference. For example, imagine the situation discussed in \cref{sameMethod}, where left changes the value of a variable on a point 1 (of the integrated revision) used by a statement on point 3, while right modifies the value of the same variable on a point 2 (between left modification from point 1 and  the statement from point 3). There is interference in such a situation because right overwrites the value also modified by left affecting the value of the variable on point 3. However, as right overwrites left, there is no information flow from right to left, from left to right or from left and right to a common point. In fact, the effect of integrating left and right is actually ``killing'' the previously existing flow from point 1 to point 3 (on left revision), so, right overwrites left with respect to point 3. Hence, with respect to code review, our strategy may be useful to speed-up the process of detection of dynamic semantic conflicts, and even to detect some conflicts that could potentially pass unnoticed. Nevertheless, we provide no guarantees: in addition to false positives, our strategy may also produce false negatives.

Finally, with respect to semantic merging, our strategy tends to have both more false positives and negatives, specially the latter. However, to achieve such accuracy, semantic merging tends to be too heavy to be used in practice. Therefore, although we may have more false positives and negatives, our approach focuses on improving performance towards obtaining a practical approach. Specifically, we deal with a single \ac{sdg}, versus four in the original work of Horwitz, Prins and Reps \cite{horwitz1989integrating}. Furthermore, while Horwitz, Prins and Reps create full program \acp{sdg}, our \ac{sdg} involves only methods in the call graph of methods edited by both contributions. Thus, with the goal of improving performance, our strategy is not sound.

\uppercase{\section{Important links}}
\label{links}
	
	In this appendix we provide important links involving this work. In particular, we provide links to our source code, and to plots and files generated during our evaluation.
	\begin{itemize}
		\item GitHub repository with source code and necessary scripts: \url{https://github.com/rsmbf/joana}
		\item Information from evaluation including resultant plots, files and details of the data: \url{https://drive.google.com/drive/folders/0B1QVNk49Q0GbOGszMnJtUVlRUnc}  
	\end{itemize}

\bibliographystyle{ACM-Reference-Format}
\bibliography{sample-base}


\begin{thebibliography}{26}

%%% the Web address.

\ifx \showCODEN    \undefined \def \showCODEN     #1{\unskip}     \fi
\ifx \showDOI      \undefined \def \showDOI       #1{#1}\fi
\ifx \showISBNx    \undefined \def \showISBNx     #1{\unskip}     \fi
\ifx \showISBNxiii \undefined \def \showISBNxiii  #1{\unskip}     \fi
\ifx \showISSN     \undefined \def \showISSN      #1{\unskip}     \fi
\ifx \showLCCN     \undefined \def \showLCCN      #1{\unskip}     \fi
\ifx \shownote     \undefined \def \shownote      #1{#1}          \fi
\ifx \showarticletitle \undefined \def \showarticletitle #1{#1}   \fi
\ifx \showURL      \undefined \def \showURL       {\relax}        \fi
% invisible to TeX
\providecommand\bibfield[2]{#2}
\providecommand\bibinfo[2]{#2}
\providecommand\natexlab[1]{#1}
\providecommand\showeprint[2][]{arXiv:#2}

\bibitem[Accioly(2017)]%
        {paolaPhd}
\bibfield{author}{\bibinfo{person}{Paola Accioly}.}
  \bibinfo{year}{2017}\natexlab{}.
\newblock \emph{\bibinfo{title}{Understanding Collaboration Conflicts
  Characteristics}}.
\newblock \bibinfo{thesistype}{Ph.\,D. Dissertation}.
  \bibinfo{school}{Universidade Federal de Pernambuco}.
\newblock


\bibitem[Accioly et~al\mbox{.}(2017)]%
        {Accioly2017}
\bibfield{author}{\bibinfo{person}{Paola Accioly}, \bibinfo{person}{Paulo
  Borba}, {and} \bibinfo{person}{Guilherme Cavalcanti}.}
  \bibinfo{year}{2017}\natexlab{}.
\newblock \showarticletitle{Understanding semi-structured merge conflict
  characteristics in open-source Java projects}.
\newblock \bibinfo{journal}{\emph{Empirical Software Engineering}}
  (\bibinfo{date}{21 Dec} \bibinfo{year}{2017}).
\newblock
\showISSN{1573-7616}
\urldef\tempurl%
\url{https://doi.org/10.1007/s10664-017-9586-1}
\showDOI{\tempurl}


\bibitem[Apel et~al\mbox{.}(2011)]%
        {apel2011semistructured}
\bibfield{author}{\bibinfo{person}{Sven Apel}, \bibinfo{person}{J{\"o}rg
  Liebig}, \bibinfo{person}{Benjamin Brandl}, \bibinfo{person}{Christian
  Lengauer}, {and} \bibinfo{person}{Christian K{\"a}stner}.}
  \bibinfo{year}{2011}\natexlab{}.
\newblock \showarticletitle{Semistructured merge: rethinking merge in revision
  control systems}.
\newblock \bibinfo{journal}{\emph{Proceedings of the 19th ACM SIGSOFT symposium
  and the 13th European conference on Foundations of software engineering}}
  (\bibinfo{year}{2011}), \bibinfo{pages}{190--200}.
\newblock


\bibitem[Binkley et~al\mbox{.}(1995)]%
        {binkley1995program}
\bibfield{author}{\bibinfo{person}{David Binkley}, \bibinfo{person}{Susan
  Horwitz}, {and} \bibinfo{person}{Thomas Reps}.}
  \bibinfo{year}{1995}\natexlab{}.
\newblock \showarticletitle{Program integration for languages with procedure
  calls}.
\newblock \bibinfo{journal}{\emph{ACM Transactions on Software Engineering and
  Methodology (TOSEM)}} \bibinfo{volume}{4}, \bibinfo{number}{1}
  (\bibinfo{year}{1995}), \bibinfo{pages}{3--35}.
\newblock


\bibitem[B{\"o}hme et~al\mbox{.}(2013)]%
        {bohme2013regression}
\bibfield{author}{\bibinfo{person}{Marcel B{\"o}hme}, \bibinfo{person}{Bruno C
  d~S Oliveira}, {and} \bibinfo{person}{Abhik Roychoudhury}.}
  \bibinfo{year}{2013}\natexlab{}.
\newblock \showarticletitle{Regression tests to expose change interaction
  errors}.
\newblock \bibinfo{journal}{\emph{Proceedings of the 2013 9th Joint Meeting on
  Foundations of Software Engineering}} (\bibinfo{year}{2013}),
  \bibinfo{pages}{334--344}.
\newblock


\bibitem[Brun et~al\mbox{.}(2013)]%
        {brun2013early}
\bibfield{author}{\bibinfo{person}{Yuriy Brun}, \bibinfo{person}{Reid Holmes},
  \bibinfo{person}{Michael~D Ernst}, {and} \bibinfo{person}{David Notkin}.}
  \bibinfo{year}{2013}\natexlab{}.
\newblock \showarticletitle{Early detection of collaboration conflicts and
  risks}.
\newblock \bibinfo{journal}{\emph{IEEE Transactions on Software Engineering}}
  \bibinfo{volume}{39}, \bibinfo{number}{10} (\bibinfo{year}{2013}),
  \bibinfo{pages}{1358--1375}.
\newblock


\bibitem[CAVALCANTI(2016)]%
        {cavalcanticomparing}
\bibfield{author}{\bibinfo{person}{Guilherme Jos{\'e}~Carvalho CAVALCANTI}.}
  \bibinfo{year}{2016}\natexlab{}.
\newblock \showarticletitle{Comparing integration effort and correctness of
  Different merge approaches in version control systems}.
\newblock \bibinfo{journal}{\emph{Master's Dissertation}}
  (\bibinfo{year}{2016}).
\newblock


\bibitem[Fowler and Beck(1999)]%
        {fowler1999refactoring}
\bibfield{author}{\bibinfo{person}{Martin Fowler} {and} \bibinfo{person}{Kent
  Beck}.} \bibinfo{year}{1999}\natexlab{}.
\newblock \bibinfo{booktitle}{\emph{Refactoring: improving the design of
  existing code}}.
\newblock \bibinfo{publisher}{Addison-Wesley Professional}.
\newblock


\bibitem[Goguen and Meseguer(1982)]%
        {goguen1982security}
\bibfield{author}{\bibinfo{person}{Joseph~A Goguen} {and}
  \bibinfo{person}{Jos{\'e} Meseguer}.} \bibinfo{year}{1982}\natexlab{}.
\newblock \showarticletitle{Security policies and security models}.
\newblock \bibinfo{journal}{\emph{IEEE Symposium on Security and Privacy}}
  \bibinfo{volume}{11} (\bibinfo{year}{1982}), \bibinfo{pages}{77}.
\newblock


\bibitem[Graf et~al\mbox{.}(2013)]%
        {graf2013using}
\bibfield{author}{\bibinfo{person}{J{\"u}rgen Graf}, \bibinfo{person}{Martin
  Hecker}, {and} \bibinfo{person}{Martin Mohr}.}
  \bibinfo{year}{2013}\natexlab{}.
\newblock \showarticletitle{Using JOANA for Information Flow Control in Java
  Programs-A Practical Guide.}
\newblock \bibinfo{journal}{\emph{Software Engineering (Workshops)}}
  \bibinfo{volume}{215} (\bibinfo{year}{2013}), \bibinfo{pages}{123--138}.
\newblock


\bibitem[Graf et~al\mbox{.}(2015)]%
        {graf2015checking}
\bibfield{author}{\bibinfo{person}{J{\"u}rgen Graf}, \bibinfo{person}{Martin
  Hecker}, \bibinfo{person}{Martin Mohr}, {and} \bibinfo{person}{Gregor
  Snelting}.} \bibinfo{year}{2015}\natexlab{}.
\newblock \showarticletitle{Checking applications using security APIs with
  JOANA}.
\newblock \bibinfo{journal}{\emph{8th International Workshop on Analysis of
  Security APIs}} (\bibinfo{year}{2015}).
\newblock


\bibitem[Grove and Chambers(2001)]%
        {grove2001framework}
\bibfield{author}{\bibinfo{person}{David Grove} {and} \bibinfo{person}{Craig
  Chambers}.} \bibinfo{year}{2001}\natexlab{}.
\newblock \showarticletitle{A framework for call graph construction
  algorithms}.
\newblock \bibinfo{journal}{\emph{ACM Transactions on Programming Languages and
  Systems (TOPLAS)}} \bibinfo{volume}{23}, \bibinfo{number}{6}
  (\bibinfo{year}{2001}), \bibinfo{pages}{685--746}.
\newblock


\bibitem[Hammer and Snelting(2009)]%
        {hammer2009flow}
\bibfield{author}{\bibinfo{person}{Christian Hammer} {and}
  \bibinfo{person}{Gregor Snelting}.} \bibinfo{year}{2009}\natexlab{}.
\newblock \showarticletitle{Flow-sensitive, context-sensitive, and
  object-sensitive information flow control based on program dependence
  graphs}.
\newblock \bibinfo{journal}{\emph{International Journal of Information
  Security}} \bibinfo{volume}{8}, \bibinfo{number}{6} (\bibinfo{year}{2009}),
  \bibinfo{pages}{399--422}.
\newblock


\bibitem[Horwitz et~al\mbox{.}(1989)]%
        {horwitz1989integrating}
\bibfield{author}{\bibinfo{person}{Susan Horwitz}, \bibinfo{person}{Jan Prins},
  {and} \bibinfo{person}{Thomas Reps}.} \bibinfo{year}{1989}\natexlab{}.
\newblock \showarticletitle{Integrating noninterfering versions of programs}.
\newblock \bibinfo{journal}{\emph{ACM Transactions on Programming Languages and
  Systems (TOPLAS)}} \bibinfo{volume}{11}, \bibinfo{number}{3}
  (\bibinfo{year}{1989}), \bibinfo{pages}{345--387}.
\newblock


\bibitem[Jackson and Ladd(1994)]%
        {jackson1994semantic}
\bibfield{author}{\bibinfo{person}{Daniel Jackson} {and}
  \bibinfo{person}{David~A Ladd}.} \bibinfo{year}{1994}\natexlab{}.
\newblock \showarticletitle{Semantic Diff: A Tool for Summarizing the Effects
  of Modifications.}
\newblock \bibinfo{journal}{\emph{ICSM}}  \bibinfo{volume}{94}
  (\bibinfo{year}{1994}), \bibinfo{pages}{243--252}.
\newblock


\bibitem[Kasi and Sarma(2013)]%
        {kasi2013cassandra}
\bibfield{author}{\bibinfo{person}{Bakhtiar~Khan Kasi} {and}
  \bibinfo{person}{Abhijit Sarma}.} \bibinfo{year}{2013}\natexlab{}.
\newblock \showarticletitle{Cassandra: Proactive conflict minimization through
  optimized task scheduling}.
\newblock \bibinfo{journal}{\emph{35th International Conference on Software
  Engineering (ICSE)}} (\bibinfo{year}{2013}), \bibinfo{pages}{732--741}.
\newblock


\bibitem[Mens(2002)]%
        {mens2002state}
\bibfield{author}{\bibinfo{person}{Tom Mens}.} \bibinfo{year}{2002}\natexlab{}.
\newblock \showarticletitle{A state-of-the-art survey on software merging}.
\newblock \bibinfo{journal}{\emph{IEEE transactions on software engineering}}
  \bibinfo{volume}{28}, \bibinfo{number}{5} (\bibinfo{year}{2002}),
  \bibinfo{pages}{449--462}.
\newblock


\bibitem[Nielson et~al\mbox{.}(2005)]%
        {nielson2005principles}
\bibfield{author}{\bibinfo{person}{Flemming Nielson}, \bibinfo{person}{Hanne~R
  Nielson}, {and} \bibinfo{person}{Chris Hankin}.}
  \bibinfo{year}{2005}\natexlab{}.
\newblock \bibinfo{booktitle}{\emph{Principles of program analysis}}.
\newblock \bibinfo{publisher}{Springer}.
\newblock


\bibitem[Ryder(1979)]%
        {ryder1979constructing}
\bibfield{author}{\bibinfo{person}{Barbara~G Ryder}.}
  \bibinfo{year}{1979}\natexlab{}.
\newblock \showarticletitle{Constructing the call graph of a program}.
\newblock \bibinfo{journal}{\emph{IEEE Transactions on Software Engineering}}
  \bibinfo{volume}{3}, \bibinfo{number}{3} (\bibinfo{year}{1979}),
  \bibinfo{pages}{216--226}.
\newblock


\bibitem[Shao et~al\mbox{.}(2007)]%
        {shao2007evaluation}
\bibfield{author}{\bibinfo{person}{Danhua Shao}, \bibinfo{person}{Sarfraz
  Khurshid}, {and} \bibinfo{person}{Dewayne~E Perry}.}
  \bibinfo{year}{2007}\natexlab{}.
\newblock \showarticletitle{Evaluation of semantic interference detection in
  parallel changes: an exploratory experiment}.
\newblock \bibinfo{journal}{\emph{IEEE International Conference on Software
  Maintenance (ICSM)}} (\bibinfo{year}{2007}), \bibinfo{pages}{74--83}.
\newblock


\bibitem[Shapiro and Wilk(1965)]%
        {shapiro1965analysis}
\bibfield{author}{\bibinfo{person}{Samuel~Sanford Shapiro} {and}
  \bibinfo{person}{Martin~B Wilk}.} \bibinfo{year}{1965}\natexlab{}.
\newblock \showarticletitle{An analysis of variance test for normality
  (complete samples)}.
\newblock \bibinfo{journal}{\emph{Biometrika}} \bibinfo{volume}{52},
  \bibinfo{number}{3-4} (\bibinfo{year}{1965}), \bibinfo{pages}{591--611}.
\newblock


\bibitem[Smaragdakis et~al\mbox{.}(2015)]%
        {smaragdakis2015pointer}
\bibfield{author}{\bibinfo{person}{Yannis Smaragdakis}, \bibinfo{person}{George
  Balatsouras}, {et~al\mbox{.}}} \bibinfo{year}{2015}\natexlab{}.
\newblock \showarticletitle{Pointer analysis}.
\newblock \bibinfo{journal}{\emph{Foundations and Trends{\textregistered} in
  Programming Languages}} \bibinfo{volume}{2}, \bibinfo{number}{1}
  (\bibinfo{year}{2015}), \bibinfo{pages}{1--69}.
\newblock


\bibitem[Snelting et~al\mbox{.}(2014)]%
        {joana14it}
\bibfield{author}{\bibinfo{person}{Gregor Snelting}, \bibinfo{person}{Dennis
  Giffhorn}, \bibinfo{person}{J{\"u}rgen Graf}, \bibinfo{person}{Christian
  Hammer}, \bibinfo{person}{Martin Hecker}, \bibinfo{person}{Martin Mohr},
  {and} \bibinfo{person}{Daniel Wasserrab}.} \bibinfo{year}{2014}\natexlab{}.
\newblock \showarticletitle{Checking Probabilistic Noninterference Using
  JOANA}.
\newblock \bibinfo{journal}{\emph{Information Technology (it)}}
  \bibinfo{volume}{56} (\bibinfo{date}{Nov.} \bibinfo{year}{2014}),
  \bibinfo{pages}{280--287}.
\newblock
\urldef\tempurl%
\url{https://doi.org/10.1515/itit-2014-1051}
\showDOI{\tempurl}


\bibitem[Soares et~al\mbox{.}(2010)]%
        {soares2010making}
\bibfield{author}{\bibinfo{person}{Gustavo Soares}, \bibinfo{person}{Rohit
  Gheyi}, \bibinfo{person}{Dalton Serey}, {and} \bibinfo{person}{Tiago
  Massoni}.} \bibinfo{year}{2010}\natexlab{}.
\newblock \showarticletitle{Making program refactoring safer}.
\newblock \bibinfo{journal}{\emph{IEEE software}} \bibinfo{volume}{27},
  \bibinfo{number}{4} (\bibinfo{year}{2010}), \bibinfo{pages}{52--57}.
\newblock


\bibitem[Wilcoxon(1945)]%
        {wilcoxon1945individual}
\bibfield{author}{\bibinfo{person}{Frank Wilcoxon}.}
  \bibinfo{year}{1945}\natexlab{}.
\newblock \showarticletitle{Individual comparisons by ranking methods}.
\newblock \bibinfo{journal}{\emph{Biometrics bulletin}} \bibinfo{volume}{1},
  \bibinfo{number}{6} (\bibinfo{year}{1945}), \bibinfo{pages}{80--83}.
\newblock


\bibitem[Yang et~al\mbox{.}(1992)]%
        {yang1992program}
\bibfield{author}{\bibinfo{person}{Wuu Yang}, \bibinfo{person}{Susan Horwitz},
  {and} \bibinfo{person}{Thomas Reps}.} \bibinfo{year}{1992}\natexlab{}.
\newblock \showarticletitle{A program integration algorithm that accommodates
  semantics-preserving transformations}.
\newblock \bibinfo{journal}{\emph{ACM Transactions on Software Engineering and
  Methodology (TOSEM)}} \bibinfo{volume}{1}, \bibinfo{number}{3}
  (\bibinfo{year}{1992}), \bibinfo{pages}{310--354}.
\newblock


\end{thebibliography}

\appendix

\end{document}